\documentclass[fleqn,usenatbib]{mnras}
\pdfoutput=1

\usepackage{newtxtext,newtxmath}
\usepackage[T1]{fontenc}
\usepackage{ae,aecompl}
\usepackage{graphicx}	
\usepackage{amsmath}	
\usepackage{amssymb}	

\usepackage{multicol}

\graphicspath{ {images/} }

\title[Automated asteroseismic peak detections]{Automated asteroseismic peak detections}

\author[A. Garc\'{i}a Saravia Ortiz de Montellano et al.]{
A. Garc\'{i}a Saravia Ortiz de Montellano$^{1,2}$\thanks{E-mail: ags3006@gmail.com},
S. Hekker$^{1,2}$,
N. Theme{\ss}l$^{1,2}$
\\
$^{1}$Max-Planck-Institut f\"{u}r Sonnensystemforschung, Justus-von-Liebig-Weg 3, 37077 G\"{o}ttingen, Germany\\
$^{2}$Stellar Astrophysics Centre, Department of Physics and Astronomy, Aarhus University, Ny Munkegade 120, DK-8000 Aarhus C, Denmark
}

\date{Accepted XXX. Received YYY; in original form ZZZ}
\pubyear{2017}

\begin{document}
\label{firstpage}
\pagerange{\pageref{firstpage}--\pageref{lastpage}}
\maketitle


\begin{abstract}
Space observatories such as \textit{Kepler} have provided data that can potentially revolutionise our understanding of stars.
Through detailed asteroseismic analyses we are capable of determining fundamental stellar parameters and reveal the stellar internal structure with unprecedented accuracy.
However, such detailed analyses, known as peak bagging, have so far been obtained for only a small percentage of the observed stars while most of the scientific potential of the available data remains unexplored.
One of the major challenges in peak bagging is identifying how many solar-like oscillation modes are visible in a power density spectrum.
Identification of oscillation modes is usually done by visual inspection which is time-consuming and has a degree of subjectivity.
Here, we present a peak detection algorithm specially suited for the detection of solar-like oscillations.
It reliably characterises the solar-like oscillations in a power density spectrum and estimates their parameters without human intervention.
Furthermore, we provide a metric to characterise the false positive and false negative rates to provide further information about the reliability of a detected oscillation mode or the significance of a lack of detected oscillation modes.
The algorithm presented here opens the possibility for detailed and automated peak bagging of the thousands of solar-like oscillators observed by \textit{Kepler}.
\end{abstract}



\section{Introduction}

With the advent of space observatories CoRoT \citep{Baglin2006} and \textit{Kepler} \citep{Borucki2010} there are high-quality, near-uninterrupted and long ($>100$ days) photometric time series  measurements for an unprecedented number of stars available.
For solar-like oscillators the power density spectrum (PDS) of time series data contains rich information that allows for a precise determination of fundamental stellar properties provided that individual oscillation mode parameters are measured accurately \citep{Christensen-Dalsgaard2004}.
However, most of the currently existing methods to analyse in detail the PDS require substantial human intervention.
For this reason the scientific potential of the large amount of available data has not been fully exploited.
Furthermore, since current methods rely on human input they carry a considerable degree of subjectivity.
Therefore, a method to reliably extract all the relevant information from the measured PDS in an automated way is greatly needed.
This issue will be fundamental to take advantage of upcoming missions such as TESS \citep{Ricker2014} and PLATO \citep{Rauer2014}.
It is expected that TESS and PLATO will detect approximately $3\times 10^5$ and $2\times 10^5$ stars with solar-like oscillations \citep{Huber2018}, respectively.
This will greatly increase the amount of available data.

The power density spectra of solar-like oscillators have a complex structure stemming from a combination of the granulation background and the stochastically excited global oscillation modes.
This makes it challenging to model the PDS accurately.
The functional form of the granulation background component has been the topic of several studies \citep[e.g.][]{Harvey1985, Michel2009, Kallinger2014} and it takes the form of a superposition of super-Lorentzian profiles to take into account granulation at different time scales.
Additionally, each individual global stochastic oscillation mode in the power excess is well described by a Lorentzian profile, in case the mode width is larger than the frequency resolution \citep{Kumar1988}, or by a $\mathrm{sinc^2}$ function otherwise \citep{Christensen-Dalsgaard2004}.
One of the main challenges in modelling the stochastic oscillations resides in identifying how many oscillation modes are visible in a given PDS realisation. 
Furthermore, in practice, it is desirable to have at least a crude estimate of all the parameters describing each oscillation mode.
For example in a Maximum Likelihood Estimation (MLE) having a set of adequate initial values for all parameters is necessary to avoid local minima \citep{Toutain1994}.
The Bayesian estimations using the Markov-Chain Monte-Carlo methods require an adequate prior probability distribution for each parameter which can be constructed from the initial estimates \citep{Handberg2011}. 
These initial estimates can be obtained by visual inspection of the PDS since the oscillation modes produce peaks with recognisable patterns in the PDS.
However, this method is not scalable to the large number of stars observed by space missions.
Additionally, a visual inspection inevitably introduces a degree of subjectivity tied to the person doing the analysis. 
This subjectivity can partially be mitigated by assessing the statistical significance of each oscillation mode found in the PDS which prevents the inclusion of non-significant peaks in the PDS model.
However, there might still be significant peaks that escape the visual inspection and are never tested for inclusion.
Therefore, a reliable peak-detection method that is free from significant human input would greatly benefit the analysis of large samples of stars.

A number of peak detection algorithms have been developed over the last decades. 
A common approach is to search for local maxima with a signal-to-noise ratio (SNR) above a certain threshold with the SNR depending only on the peak height \citep[e.g.][]{Appourchaux2012}. 
A major problem in this approach is that noise can have larger heights than some peaks caused by oscillations, this could produce a large number of either false positives or false negatives depending on the chosen SNR threshold. 
Furthermore, tests on the peak height would miss significant peaks that are wide but have a small height.
Since most peaks in the PDS of solar-like oscillators have a width larger than the frequency resolution, this issue can be partially mitigated by smoothing before attempting a peak detection.
However this approach is very sensitive to the width of the mode (i.e. its lifetime) and the amount of smoothing applied; peaks with different widths are more prominent with different amounts of smoothing.
So with this approach it is impossible to choose a unique best strategy to correctly identify all features of interest.

In the current work we present a different approach based on an extension of the peak-detection algorithm proposed by \citet{Du2006} in the context of mass spectrometry. 
This algorithm uses a continuous wavelet transform (CWT)-based pattern-matching algorithm where there is no need for smoothing and most features are correctly identified while keeping the false positive rate low \citep{Cruz-Marcelo2008}. 
The CWT serves as a pattern-matching function where the signal is compared to a wavelet function specifically chosen to have similar features as the most common peaks that contain signal. 
The CWT has two parameters, location and scale, that regulate the position on which the matching is being calculated and the width of the feature being matched, respectively. 
The CWT has a similar effect as a smoothing where the amount of smoothing is variable and controlled by the scale parameter. 
This approach is similar to searching for narrow features with little smoothing and wide features with a larger amount of smoothing simultaneously.

Solar-like oscillators produce peaks in their PDS that are similar to the peaks studied by \citet{Du2006} which makes the CWT-based pattern-matching algorithm adequate for this context. 
We find that the original formulation of the algorithm is reliable when the peaks are well separated, however it fails to correctly identify them when the modes have significant overlap, which is a common scenario in the PDS of solar-like oscillators. 
Furthermore, the original algorithm by \citet{Du2006} only estimates the location of the peak and not its height and width. 
We propose an extension which is more tolerant to peak overlap and also estimates the height and width of the peaks.
To achieve these improvements we incorporate the additional assumption that the individual peaks are described by Lorentzian functions, as is the case for solar-like oscillations \citep{Anderson1990}.
Additionally, we require that the PDS contains only information from the stochastically excited oscillation modes, i.e. that the PDS has been normalised to remove the granulation contributions.

In Section \ref{sec:PeakDetection} we describe the algorithm by \citet{Du2006} and the adaptations we made for the case of analysing the PDS of solar-like oscillators.
We then characterise the performance of the proposed algorithm by looking at the number false positive (Section \ref{sec:EFP}) and false negative (Section \ref{sec:DetProb}) peak detections.
In Section \ref{sec:BiSON} we compare our estimation with previous studies of helioseismic data from the Birmingham Solar-Oscillations Network (BiSON) \citep{Davies2014,Hale2016} and in Section \ref{sec:Kepler} we compare our results as obtained from photometric measurements made by the \textit{Kepler} space observatory with the results from \citet{Corsaro2015} for 19 red-giants.
In each case we find similar values for the number of peaks present in the background-normalised PDS as well as their location, height and width albeit with some minor discrepancies.
In contrast to previous approaches, the procedure presented here requires no human intervention which makes it more objective and suitable for the analysis of present and upcoming large data sets.\footnote{The source code implementing the algorithm presented here will be publicly available in the future as part of an automated peak-bagging pipeline.}

To study the internal structure of solar-like oscillators, the frequency of the oscillations are not sufficient.
We also need to characterise each mode by its spherical degree and azimutal order.
An automated procedure to obtain this mode identification will be presented in a forthcoming paper.

\section{Peak detection method}\label{sec:PeakDetection}\label{sec:peakMethod}

\begin{figure}
\includegraphics[width=0.47\textwidth]{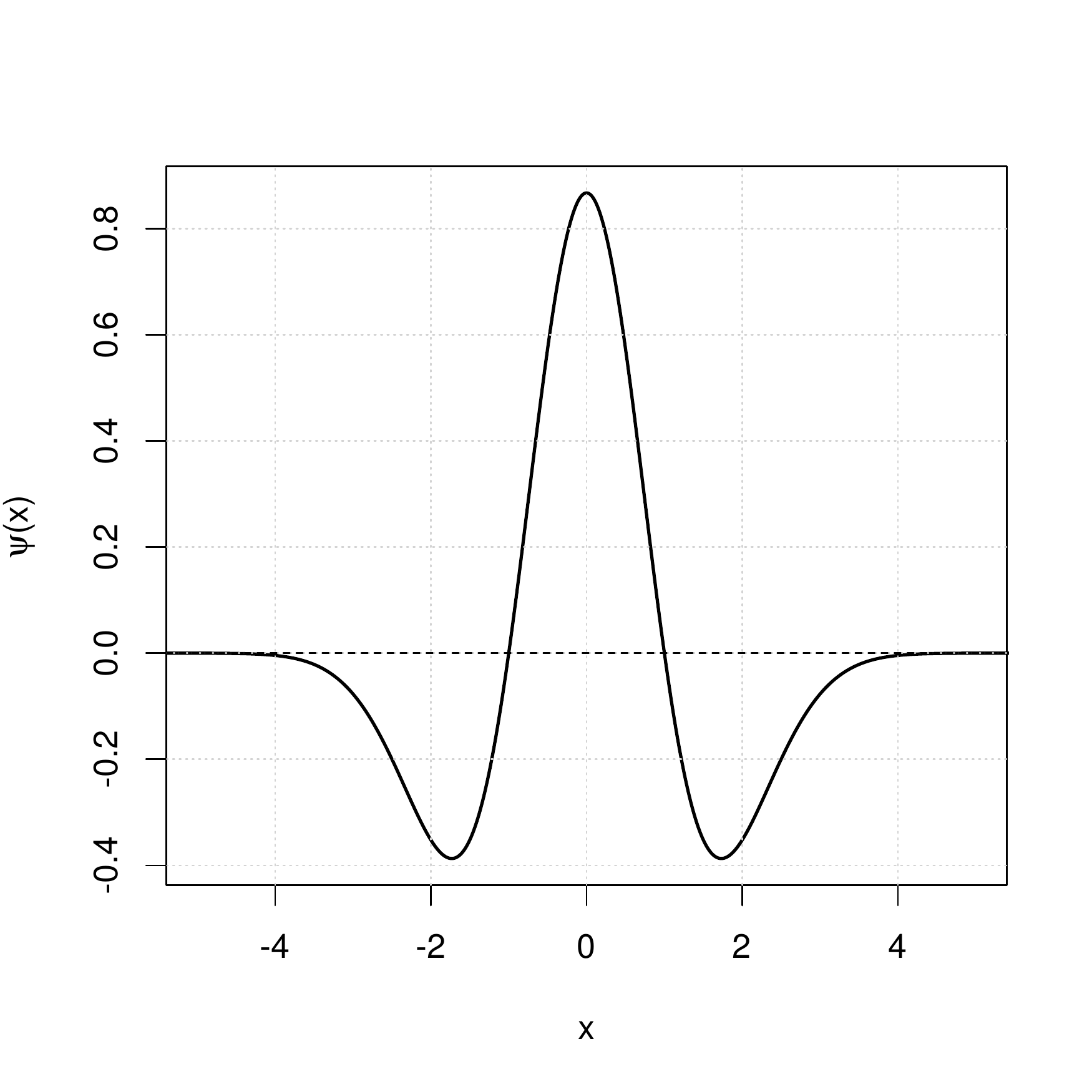}
\caption{Mexican-hat mother wavelet as defined by (\ref{eq:mh-wav}) with $\sigma=1$.}
\label{fig:mhWavelet}
\end{figure}

The peak detection method proposed by \citet{Du2006} is based on using the continuous wavelet transform (CWT) of a signal as a pattern-matching function. 
The CWT of a function $s(x)$, denoted as $\mathcal{W}_\psi[s]$, is an integral transform that depends on two parameters, $(a,b)$, usually referred to as scale and location, respectively. 
It is defined as
\begin{equation}\label{eq:CWTdef}
\mathcal{W}_\psi[s]\ (a,b) = \frac{1}{\sqrt{a}} \int_{-\infty}^\infty s(x')\ \overline{\psi} \left( \frac{x'-b}{a} \right) dx'\ ,
\end{equation}
where $\psi(x)$ is a continuous function called the mother wavelet and the overline denotes complex conjugation. 
Intuitively, the CWT value reflects the pattern matching between the signal $s(x')$ and $\overline{\psi} \left( \frac{x'-b}{a} \right)$ with larger values representing a better match. 
The mother wavelet is intentionally chosen to have properties useful for a specific analysis. 
An appropriate mother wavelet for peak detection is the mexican-hat wavelet, also know as Ricker wavelet \citep{Ricker1944}, given by
\begin{equation}\label{eq:mh-wav}
\psi(x)=\frac{2}{\sqrt{3\sigma}\pi^{1/4}}\left(1 - \frac{x^2}{\sigma^2}\right)e^{-x^2/2\sigma^2}\ .
\end{equation}
It is the second derivative of a Gaussian function with variance $\sigma^2$ and a normalization factor such that $\int_\infty^\infty |\psi(x')|^2 dx'=1$ (see Fig. \ref{fig:mhWavelet}).
It was shown by \citet{Du2006} that this wavelet is useful for peak finding in spectra since it has the basic features of the most common peaks in a spectrum: approximate symmetry, a major positive peak and finite width.

\subsection{Single peak}\label{sec:singlePeak}

\begin{figure}
\includegraphics[width=0.47\textwidth]{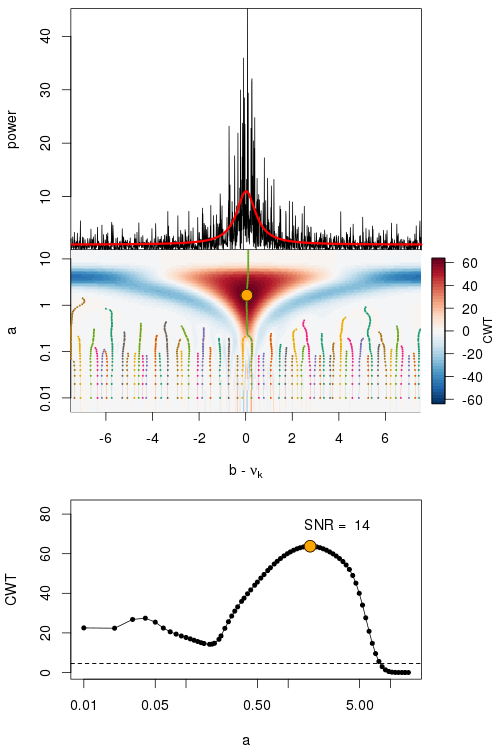}
\caption{\textit{Top panel}: Simulated PDS (black) realisation for a limit PDS (red) described by a Lorentzian profile with a half width at half maximum $\gamma_k=0.5$ and height $I_k=10$ (arbitrary units). \textit{Middle panel}: Mexican-hat CWT of the simulated PDS realisation in the top panel as a colour map with values indicated in the side bar. The small dots are at the local maxima across each scale ($a$) with nearby points of the same colour being identified as belonging to the same ridge. The larger orange dot indicated the maximum CWT value for the longest ridge. \textit{Bottom panel}: CWT as a function of scale ($a$) for points in the longest ridge from the middle panel. The horizontal dashed line is placed at a SNR of 1 as defined by \citet{Du2006}. The orange dot is the maximum of the ridge which has a SNR close to 14.}
\label{fig:singlePeakCWT}
\end{figure}

We describe the peak detection method by \cite{Du2006} by illustrating it on a PDS realisation originating from a single stochastically-excited global oscillation mode. 
We assume that there is no contribution to the PDS other than the oscillation mode, i.e. we have a background-normalised PDS. 
The limit PDS in this case can be described by a Lorentzian profile.
As an example we consider a Lorentzian profile with a central frequency $\nu_k$, a height of $I_k=10$ and a half-width at half-maximum of $\gamma_k=0.5$, all parameters are given in arbitrary units. 
The frequency resolution is taken as $\delta\nu=0.01$ also in arbitrary units.
We construct a possible PDS realisation by multiplying each frequency bin by a random number drawn from an exponentially decaying probability distribution (see top panel of Fig. \ref{fig:singlePeakCWT}).
This corresponds to a scaled $\chi^2$ distribution with two degrees of freedom which is the probability density function of noise in a PDS.
This example PDS is similar to the PDS of a single stochastically excited and damped oscillation mode.
We then compute the CWT of the PDS using a mexican-hat wavelet as the mother wavelet with scale values ranging from $\delta\nu$ to the frequency range of the PDS (see middle panel of Fig. \ref{fig:singlePeakCWT}).
Finally we look at all the local maxima in the CWT values as a function of location ($b$) for each scale ($a$).
At small scales the CWT is sensitive to narrow features, which can be seen by the large number of local maxima at small values of $a$. 
Towards larger scales the CWT map becomes smoother and more sensitive to wider features.

\citet{Du2006} noted that the local maxima can be connected in ridges. 
The ridges produced by noise are short and have low absolute CWT values while peaks with a resolved width produce longer ridges with larger CWT values. 
Thus, peaks are identified by looking at ridges longer than a certain threshold which have a maximum CWT value larger than a chosen signal-to-noise ratio (SNR). 
To take into account the possibility of a different baseline in each peak, \citet{Du2006} proposed a SNR definition that is local for each ridge.
It is defined as the maximum CWT value on the ridge divided by the 95 percentile of the absolute value of all the CWT values at the smallest scale in a frequency range close to the ridge location.
The optimal frequency range for this local noise definition depends on the problem.
Here we used the implementation by \citet{wmtsa} in which the range for the local noise is ten times the frequency resolution (see bottom panel of Fig. \ref{fig:singlePeakCWT}).

As already mentioned, the algorithm by \citet{Du2006} was proposed in the context of mass spectrometry with no assumption of a particular peak shape other than being positive, approximately symmetric and with a finite width.
For the particular case of finding peaks in the background-normalised PDS of solar-like oscillators we can assume that resolved oscillation modes can be modelled by Lorentzian profiles. 
With this assumption we can find an approximate relationship between the three parameters of the limit PDS Lorentzian $(\nu_k, I_k, \gamma_k)$ and the value of the CWT maximum $\mathcal{W}_{\max}$, its scale $a_{\max}$ and location $b_{\max})$. 
By making 5000 simulations of this single-peak scenario with representative values, in arbitrary units, of $I_k\in(10,200)$ and $\gamma_k\in(5,30)$ we derived the following empirical relationships:
\begin{align}
    \nu_k    &\simeq b_{\max}\label{eq:nuk}\\
    \gamma_k &\simeq 1.26 + 0.32\ a_{\max}\label{eq:gammak}\\
    I_k      &\simeq 2.30 \left(\mathcal{W}_{\max}\sqrt{\delta\nu / a_{\max}}\right)^{0.93}\label{eq:Ik}\ ,
\end{align}
where $\delta\nu$ is the PDS frequency resolution simulated to be in the range $\delta\nu\in(0.01,1)$. 
Equations (\ref{eq:nuk})--(\ref{eq:Ik}) are not relevant for the algorithm as formulated by \citet{Du2006}, we will however use them later as a first characterisation of the oscillation modes in solar-like oscillators.

\subsection{Multiple peaks}\label{sec:MultiplePeaks}

\begin{figure*}
\includegraphics[width=0.75\textwidth]{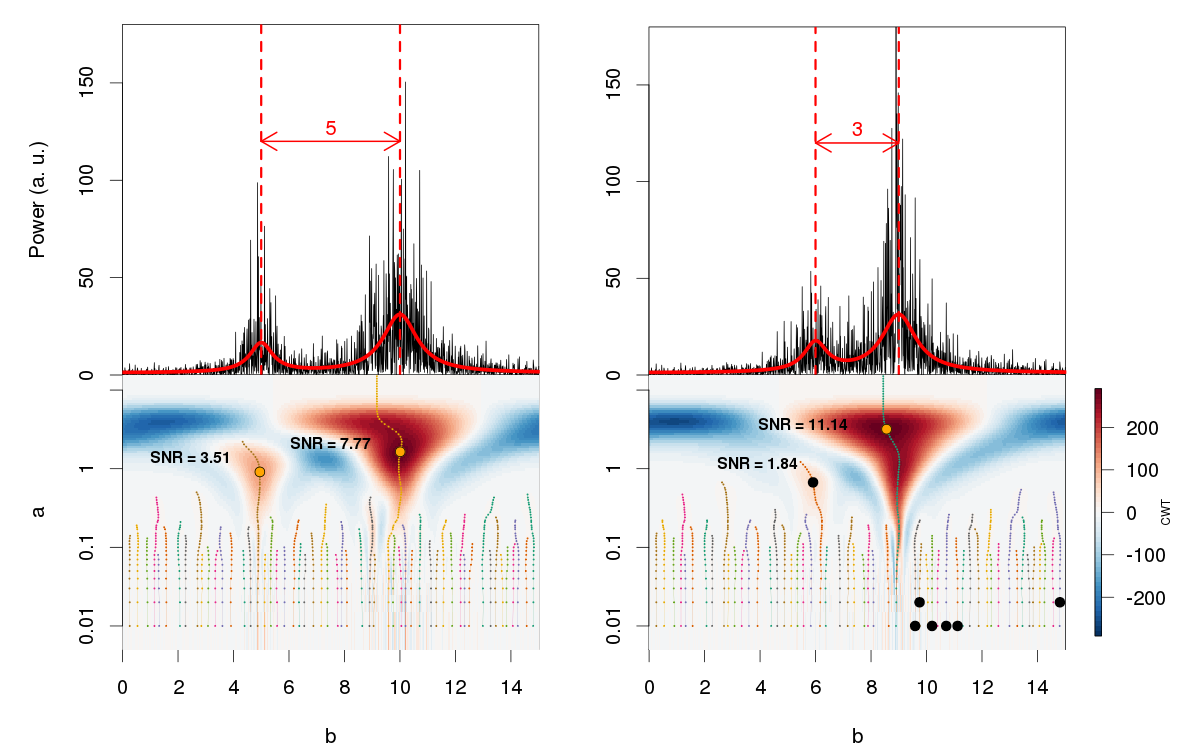}
\caption{\textit{Left panels}: The upper part shows a limit PDS represented by two Lorentzian profiles with heights $I_1=120$ and $I_2=150$, half widths at half maximum $\gamma_1=0.5$, $\gamma_2=0.75$ and central frequencies $\nu_2 - \nu_1=5$ (red) and the simulated PDS realisation (black). The bottom part shows a colour map of the CWT of the simulated PDS realisations with the identified ridges as small dots. The bigger orange dots are at the maximum CWT values on each ridge that have a SNR greater than 3 as defined by \citet{Du2006}. \textit{Right panels}: Same as before with a reduced distance between the central frequencies to $\nu_2 - \nu_1=3$. The black dots are at the maximum CWT value for ridges with SNR greater or equal than 1.84 and smaller than 3.}
\label{fig:doublePeakCWT}
\end{figure*}

Unlike the example in Fig. \ref{fig:singlePeakCWT} with one mode, solar-like oscillators show several oscillation modes in their PDS. 
The limit PDS is thus a superposition of several Lorentzian profiles. 
The peak detection method by \citet{Du2006} is adequate when the overlap between these Lorentzian profiles is small, however it is less reliable when the overlap is significant.
When the separation between two peaks is large there are two ridges in the CWT each one with a large SNR (see left panel of Fig. \ref{fig:doublePeakCWT}).
As the central frequencies get closer the SNR of the ridge corresponding to the peak with smaller amplitude decreases considerably until it has a SNR similar to the noise even though it is visible by eye in the PDS realisation (see right panel of Fig. \ref{fig:doublePeakCWT}).
Reducing the SNR threshold for such cases usually produces several false positive identifications. 

To overcome this limitation we now propose an adjustment to the peak-detection algorithm by \citet{Du2006} for the context of solar-like oscillators.
We use the same CWT with the mexican-hat mother wavelet and use the same approach to find the ridges of local CWT maxima.
We modify the SNR definition using our knowledge about the statistical distribution of noise in a PDS realisation. 
In this modified SNR definition we consider a global instead of a local noise level since a background-normalised PDS has a constant baseline. 
We simulated a PDS realisation of pure white noise, which has a constant limit PDS with value 1, and defined as noise the 95 percentile of the absolute values of all the CWT coefficients at the lowest scale, which we choose as the frequency resolution $\delta\nu$.
With this SNR definition the noise has a value of approximately 2 and thus the SNR is defined as half the maximum CWT value in a ridge. 
This definition is adopted throughout this work.

\begin{figure}
\includegraphics[width=0.47\textwidth]{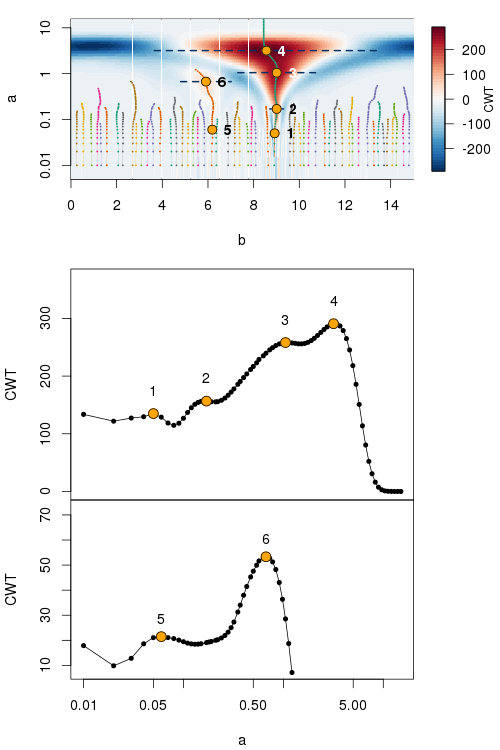}
\caption{\textit{Top panel}: CWT of the simulated PDS realisation from the right panel of Fig. \ref{fig:doublePeakCWT} as a colour map with the identified ridges as small dots. 
The larger orange dots are at the local CWT maxima of the two longest ridges. The horizontal dashed lines span $4\gamma_k$ for each point as inferred from Eq. (\ref{eq:gammak}). 
\textit{Middle and bottom panels}: CWT as a function of scale ($a$) for the longest and second longest ridges in the CWT map from the top panel, respectively. The orange points denote the local maxima with the labels being the same as in the top panel.}
\label{fig:doublePeakCWTbis}
\end{figure}

\begin{figure}
\includegraphics[width=0.47\textwidth]{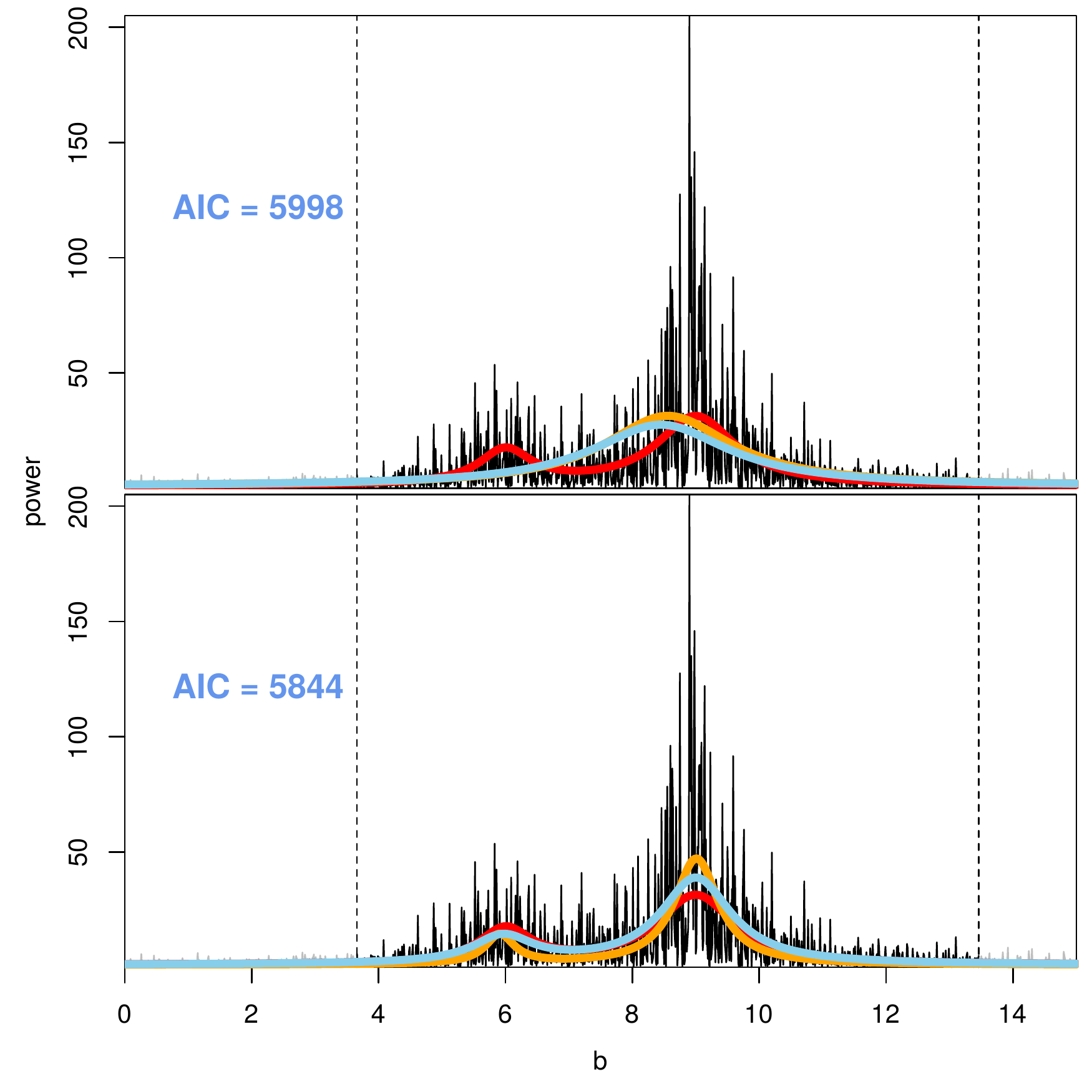}
\caption{Same simulated PDS realisation (black) and limit PDS (red) as the right panel of Fig. \ref{fig:doublePeakCWT}. The vertical dashed lines are at $\nu_k\pm2\gamma_k$ as obtained from Eqs. (\ref{eq:nuk})--(\ref{eq:Ik}) for point 4 from Fig. \ref{fig:doublePeakCWTbis}. In orange a PDS model is shown based on Eqs. (\ref{eq:nuk})--(\ref{eq:Ik}) using point 4 (\textit{top panel}) and points 3 and 6 (\textit{bottom panel}). In blue is the model obtained after an MLE parameter optimisation using the previous values as initial guesses in the range $\nu_k \pm 2\gamma_k$. The AIC values of the MLE are calculated in the PDS region delimited by the vertical dashed lines. A lower AIC indicates a better model.}
\label{fig:doublePeakFits}
\end{figure}

Furthermore, in contrast to the approach taken by \citet{Du2006} which considers only the global maximum on each ridge, we consider all local maxima since their occurrence frequently indicate a significant peak overlap.
By using Eqs. (\ref{eq:nuk})--(\ref{eq:Ik}) each one of these local maxima represents a possible Lorentzian profile in the limit PDS (see Fig. \ref{fig:doublePeakCWTbis}).
Thus, each combination of local maxima in the CWT ridges is considered as a possible PDS model.
However, when there are multiple peaks partially overlapping the CWT values close to the location of a peak are reduced by a negative contribution from neighbouring peaks.
This usually results in a displacement of some local maxima in a ridge to lower scales and, subsequently, an underestimation of $\gamma_k$ for some peaks when using Eqs. (\ref{eq:nuk})--(\ref{eq:Ik}).
We correct for this underestimation by an MLE parameter optimisation using the values obtained from Eqs. (\ref{eq:nuk})--(\ref{eq:Ik}) as initial estimates (see Fig. \ref{fig:doublePeakFits}).
Finally, to select the most appropriate model for the PDS from all the possible combinations we use the Akaike Information Criterion (AIC) which penalises the likelihood of the model with its complexity to avoid over-fitting \citep{Akaike1998}.
Since a lower AIC value is indicative of a better model, we select the most appropriate model as the one with the lowest AIC.

Since the total number of local maxima in all the ridges is usually of the order $10^2-10^3$, depending on the chosen SNR threshold, comparing all combinations of these points as possible PDS models is computationally expensive.
We reduce the computational time by considering different segments of the PDS separately.
In the first step we select the local CWT maxima that has the highest scale.
According to Eqs. (\ref{eq:nuk})--(\ref{eq:Ik}) this corresponds to a peak located at a certain frequency $\nu_k$ and a line width $\gamma_k$.
We define a PDS region local to this peak as the PDS in the frequency range $\nu_k\pm 2\gamma_k$.
Finally, we consider all the local maxima in this region and select the best model as described before.

The algorithm presented here works optimally for oscillations that have a width larger than the frequency resolution of the PDS. This might not be the case for some of the oscillations. In such circumstances the peak is best modelled by a $\mathrm{sinc}^2$ profile. We find these oscillations by performing the peak detection as described above and looking at the residuals of this fit with a false-alarm probability test for a single frequency bin \citep{Appourchaux2012}.

\section{False positive peak detections}\label{sec:EFP}

\begin{figure*}
\includegraphics[width=0.93\textwidth]{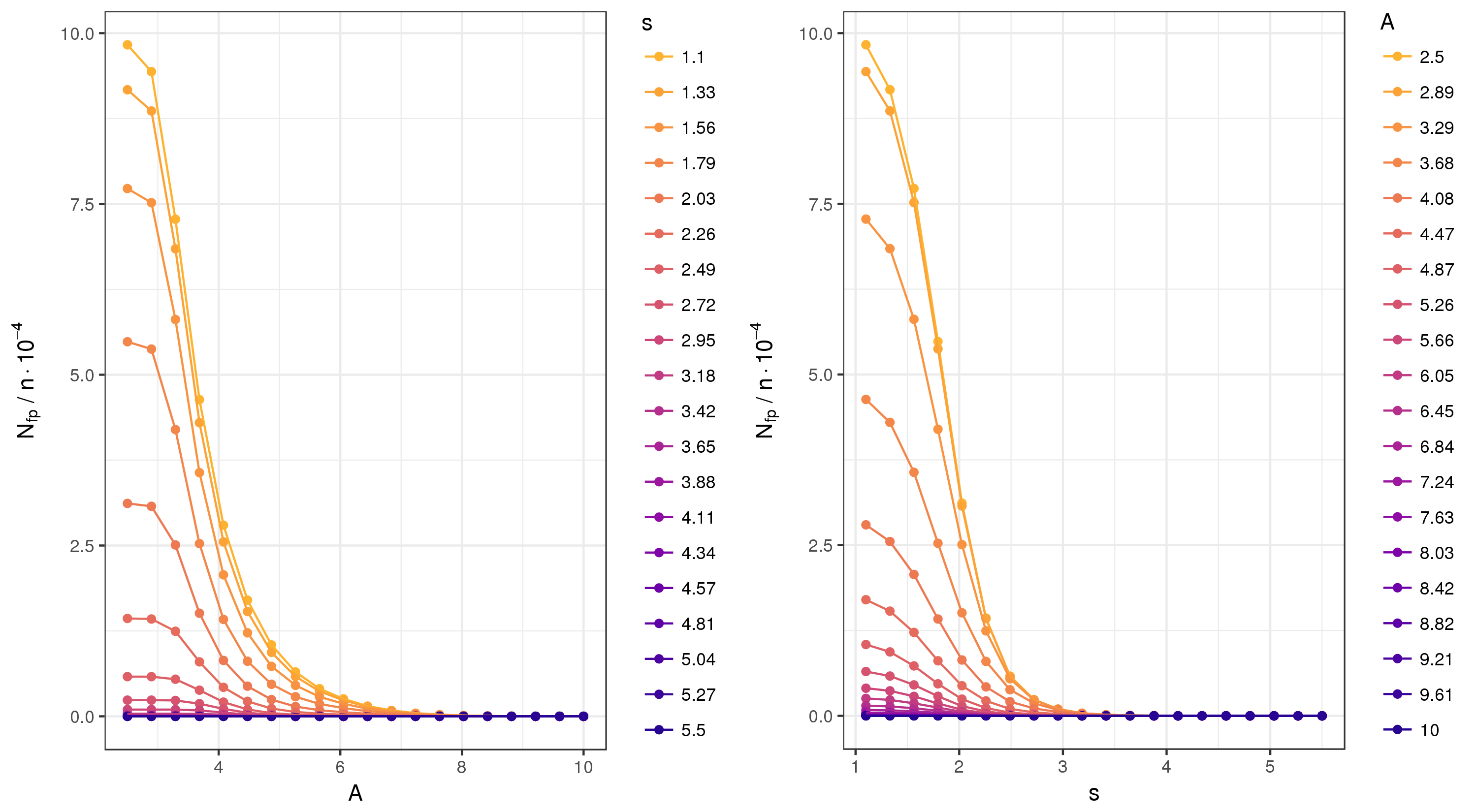}
\caption{
Expected number of false-positive detections $N_\mathrm{fp}$ per number of frequency bins $n$ of peaks having a SNR greater than or equal to $s$ and an amplitude greater than or equal to $A$.
The \textit{left panel} shows $A$ as the independent variable with $s$ as a colour code while in the \textit{right panel} the roles are inverted to show the symmetry between $A$ and $s$.
}
\label{fig:EFP}
\end{figure*}

Since noise in the PDS can generate false positive detections, it is useful to asses the probability that a given peak can be generated only by random chance.
A significance test can be made by the ratio between the likelihood of the observed PDS assuming a model with the peak included and its likelihood if the peak is omitted from the model.
Alternatively, the AIC difference between both models provides a similar and more robust test that penalises for the number of degrees of freedom in the model.
Complementary to these tests, the wavelet-based SNR described previously can also be used as a significance test that is sensitive to different features of the PDS.
Since our algorithm uses a mexican-hat mother wavelet, which has the general shape of the most common peaks, the SNR is particularly suited to assess the significance of solar-like oscillations.
The wavelet-based SNR is, on average, a monotonically increasing function of the AIC difference between a PDS model with the peak and without it. 
However, their relationship is non linear and has considerable spread so they provide to some extent complementary information.
In this section we describe the statistical properties of the SNR for peaks generated purely by random noise in the PDS and quantify the number of such peaks that we can expect to detect in any given PDS realisation.

To estimate the chance of false positive peak detections we generated $10^5$ PDS each one containing $10^4$ frequency bins with only noise having the same distribution as the noise in a background-normalised PDS, i.e. a scaled $\chi^2$ distribution with 2 degrees of freedom.
We applied the wavelet-based peak detection method described here to all the generated PDS and aggregated the results. 
We found that there is a $10^{-3}$ chance per frequency bin in a PDS of having a false positive detection with a SNR greater than 1.1 and an AIC difference greater than 0.
For a typical \textit{Kepler} PDS this amounts to approximately 30 false positives in the whole PDS.
However, the number is reduced when considering only the power excess region (see below).

The false-positive detections have different amplitudes and their number depends on the SNR threshold. 
To quantitatively characterise this dependence we define $N_\mathrm{fp} (A,s)$ to be the number of false-positive peak detections that have an amplitude greater than or equal to $A$ and a SNR greater than or equal to $s$.
Since the number $N_\mathrm{fp} (A,s)$ depends also on the number of frequency bins, $n$, in the PDS it is more convenient to normalise it by $n$ (see Fig. \ref{fig:EFP} $N_\mathrm{fp}/n$).
Additionally, to take into account the PDS frequency resolution, $\delta\nu$. we calculate $N_\mathrm{fp}$ as a function of $A' = A / \sqrt{\delta\nu}$
By using symbolic regression \citep{GPTIPS} we discovered that $N_\mathrm{fp}$ can be described by a function that is symmetric in $A'$ and $s$ and has the form
\begin{align}
    \log \left(N_\mathrm{fp} \right) = 
    & c_0 + c_1 s + c_2 A' \nonumber \\
    & + c_3 s^2 + c_4 A'^2 + c_5 s A' \label{eq:EFP}\\
    & + c_6 (sA')^2 + c_7 (A'/s)^2 + c_8 (s/A')^2 + \log (n)\nonumber \ .
\end{align}
We further refined the symbolic regression estimate of the coefficients $c_i$ by using a multiple linear regression with the same model. 
That is, we considered $\left\{s, A', s^2, A'^2, s A', (sA')^2, (A'/s)^2,(s/A')^2\right\}$ as linear predictors for $\log \left(N_\mathrm{fp}/n\right)$ and performed a least squares regression to estimate $c_i$.
The obtained coefficient values, their standard error and the null hypothesis probability for each term ($p$-value) are reported in Table \ref{tab:coefs1}.

\begin{table}
    \centering
    \caption{Least-squares linear regression estimate for the parameter values, standard errors and null-hypothesis probability ($p$-value) for each parameter in Eq. \ref{eq:EFP}.}
    \label{tab:coefs1}
    \begin{tabular}{r|c|c|c|}
               & estimate & standard error & $p$-value \\
        \hline
        $c_0$  &  -9.63   &     1.18       & 2.67$\times 10^{-14}$ \\
        $c_1$  &   3.99   &     0.75       & 2.13$\times 10^{-7}$  \\
        $c_2$  &   2.24	  &     0.36       & 1.88$\times 10^{-9}$  \\
        $c_3$  &  -0.21	  &     0.06       & 1.94$\times 10^{-4}$  \\
        $c_4$  &  -0.07	  &     0.01       & 1.49$\times 10^{-6}$  \\
        $c_5$  &  -1.93	  &     0.19       & 7.99$\times 10^{-20}$ \\
        $c_6$  &   0.027  &     0.003      & 1.73$\times 10^{-15}$ \\
        $c_7$  &  -0.14	  &     0.01       & 4.30$\times 10^{-32}$ \\
        $c_8$  &  -2.78	  &     0.24       & 1.71$\times 10^{-25}$ \\
        \hline
    \end{tabular}
\end{table}

To calculate $N_\mathrm{fp}$ we must also provide the number of frequency bins $n$ under consideration.
The value of $N_\mathrm{fp}$ will change considerably whether we take $n$ as the number of frequency bins in the whole PDS or only in the power excess region.
Since we only expect stochastically excited global oscillations in the power excess region we adopt $n$ as the number of frequency bins in the power excess region.
To define this region precisely we adopt the global description of the power excess as a gaussian function centred at $\nu_\mathrm{max}$ with a width of the gaussian $\sigma_\mathrm{env}$ and take $n$ as the number of frequency bins in $\nu_\mathrm{max} \pm 4\sigma_\mathrm{env}$.
This definition is adopted throughout this work.

In summary, given a peak amplitude $A$ and SNR $s$ we can estimate $N_\mathrm{fp}$ using Eq. (\ref{eq:EFP}) using $A' = A/\sqrt{\delta\nu}$.
The number $N_\mathrm{fp}$ is the expected number of false positives in the power excess region that have an amplitude equal or greater than $A$ and a SNR equal or greater than $s$.
A value of $N_\mathrm{fp}$ greater than 1 indicates that the peak is more likely to have been generated by noise than from a real signal.
Conversely, if $N_\mathrm{fp}$ is smaller than 1, it is more probable that the peak originates from a process different than noise.
Smaller values of $N_\mathrm{fp}$ denote that a peak is less likely to be a false positive detection.

\section{Detection probability}\label{sec:DetProb}

\begin{figure}
\includegraphics[width=0.47\textwidth]{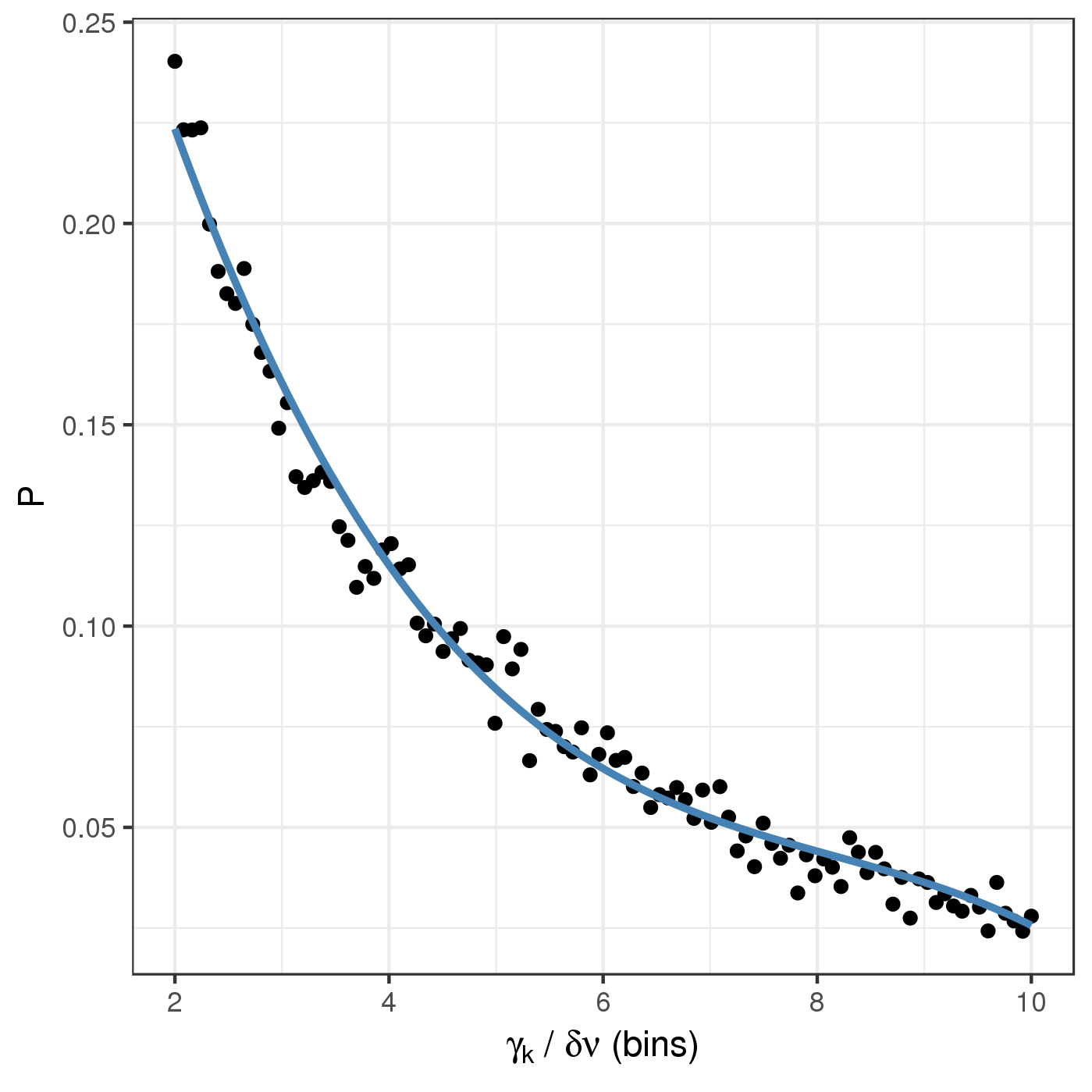}
\caption{Probability ($P$) of not finding a peak of a given linewidth $\gamma_k$ where $\delta\nu$ is the PDS frequency resolution. The blue line is the fit using Eq. \ref{eq:DetProb}.}
\label{fig:DetProb}
\end{figure}

Complementary to assessing the significance of a peak detection we now address the significance of the lack of detections, that is, for a given peak amplitude we estimate the probability that the algorithm presented here can recognise it among the noise.
To estimate this detection probability we simulated $10^5$ different PDS each one with a Lorentzian peak multiplied by noise as described in section \ref{sec:singlePeak}.
The Lorentzian peaks were generated with parameters in the ranges $\gamma_k / \delta\nu \in(2,10)$ and $I\in(1,10)$ in arbitrary units.
We then applied the algorithm presented here to the PDS using a SNR of 1.1 and attempted to recover the input peak.
From these simulations we estimated the proportion of times $P$ that a peak with linewidth $\gamma_k / \delta\nu$ was not successfully recovered (see Fig. \ref{fig:DetProb}).
It can be seen from Fig. \ref{fig:DetProb} that $P$ can be described as a function of $\gamma_k / \delta\nu$ alone.
In this case it is not necessary to use a model discovery technique, like symbolic regression. Instead we propose a polynomial model in the form.
\begin{equation}\label{eq:DetProb}
P = \sum_{i=0}^N d_i \left( \frac{\gamma_k}{\delta\nu} \right)^i
\end{equation}
for some coefficients $d_i$ with $N$ being the degree of the polynomial.
We estimated the coefficients $d_i$ by a multiple linear regression, similar to Sec. \ref{sec:EFP}, for increasing values of $N$ and found that the model with the lowest AIC is a third degree polynomial ($N=3$).
We adopt this model throughout this work and give in Table \ref{tab:coefs2} the estimated coefficients $d_i$, their standard error and null hypothesis probability for each term ($p$-value).

\begin{table}
    \centering
    \caption{Same as Table \ref{tab:coefs1} for Eq. \ref{eq:DetProb}.}
    \label{tab:coefs2}
    \begin{tabular}{r|c|c|c|}
               & estimate & standard error  & $p$-value \\
        \hline
        $d_0$  &    0.09  &      $<10^{-4}$ & 3.11$\times 10^{-130}$ \\
        $d_1$  &   -0.49  &      $<10^{-4}$ & 1.34$\times 10^{-107}$ \\
        $d_2$  &    0.18  &      $<10^{-4}$ & 5.94$\times 10^{-67}$ \\
        $d_3$  &   -0.06  &      $<10^{-4}$ & 6.71$\times 10^{-26}$ \\
        \hline
    \end{tabular}
\end{table}

In summary, for a predicted peak with linewidth $\gamma_k$ and a PDS frequency resolution $\delta\nu$ we can calculate the probability that the algorithm will not find the peak, $P$, using Eq. \ref{eq:DetProb}.

\section{Solar frequencies observed by \textit{BiSON}}\label{sec:BiSON}

\begin{figure}
\includegraphics[width=0.47\textwidth]{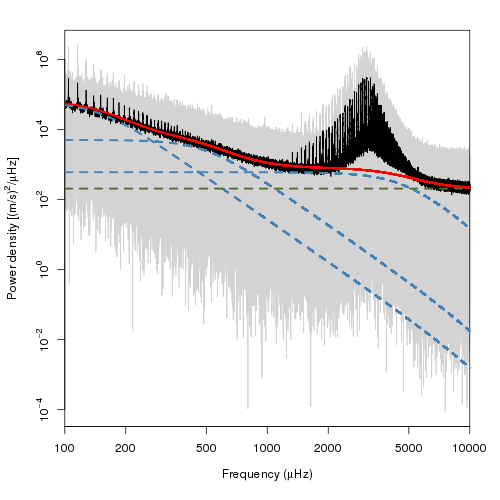}
\caption{PDS (grey) of the radial velocity measurements by BiSON with a $0.1\mu$Hz binned version (black). The background function is shown in red with the individual granulation components in blue and the white noise in green.}
\label{fig:BiSON-bg-fit}
\end{figure}

As a proof of concept for the peak detection algorithm presented here we analyse the solar stochastically excited global oscillations modes obtained from radial velocity variations measured by the Birmingham Solar Observatory Network (BiSON).
We used the time series from the 1st January 1991 to the 31st December 2015 optimised for fill \citep{Davies2014,Hale2016}.
We calculated the PDS from the discrete Fourier transform of the time-series normalised using the spectral window function \citep{Kallinger2014} (see Fig. \ref{fig:BiSON-bg-fit}).

For the CWT-based peak detection algorithm we need to normalise the PDS by an estimation of the granulation background.
As an estimation of this background we used a superposition of three granulation components in the form of super-Lorentzians $A_i/[1+(\nu/b_i)^4]$ and a white noise $P_n'$.
The oscillation power excess region in this dataset is difficult to model accurately so we excluded it from our fit.
Specifically, during the parameter estimations we computed the likelihood of the PDS model only for the region outside the frequency interval ranging from 1200 to 5000 $\mu$Hz.
Thus we fitted a PDS model containing only the granulation background in the form
\begin{equation}\label{eq:PDSBgModel}
    P(\nu) = P_n' + \eta(\nu)^2\left[\sum_{i=1}^3 \frac{A_i}{1+(\nu/b_i)^4} \right],
\end{equation}
where $\eta(\nu) = \mathrm{sinc}(\pi\nu/2\nu_\mathrm{nq})$ is a frequency-dependent damping, consequence of the measurement discretisation process, \citep{Chaplin2011} and $\nu_\mathrm{nq}$ is the Nyquist frequency.

To estimate the model parameters we used the affine invariant Markov-chain Monte-Carlo algorithm by \citet{Goodman2010}, as implemented by \citet{emcee}.
We estimated the posterior probability density function (PDF) for each parameter using the chains after convergence was reached.
The expectation value for each parameter was estimated as the median of its PDF.
Additionally, we estimated the uncertainties by the 16-84 percentiles of each PDF.
The obtained parameter estimations with their uncertainties are:
\begin{align*}
    P_n &= 2.06_{-0.02}^{+0.02}\times 10^2 (m/s)^2/\mu\mathrm{Hz},\\
    A_1 &= 6.09_{-0.60}^{+0.77}\times 10^4 (m/s)^2/\mu\mathrm{Hz},\\
    b_1 &= 1.45_{-0.08}^{+0.08}\times 10^2 \mu\mathrm{Hz},\\
    A_2 &= 5.04_{-0.59}^{+0.61}\times 10^3 (m/s)^2/\mu\mathrm{Hz},\\
    b_2 &= 4.95_{-0.22}^{+0.26}\times 10^2 \mu\mathrm{Hz},\\
    A_3 &= 6.06_{-0.38}^{+0.35}\times 10^2 (m/s)^2/\mu\mathrm{Hz},\\
    b_3 &= 4.60_{-0.10}^{+0.12}\times 10^3 \mu\mathrm{Hz}.
\end{align*}
Even though this granulation background description is too simplistic to accurately describe all phenomena contributing to the BiSON PDS, it is sufficiently accurate for our purpose of finding the most relevant signatures of the stochastically excited global oscillations (see Fig. \ref{fig:BiSON-bg-fit}).

\begin{figure}
\includegraphics[width=0.47\textwidth]{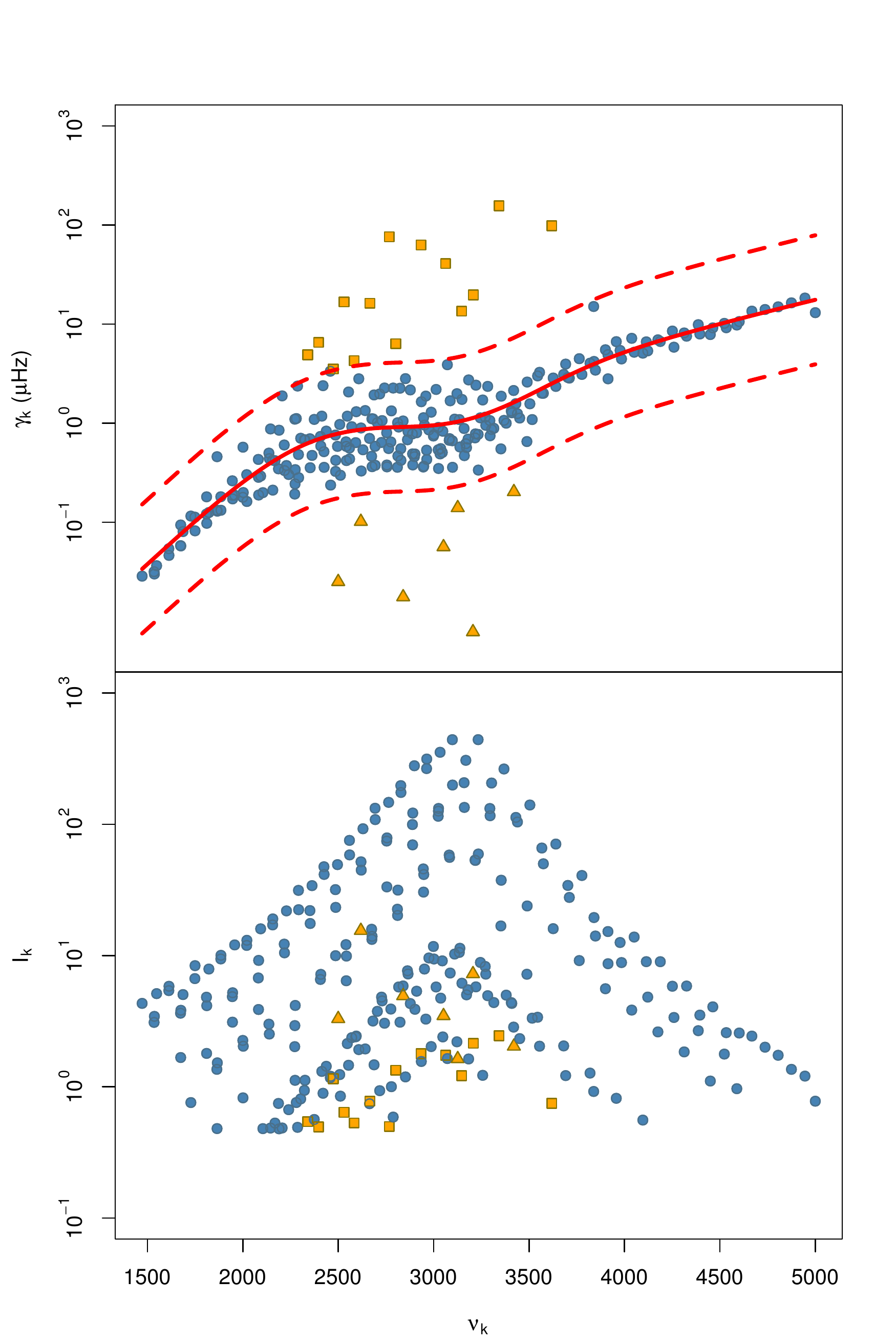}
\caption{
Half-width at half maxima ($\gamma_k$, \textit{top}) and heights ($I_k$ \textit{bottom}) as functions of frequency for the peaks identified in the PDS of the solar radial velocity measurements by BiSON. 
The solid red line in the top panel is a smoothing cubic spline to the data; the dashed red lines are the spline multiplied by $e^{\pm1.5}$. 
The points outside the region delimited by the dashed lines are coloured orange.
The square symbols for the outliers denote peaks with a larger width while the triangular symbols denote a smaller width.}
\label{fig:BiSONpeaks}
\end{figure}

Having obtained a background-normalised PDS we proceeded to apply our peak detection algorithm to it.
However, due to the large size of the BiSON data set, it was not computationally feasible to use the peak detection algorithm on the full PDS.
Instead, we analysed segments of the PDS spanning 500 $\mu$Hz each in steps of 150 $\mu$Hz so that there is a 350 $\mu$Hz overlap between each consecutive segment, to avoid possible edge effects in the CWT, and combined the resulting peaks.
Figure \ref{fig:BiSONpeaks} shows the frequencies of the identified peaks, $\nu_k$, as a function of $\gamma_k$ (top) and $I_k$ (bottom).
The points follow the same frequency-dependant $\gamma_k$ as found by \citet{Chaplin1997}.
The red solid line in the top panel of Fig. \ref{fig:BiSONpeaks} is a smoothing cubic spline that shows the overall trend; the dashed red lines are the spline fit multiplied by $e^{\pm1.5}$ and delimit a region where most points lie.
The outliers from this trend have been coloured in orange in order to distinguish them from the rest.
Most outliers from this trend have a relatively small height (see bottom panel of Fig. \ref{fig:BiSONpeaks}).
We interpret the wide outliers as not being stochastically excited global oscillations but originating as a compensation for an incomplete description of the granulation background.

\begin{figure}
\includegraphics[width=0.47\textwidth]{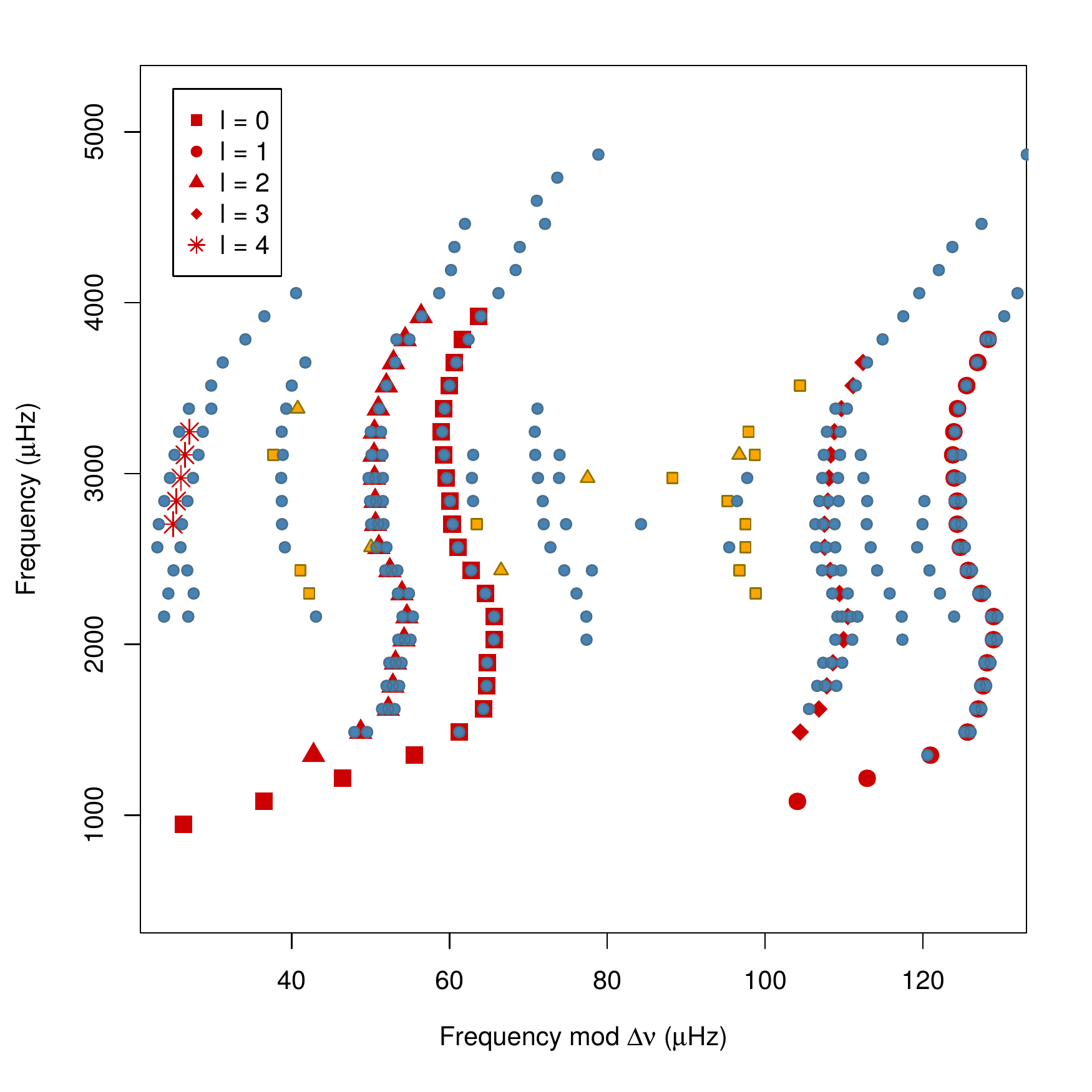}
\caption{\'{E}chelle diagram of the peaks shown in Fig. \ref{fig:BiSONpeaks} with the same colouring and symbol scheme. In red are a combination of the oscillations found by \citet{Chaplin1996, Broomhall2009} and \citet{Davies2014}; the symbol identifies the spherical harmonic degree $l$ according to the legend.}
\label{fig:BiSONechelle}
\end{figure}

\begin{figure}
\includegraphics[width=0.47\textwidth]{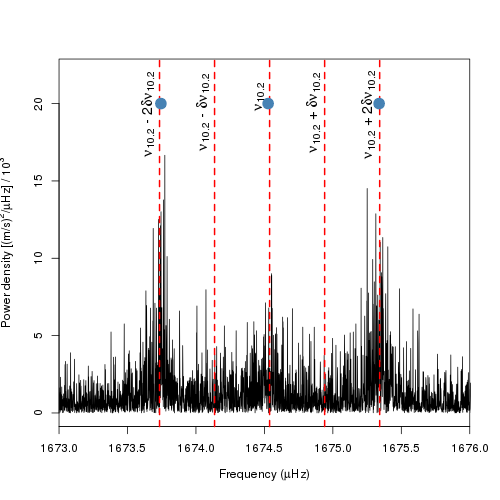}
\caption{PDS (black) of the radial velocity measurements by BiSON. The dashed red lines are located at the frequencies $\delta_{n,l}+m\delta\nu_{n,l}$ for $n=10$ and $l=2$ as reported by \citet{Davies2014}.
The blue dots are located at the frequencies of the peaks identified in this work.}
\label{fig:BiSONsplit}
\end{figure}

In Fig. \ref{fig:BiSONechelle} we use an \'{e}chelle diagram to compare the frequency of the solar-like oscillation modes found with the peak-detection method presented here with the results obtained by \citet{Chaplin1996}, \citet{Broomhall2009} and \citet{Davies2014}.
All oscillations with a frequency higher than $\sim 1400 \mu$Hz are correctly identified.
The lowest frequency modes are not found with the CWT-based method because their width is small and the CWT-based pattern-matching is optimal for peaks with a width that can be resolved.
It is also visible that, due to rotational splitting, most oscillation modes with $l>0$ are detected as having more than one peak.
Figure \ref{fig:BiSONsplit} shows an example of a rotational split $l=2$ mode compared with the reported values by \citet{Davies2014}.
It should be noted that \citet{Davies2014} fit for each spherical degree, $l$, a model described by the central frequency $\nu_{n,l}$ and the rotational splitting $\delta\nu_{n,l}$ whereas the method presented here finds each oscillation frequency individually.

There are also several ridges detected that are not reported in the literature as stochastically excited oscillation modes.
The origin of these peaks can be traced back to the window function of BiSON observations which produces prominent aliases of the oscillation modes.
We find more oscillation modes at higher frequencies than reported in the literature.

\section{Red-giants observed by \textit{Kepler}}\label{sec:Kepler}

The CWT-based peak detection presented here is capable of identifying the solar-like oscillations in the red-giant stars observed by Kepler.
We use the set of 19 red-giant stars studied in detail by \cite{Corsaro2015} to compare our results.
The Kepler light curves used in this work have been extracted using the pixel data following the methods described in Mathur et al. (in preparation) and corrected following \cite{Garcia2011}\footnote{This data preparation procedure includes concatenating the light curves from different quarters, correcting for the levels of each quarter, thermal drifts and jumps and removing a second order polynomial.}.
We calculated the PDS from the corrected time-series using a Lomb-Scargle periodogram \citep{Scargle1982} normalised using the spectral window function \citep{Kallinger2014}.

\begin{figure}
\includegraphics[width=0.47\textwidth]{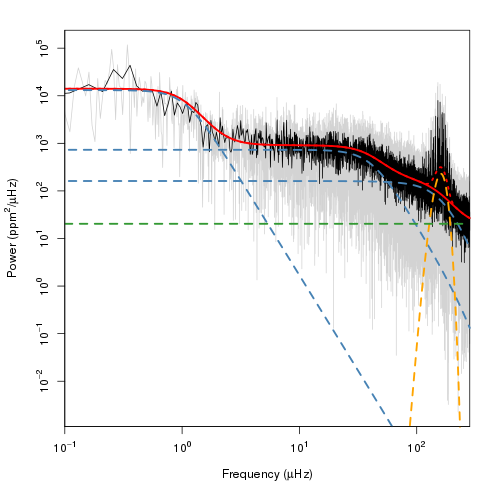}
\caption{Background fit of KIC 12008916. The PDS is shown in gray with a binned version in black. The solid red line is the background without the power excess while the dotted red line is the background fit plus the power excess. The individual granulation components are shown in blue, the power excess in orange and the white noise in green dashed lines.}
\label{fig:KIC-12008916-bg-fit}
\end{figure}

To obtain a background-normalised PDS we fit the following model
\begin{equation}\label{eq:PDSBgModelKplr}
    P(\nu) = P_n' + \eta(\nu)^2\left[\sum_{i=1}^3 \frac{A_i}{1+(\nu/b_i)^4} + 
    P_g\exp\left(\frac{\nu-\nu_\mathrm{max}}{\sigma_\mathrm{env}}\right)^2 \right]
\end{equation}
to the observed PDS.
To estimate the model parameters we adopt a Bayesian framework and use the MCMC algorithm of \citet{Goodman2010} as implemented by \cite{emcee} to estimate the posterior density function of each parameter.
The expectation value for each parameter is approximated as the median of its posterior density function and the standard errors are approximated as the 16-84 percentiles.
An example of the resulting background fit is shown in Fig. \ref{fig:KIC-12008916-bg-fit} for KIC 12008916. 
The estimated background parameter values and their uncertainties are given in Appendix \ref{sec:KeplerPDSfits}.

\begin{figure}
\includegraphics[width=0.47\textwidth]{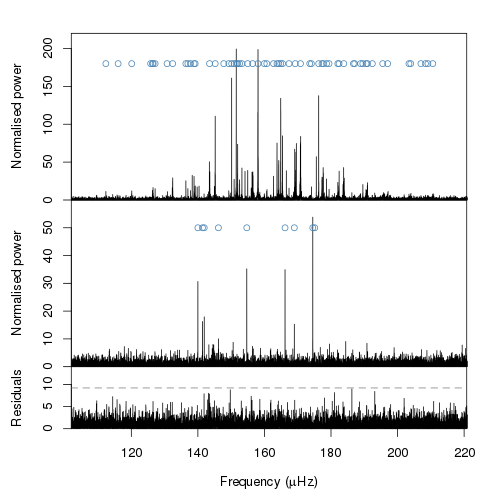}
\caption{
\textit{Top panel:} Background-normalised PDS for KIC 12008916. The blue circles indicate the location of the oscillation modes identified with the peak detection algorithm presented here.
\textit{Middle panel:} PDS from the top panel divided by a PDS model containing the peaks identified previously. The blue circles indicate the location of points having a false-alarm probability lower than $10^{-4}$.
\textit{Bottom panel}: Residual power obtained by dividing the PDS from the top panel by a model containing all the resolved and unresolved oscillations identified with their parameters optimised with a MLE on the whole background-normalised PDS. The dashed horizontal line is placed at a level corresponding to a false-alarm probability of $10^{-4}$.
}
\label{fig:KIC12008916peakFinding}
\end{figure}

We subsequently identified the resolved oscillation modes in the background-normalised PDS using the peak-detection algorithm presented here with a SNR threshold of 1.1 discarding any detections with an AIC smaller than 0 and a $N_\mathrm{fp}$ greater than 1.
As discussed in section \ref{sec:peakMethod}, the CWT is suitable to detect peaks with a width larger than the frequency resolution.
However, in some red-giant stars there are gravity-dominated mixed modes visible which can have peaks narrower than the frequency resolution , i.e. they are unresolved.
To highlight the unresolved oscillation modes we make a PDS model with the peaks detected with the CWT and divide the background-normalised PDS by this model.
The resulting PDS (see the middle panel of Fig. \ref{fig:KIC12008916peakFinding}) contains only the unresolved oscillations on top of the noise.
The peaks that have a false-alarm probability lower than a certain threshold, which we choose to be $10^{-4}$, are deemed significant and fitted with a $\mathrm{sinc}^2$ function.

To mitigate possible power leakage between the peaks in the PDS we apply a final MLE to the full background-normalised PDS taking the detections of both resolved and unresolved oscillation modes into account.
This provides our final identification of the solar-like oscillations visible in the PDS.

\begin{figure}
\includegraphics[width=0.47\textwidth]{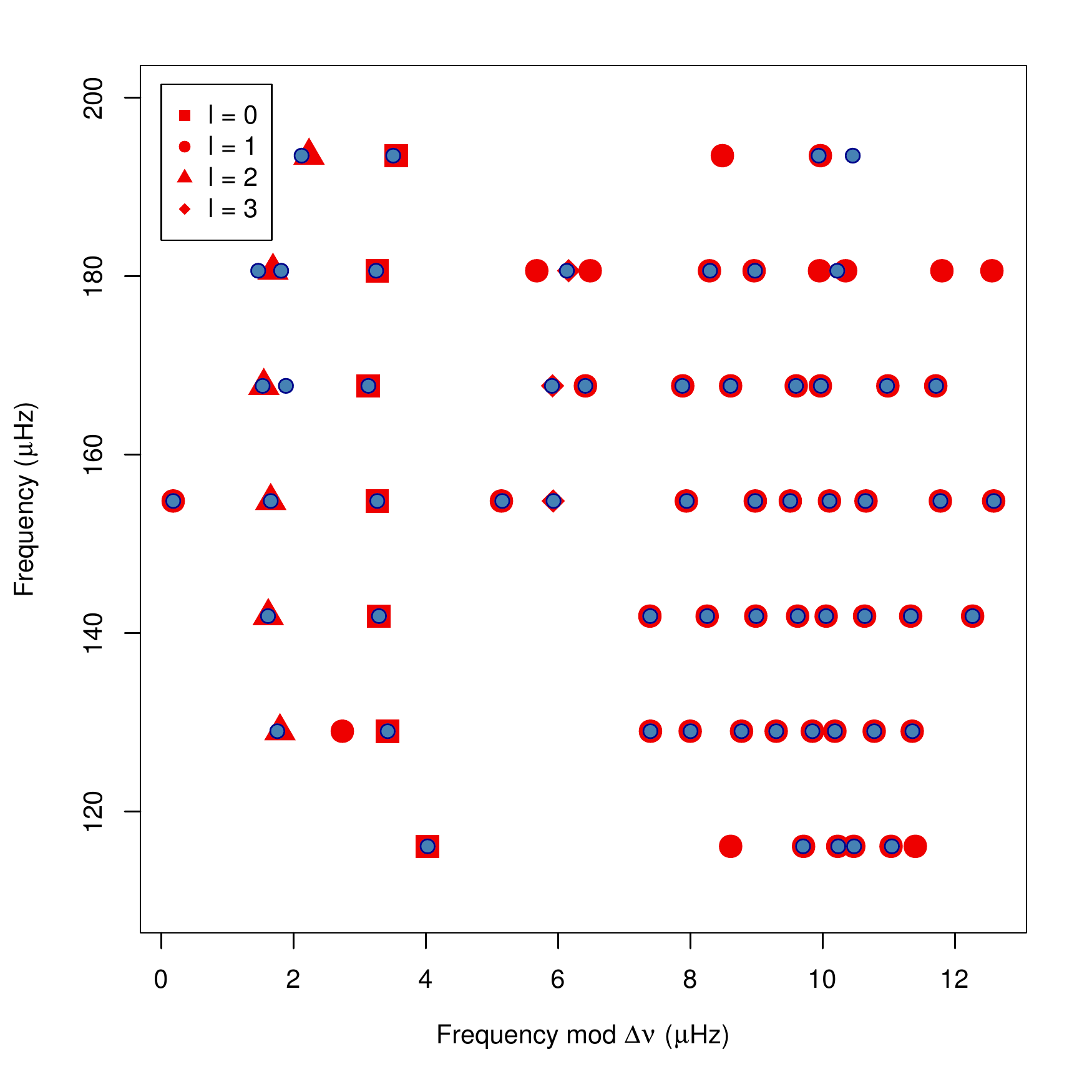}
\caption{
\'{E}chelle diagram with the frequencies of the solar-like oscillation modes of KIC 12008916 using $\Delta \nu=12.9\ \mu$Hz. 
Red symbols by \citet{Corsaro2015} with their shape identifying the spherical degree. 
The blue circles are the frequencies detected by the method presented here. 
}
\label{fig:KIC012008916echelle}
\end{figure}

\begin{figure*}
\includegraphics[width=0.89\textwidth]{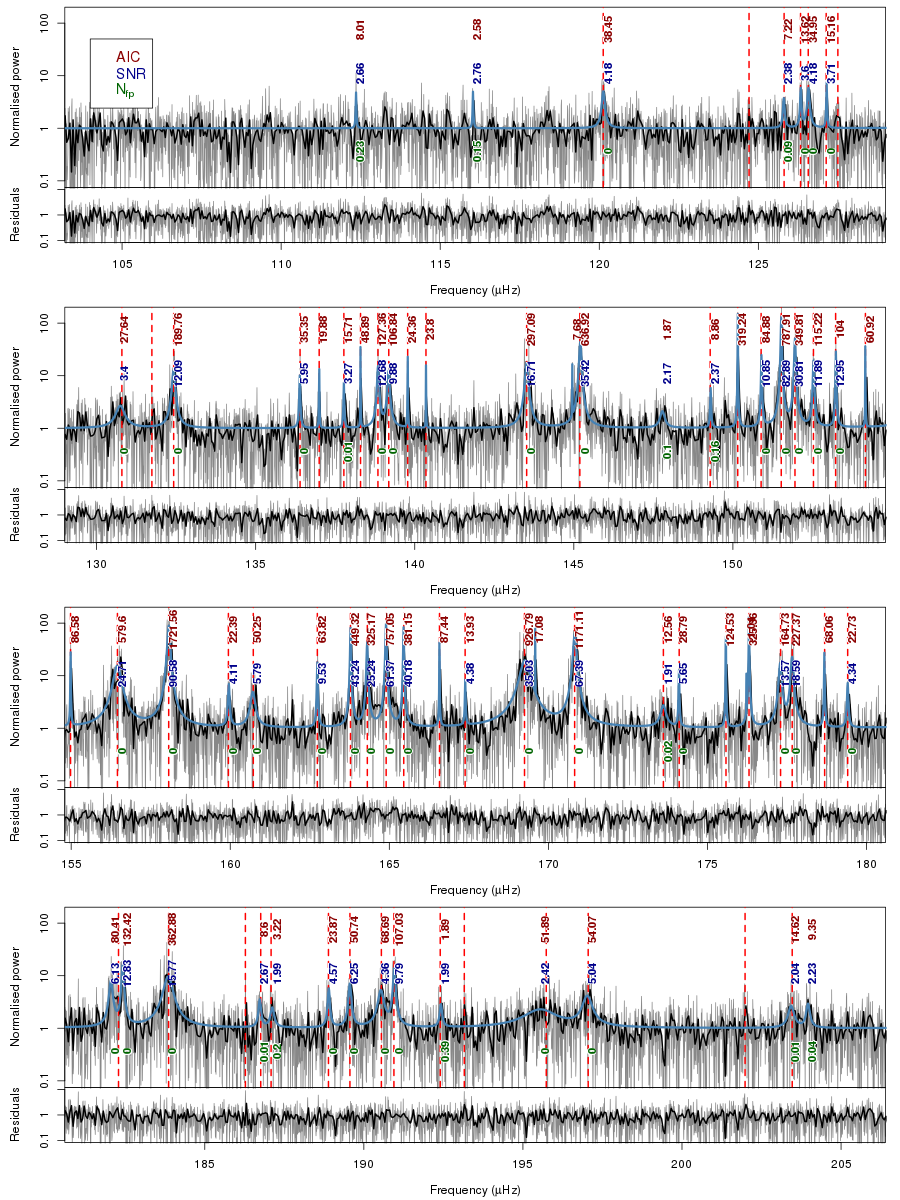}
\caption{
Background-normalised PDS (grey) for \textit{Kepler} target KIC 12008916 with a 0.05 $\mu$Hz binned version (black) in the region where the solar-like oscillations are located. 
The vertical red lines are placed at the frequencies of the oscillation modes obtained by \citet{Corsaro2015} with a detection probability greater than 0.9.
The blue line is an MLE fit using the heights, locations and line widths obtained with the peak detection results as input parameters to build a PDS model. 
We also indicate the AIC difference with the null hypothesis model (red), the SNR (blue) and $N_\mathrm{fp}$ (green) as described in Section \ref{sec:EFP} for each oscillation.
The residuals are calculated by dividing the background-normalised PDS by the fitted model.}
\label{fig:KIC12008916details}
\end{figure*}

We compared our peak identifications with the values reported by \cite{Corsaro2015}.
Both methods agree on most of the detections although there are some discrepancies specially when the peaks have low amplitude or significant overlap (see Figs. \ref{fig:KIC012008916echelle} and \ref{fig:KIC12008916details}).
The resolved and unresolved oscillation modes detected for KIC 12008916 are shown in Tables \protect\ref{tab:012008916resolved} and \protect\ref{tab:012008916unresolved} respectively.
We provide the parameters of the detected oscillations for the remaining stars in the sample analysed by \cite{Corsaro2015} in Appendix \ref{sec:KeplerOscillations}.

\section{Discussion and conclusions}

Here we have addressed the problem of finding the significant solar-like oscillations in a background normalised power density spectra (PDS).
We presented a peak finding algorithm capable of finding the oscillation modes in a PDS and provide an estimate of their model parameters provided that the PDS background description is adequate.
Furthermore, for each detected mode we asses its probability of being a false positive detection.
Additionally, for a predicted peak we can estimate the probability of the algorithm being able to detect it.
In contrast to other approaches the algorithm is fast (less than a minute for a typical \textit{Kepler} long-cadence PDS) and does not require human intervention making it suitable for the analysis of large samples of stars.
Since the algorithm does not rely on human intervention, it provides an objective criteria for peak selection in the fitted PDS model.
We have shown that the results obtained from the peak detection algorithm presented here are comparable to previous approaches.
The peak detection algorithm presented here together with an adequate background description, and the spherical degree, and azimutal order determination for each mode enables the study of the internal structure of the large number of solar-like oscillators that are already observed and yet remain unexplored.

\section*{Acknowledgements}

The research leading to the presented results has received funding from the European Research Conuncil under the European Commumity's Seventh Framework Programme (FP7/2007-2013)/ERC grant agreement no 338251 (StellarAges).
This work partially used data analyzed under the NASA grant NNX12AE17.
The authors also aknowledge the anonymous referee for the helpful and constructive comments.


\bibliographystyle{mnras}
\bibliography{biblio}


\appendix

\section{Results for the gobal PDS fits}\label{sec:KeplerPDSfits}

We performed a global PDS fit for the 19 red-giant stars analysed by \cite{Corsaro2015} using Eq. (\ref{eq:PDSBgModelKplr}) and the procedure described in Section \ref{sec:Kepler}.
The obtained parameter values for the white noise and granulation background are reported in Table \ref{tab:KeplerBgs} and for the power excess in Table \ref{tab:KeplerPE}.
The expected value for each parameter is estimated by the median of the posterior density function while the lower and upper uncertainties are estimated as the 18 and 64 percentiles of the posterior density function.

\begin{table*}
  \centering
  \caption{Estimated PDS fit parameters for the white noise and granulation background in the 19 red-giant stars analysed by \protect\cite{Corsaro2015} using Eq. \ref{eq:PDSBgModelKplr}.
  The estimated parameter value is the median of the posterior distribution.
  The lower and upper uncertainties for each parameter are the 16 and 84 percentiles of its posterior distribution.}
  \label{tab:KeplerBgs}
  \begin{tabular}{r|c|c|c|c|c|c|c|}
    KIC 
    & $P_n$ (ppm$^2/\mu$Hz)
    & $A_1$ (ppm$^2/\mu$Hz)
    & $b_1$ ($\mu$Hz)
    & $A_2$ (ppm$^2/\mu$Hz)
    & $b_2$ ($\mu$Hz)
    & $A_3$ (ppm$^2/\mu$Hz)
    & $b_3$ ($\mu$Hz) \\
    \hline

    3744043 &
    $8.23_{-0.22}^{+0.20}\times 10^{0}$ &
    $1.73_{-0.05}^{+0.06}\times 10^{3}$ &
    $2.41_{-0.06}^{+0.06}\times 10^{1}$ &
    $3.00_{-0.74}^{+0.27}\times 10^{2}$ &
    $9.44_{-0.52}^{+0.24}\times 10^{1}$ &
    $3.73_{-2.58}^{+8.47}\times 10^{1}$ &
    $1.10_{-0.08}^{+0.18}\times 10^{2}$ \\

    6117517 &
    $1.20_{-0.03}^{+0.03}\times 10^{1}$ &
    $4.03_{-0.56}^{+0.66}\times 10^{3}$ &
    $1.77_{-0.22}^{+0.26}\times 10^{0}$ &
    $1.67_{-0.05}^{+0.06}\times 10^{3}$ &
    $2.94_{-0.07}^{+0.07}\times 10^{1}$ &
    $4.01_{-0.14}^{+0.13}\times 10^{2}$ &
    $1.04_{-0.01}^{+0.01}\times 10^{2}$ \\

    6144777 &
    $1.05_{-0.03}^{+0.03}\times 10^{1}$ &
    $4.06_{-0.56}^{+0.61}\times 10^{3}$ &
    $1.32_{-0.13}^{+0.15}\times 10^{0}$ &
    $1.16_{-0.03}^{+0.04}\times 10^{3}$ &
    $3.31_{-0.08}^{+0.08}\times 10^{1}$ &
    $2.67_{-0.09}^{+0.09}\times 10^{2}$ &
    $1.20_{-0.02}^{+0.02}\times 10^{2}$ \\

    7060732 &
    $3.12_{-0.07}^{+0.07}\times 10^{1}$ &
    $8.03_{-0.99}^{+1.12}\times 10^{3}$ &
    $1.86_{-0.19}^{+0.23}\times 10^{0}$ &
    $1.07_{-0.04}^{+0.04}\times 10^{3}$ &
    $3.14_{-0.08}^{+0.09}\times 10^{1}$ &
    $2.60_{-0.09}^{+0.10}\times 10^{2}$ &
    $1.18_{-0.03}^{+0.03}\times 10^{2}$ \\

    7619745 &
    $1.78_{-0.10}^{+0.11}\times 10^{1}$ &
    $3.60_{-0.44}^{+0.52}\times 10^{3}$ &
    $1.23_{-0.10}^{+0.11}\times 10^{0}$ &
    $5.46_{-0.13}^{+0.13}\times 10^{2}$ &
    $4.31_{-0.08}^{+0.08}\times 10^{1}$ &
    $1.14_{-0.04}^{+0.04}\times 10^{2}$ &
    $1.70_{-0.06}^{+0.06}\times 10^{2}$ \\

    8366239 &
    $1.98_{-0.18}^{+0.15}\times 10^{1}$ &
    $1.68_{-0.19}^{+0.22}\times 10^{3}$ &
    $2.18_{-0.22}^{+0.24}\times 10^{0}$ &
    $4.07_{-0.11}^{+0.11}\times 10^{2}$ &
    $4.69_{-0.11}^{+0.11}\times 10^{1}$ &
    $9.64_{-0.35}^{+0.40}\times 10^{1}$ &
    $1.77_{-0.12}^{+0.13}\times 10^{2}$ \\

    8475025 &
    $2.49_{-0.05}^{+0.05}\times 10^{1}$ &
    $1.76_{-0.18}^{+0.22}\times 10^{4}$ &
    $1.34_{-0.08}^{+0.08}\times 10^{0}$ &
    $1.55_{-0.05}^{+0.05}\times 10^{3}$ &
    $2.81_{-0.07}^{+0.07}\times 10^{1}$ &
    $3.33_{-0.13}^{+0.13}\times 10^{2}$ &
    $1.03_{-0.02}^{+0.02}\times 10^{2}$ \\

    8718745 &
    $2.10_{-0.06}^{+0.06}\times 10^{1}$ &
    $9.19_{-0.97}^{+1.14}\times 10^{4}$ &
    $1.08_{-0.06}^{+0.06}\times 10^{0}$ &
    $1.57_{-0.05}^{+0.05}\times 10^{3}$ &
    $3.03_{-0.07}^{+0.07}\times 10^{1}$ &
    $3.05_{-0.10}^{+0.10}\times 10^{2}$ &
    $1.19_{-0.02}^{+0.02}\times 10^{2}$ \\

    9145955 &
    $7.96_{-0.26}^{+0.25}\times 10^{0}$ &
    $4.84_{-0.45}^{+0.55}\times 10^{3}$ &
    $2.16_{-0.15}^{+0.15}\times 10^{0}$ &
    $8.52_{-0.27}^{+0.27}\times 10^{2}$ &
    $3.33_{-0.08}^{+0.08}\times 10^{1}$ &
    $2.00_{-0.07}^{+0.08}\times 10^{2}$ &
    $1.15_{-0.02}^{+0.02}\times 10^{2}$ \\

    9267654 &
    $2.20_{-0.04}^{+0.04}\times 10^{1}$ &
    $5.12_{-0.65}^{+0.74}\times 10^{3}$ &
    $1.28_{-0.10}^{+0.12}\times 10^{0}$ &
    $9.87_{-0.31}^{+0.31}\times 10^{2}$ &
    $2.95_{-0.08}^{+0.07}\times 10^{1}$ &
    $2.47_{-0.09}^{+0.09}\times 10^{2}$ &
    $1.05_{-0.02}^{+0.02}\times 10^{2}$ \\

    9475697 &
    $1.50_{-0.03}^{+0.03}\times 10^{1}$ &
    $2.00_{-0.07}^{+0.06}\times 10^{3}$ &
    $2.60_{-0.08}^{+0.06}\times 10^{1}$ &
    $3.04_{-0.72}^{+0.52}\times 10^{2}$ &
    $9.61_{-0.97}^{+0.30}\times 10^{1}$ &
    $1.05_{-0.65}^{+0.79}\times 10^{2}$ &
    $1.08_{-0.05}^{+0.08}\times 10^{2}$ \\

    9882316 &
    $1.60_{-0.19}^{+0.17}\times 10^{1}$ &
    $1.88_{-0.19}^{+0.22}\times 10^{3}$ &
    $1.99_{-0.16}^{+0.18}\times 10^{0}$ &
    $3.72_{-0.09}^{+0.09}\times 10^{2}$ &
    $4.53_{-0.09}^{+0.09}\times 10^{1}$ &
    $8.20_{-0.24}^{+0.23}\times 10^{1}$ &
    $2.10_{-0.11}^{+0.11}\times 10^{2}$ \\

    10123207 &
    $2.16_{-0.10}^{+0.10}\times 10^{1}$ &
    $9.36_{-1.07}^{+1.24}\times 10^{3}$ &
    $1.06_{-0.08}^{+0.08}\times 10^{0}$ &
    $8.84_{-0.24}^{+0.23}\times 10^{2}$ &
    $3.83_{-0.08}^{+0.08}\times 10^{1}$ &
    $1.86_{-0.06}^{+0.06}\times 10^{2}$ &
    $1.55_{-0.04}^{+0.04}\times 10^{2}$ \\

    10200377 &
    $2.61_{-0.07}^{+0.07}\times 10^{1}$ &
    $7.52_{-0.89}^{+1.00}\times 10^{3}$ &
    $1.37_{-0.12}^{+0.14}\times 10^{0}$ &
    $8.77_{-0.26}^{+0.27}\times 10^{2}$ &
    $3.54_{-0.09}^{+0.09}\times 10^{1}$ &
    $2.25_{-0.10}^{+0.10}\times 10^{2}$ &
    $1.17_{-0.03}^{+0.03}\times 10^{2}$ \\

    10257278 &
    $2.92_{-0.09}^{+0.10}\times 10^{1}$ &
    $5.75_{-0.73}^{+0.87}\times 10^{3}$ &
    $1.05_{-0.09}^{+0.10}\times 10^{0}$ &
    $9.58_{-0.25}^{+0.27}\times 10^{2}$ &
    $3.61_{-0.09}^{+0.09}\times 10^{1}$ &
    $2.27_{-0.08}^{+0.08}\times 10^{2}$ &
    $1.36_{-0.04}^{+0.04}\times 10^{2}$ \\

    11353313 &
    $2.44_{-0.05}^{+0.05}\times 10^{1}$ &
    $7.63_{-0.88}^{+1.04}\times 10^{3}$ &
    $1.36_{-0.11}^{+0.12}\times 10^{0}$ &
    $9.74_{-0.30}^{+0.31}\times 10^{2}$ &
    $3.17_{-0.07}^{+0.08}\times 10^{1}$ &
    $2.24_{-0.08}^{+0.08}\times 10^{2}$ &
    $1.13_{-0.02}^{+0.02}\times 10^{2}$ \\

    11913545 &
    $2.39_{-0.05}^{+0.05}\times 10^{1}$ &
    $2.05_{-0.06}^{+0.06}\times 10^{3}$ &
    $2.63_{-0.06}^{+0.06}\times 10^{1}$ &
    $3.51_{-0.86}^{+0.51}\times 10^{2}$ &
    $1.04_{-0.03}^{+0.02}\times 10^{2}$ &
    $8.88_{-5.05}^{+8.99}\times 10^{1}$ &
    $1.14_{-0.05}^{+0.09}\times 10^{2}$ \\

    11968334 &
    $2.92_{-0.08}^{+0.08}\times 10^{1}$ &
    $7.60_{-0.88}^{+0.99}\times 10^{3}$ &
    $1.16_{-0.08}^{+0.09}\times 10^{0}$ &
    $9.92_{-0.27}^{+0.27}\times 10^{2}$ &
    $3.62_{-0.07}^{+0.07}\times 10^{1}$ &
    $2.17_{-0.07}^{+0.07}\times 10^{2}$ &
    $1.42_{-0.03}^{+0.03}\times 10^{2}$ \\

    12008916 &
    $2.04_{-0.11}^{+0.11}\times 10^{1}$ &
    $1.31_{-0.16}^{+0.18}\times 10^{4}$ &
    $1.06_{-0.07}^{+0.07}\times 10^{0}$ &
    $7.31_{-0.18}^{+0.19}\times 10^{2}$ &
    $4.11_{-0.09}^{+0.08}\times 10^{1}$ &
    $1.61_{-0.05}^{+0.05}\times 10^{2}$ &
    $1.62_{-0.05}^{+0.05}\times 10^{2}$ \\

    \hline
  \end{tabular}
\end{table*}

\begin{table*}
  \centering
  \caption{Same as Table \protect\ref{tab:KeplerBgs} for the oscillations power excess parameters.}
  \label{tab:KeplerPE}
  \begin{tabular}{r|c|c|c|}
    KIC &
    $P_g$ (ppm$^2/\mu$Hz) &
    $\nu_\mathrm{max}$ ($\mu$Hz) &
    $\sigma_\mathrm{env}$ ($\mu$Hz) \\
    \hline
    3744043 &
    $5.28_{-0.14}^{0.15} \times 10^{2}$ &
    $1.125_{-0.003}^{0.003} \times 10^{2}$ &
    $1.21_{-0.03}^{0.03} \times 10^{1}$ \\

    6117517 &
    $6.14_{-0.17}^{0.17} \times 10^{2}$ &
    $1.203_{-0.003}^{0.003} \times 10^{2}$ &
    $1.33_{-0.03}^{0.03} \times 10^{1}$ \\

    6144777 &
    $5.65_{-0.15}^{0.15} \times 10^{2}$ &
    $1.297_{-0.003}^{0.003} \times 10^{2}$ &
    $1.27_{-0.03}^{0.03} \times 10^{1}$ \\

    7060732 &
    $4.23_{-0.13}^{0.13} \times 10^{2}$ &
    $1.323_{-0.003}^{0.003} \times 10^{2}$ &
    $1.27_{-0.03}^{0.04} \times 10^{1}$ \\

    7619745 &
    $2.17_{-0.06}^{0.06} \times 10^{2}$ &
    $1.707_{-0.004}^{0.004} \times 10^{2}$ &
    $1.49_{-0.04}^{0.04} \times 10^{1}$ \\

    8366239 &
    $1.43_{-0.04}^{0.05} \times 10^{2}$ &
    $1.857_{-0.005}^{0.005} \times 10^{2}$ &
    $1.69_{-0.07}^{0.07} \times 10^{1}$ \\

    8475025 &
    $5.81_{-0.17}^{0.18} \times 10^{2}$ &
    $1.129_{-0.003}^{0.003} \times 10^{2}$ &
    $1.09_{-0.03}^{0.03} \times 10^{1}$ \\

    8718745 &
    $5.72_{-0.16}^{0.17} \times 10^{2}$ &
    $1.296_{-0.003}^{0.003} \times 10^{2}$ &
    $1.15_{-0.03}^{0.03} \times 10^{1}$ \\

    9145955 &
    $2.84_{-0.07}^{0.07} \times 10^{2}$ &
    $1.320_{-0.004}^{0.004} \times 10^{2}$ &
    $1.50_{-0.03}^{0.04} \times 10^{1}$ \\

    9267654 &
    $4.65_{-0.13}^{0.14} \times 10^{2}$ &
    $1.184_{-0.003}^{0.003} \times 10^{2}$ &
    $1.14_{-0.03}^{0.03} \times 10^{1}$ \\

    9475697 &
    $5.82_{-0.16}^{0.15} \times 10^{2}$ &
    $1.150_{-0.003}^{0.003} \times 10^{2}$ &
    $1.29_{-0.03}^{0.04} \times 10^{1}$ \\

    9882316 &
    $1.06_{-0.04}^{0.04} \times 10^{2}$ &
    $1.823_{-0.005}^{0.005} \times 10^{2}$ &
    $1.52_{-0.07}^{0.07} \times 10^{1}$ \\

    10123207 &
    $4.46_{-0.12}^{0.12} \times 10^{2}$ &
    $1.607_{-0.003}^{0.003} \times 10^{2}$ &
    $1.26_{-0.03}^{0.03} \times 10^{1}$ \\

    10200377 &
    $3.56_{-0.09}^{0.09} \times 10^{2}$ &
    $1.433_{-0.003}^{0.003} \times 10^{2}$ &
    $1.47_{-0.31}^{0.04} \times 10^{1}$ \\

    10257278 &
    $4.11_{-0.12}^{0.12} \times 10^{2}$ &
    $1.503_{-0.003}^{0.003} \times 10^{2}$ &
    $1.24_{-0.04}^{0.04} \times 10^{1}$ \\

    11353313 &
    $3.75_{-0.10}^{0.11} \times 10^{2}$ &
    $1.264_{-0.003}^{0.003} \times 10^{2}$ &
    $1.25_{-0.03}^{0.03} \times 10^{1}$ \\

    11913545 &
    $7.85_{-0.24}^{0.26} \times 10^{2}$ &
    $1.173_{-0.003}^{0.003} \times 10^{2}$ &
    $1.08_{-0.02}^{0.03} \times 10^{1}$ \\

    11968334 &
    $4.55_{-0.14}^{0.15} \times 10^{2}$ &
    $1.413_{-0.003}^{0.003} \times 10^{2}$ &
    $1.10_{-0.03}^{0.03} \times 10^{1}$ \\

    12008916 &
    $3.02_{-0.08}^{0.08} \times 10^{2}$ &
    $1.613_{-0.003}^{0.004} \times 10^{2}$ &
    $1.49_{-0.04}^{0.04} \times 10^{1}$ \\

    \hline
  \end{tabular}
\end{table*}

\section{Results for the oscillation modes}\label{sec:KeplerOscillations}

We applied the peak finding procedure to the 19 red-giant stars analysed by \cite{Corsaro2015} as described in Section \ref{sec:Kepler}.
We then used the identified peaks as initial values for a parameter optimisation using MLE.
When fitting a Lorentzian profile in a PDS the height and line width show high correlations which could hider the MLE optimisation.
For this reason it is more stable to use optimise the amplitude instead of the height.
That is, during the MLE optimisation each Lorentzian profile is described by
\begin{equation}\label{eq:ResProfile}
O_k^\mathrm{resolved} = \frac{A_k^2}{\pi \gamma_k\left[1 + \left(\frac{\nu-\nu_k}{\gamma_k}\right)^2\right]}
\end{equation}
where $A_k=\sqrt{\pi I_k \gamma_k}$ is the amplitude of a Lorentzian profile centered at $\nu_k$, with height $I_k$ and half-width at half-maximum $\gamma_k$.

We described the unresolved oscillations in a PDS having frequency resolution of $\delta\nu$ as
\begin{equation}\label{eq:UnrProfile}
O_k^\mathrm{unresolved} = H_k \mathrm{sinc}^2\left(\frac{\nu - \nu_k}{\delta\nu}\right)
\end{equation}
where $H_k$ is the height and $\nu_k$ is the location of the oscillation.
The unresolved oscillations are detected by identified single points in the PDS with a false-alarm probability lower than $10^{-4}$.
When a significant oscillation is deteted in this way it is fitted with both Eq. \ref{eq:ResProfile} and \ref{eq:UnrProfile} and the model with the lower AIC is chosen. 
If model \ref{eq:ResProfile} is preferred, it is reported as a resolved oscillation, however, the SNR value is missing in the table.

In the following tables we report the obtained parameteres for the resolved and unresolved oscillations using Eqs. (\ref{eq:ResProfile}) and (\ref{eq:UnrProfile}) respectively.
The uncertainties are derived from the covariance matrix as estimated by the inverse of the Hessian matrix evaluated at the MLE.
The reported \textit{AIC} value corresponds to the difference in AIC between a PDS model having and omiting that oscillation.

\clearpage
\begin{table}
\centering
\begin{tabular}{ccccc}
  \hline
$\nu_k$ & $A_k$ & $\gamma_k$ & SNR & AIC \\ 
  \hline
77.346$\pm$0.013 & 0.408$\pm$0.117 & 0.016$\pm$0.013 & 2.23 & 5.58 \\ 
  80.273$\pm$0.016 & 0.424$\pm$0.113 & 0.02$\pm$0.015 & 2.24 & 5.35 \\ 
  81.711$\pm$0.02 & 0.888$\pm$0.118 & 0.051$\pm$0.018 & 6.30 & 59.13 \\ 
  86.586$\pm$0.005 & 0.735$\pm$0.174 & 0.009$\pm$0.006 & 9.09 & 62.09 \\ 
  89.771$\pm$0.036 & 0.919$\pm$0.121 & 0.107$\pm$0.045 & 3.87 & 42.59 \\ 
  91.109$\pm$0.014 & 1.323$\pm$0.147 & 0.048$\pm$0.015 & 11.96 & 178.70 \\ 
  96.065$\pm$0.142 & 1.458$\pm$0.182 & 0.4$\pm$0.158 & 2.08 & 25.66 \\
\hline
\end{tabular}
\caption{Resolved oscillations for KIC 3744043 .} 
\label{tab:003744043resolved}
\end{table}
\begin{table}
\centering
\begin{tabular}{ccc}
  \hline
$\nu_k$ & $H_k$ & AIC \\ 
  \hline
105.89$\pm$0.002 & 32.931$\pm$31.542 & 9.79 \\ 
  105.951$\pm$0.002 & 69.363$\pm$51.538 & 17.41 \\ 
  106.737$\pm$0.002 & 63.729$\pm$46.161 & 36.35 \\ 
  107.529$\pm$0.002 & 84.342$\pm$43.169 & 82.07 \\ 
  108.382$\pm$0.002 & 37.01$\pm$23.872 & 23.61 \\ 
  113.904$\pm$0.001 & 50.605$\pm$40.142 & 63.95 \\ 
  116.848$\pm$0.002 & 84.986$\pm$46.947 & 54.30 \\ 
  118.102$\pm$0.001 & 31.681$\pm$21.444 & 41.29 \\ 
  124.615$\pm$0.003 & 49.973$\pm$56.838 & 0.73 \\ 
  124.632$\pm$0.002 & 172.94$\pm$93.903 & 133.71 \\ 
  128.467$\pm$0.002 & 52.997$\pm$34.592 & 53.68 \\ 
   \hline
\end{tabular}
\caption{Unresolved oscillations for KIC 3744043 .} 
\label{tab:003744043unresolved}
\end{table}
\begin{figure}\includegraphics[width=0.47\textwidth]{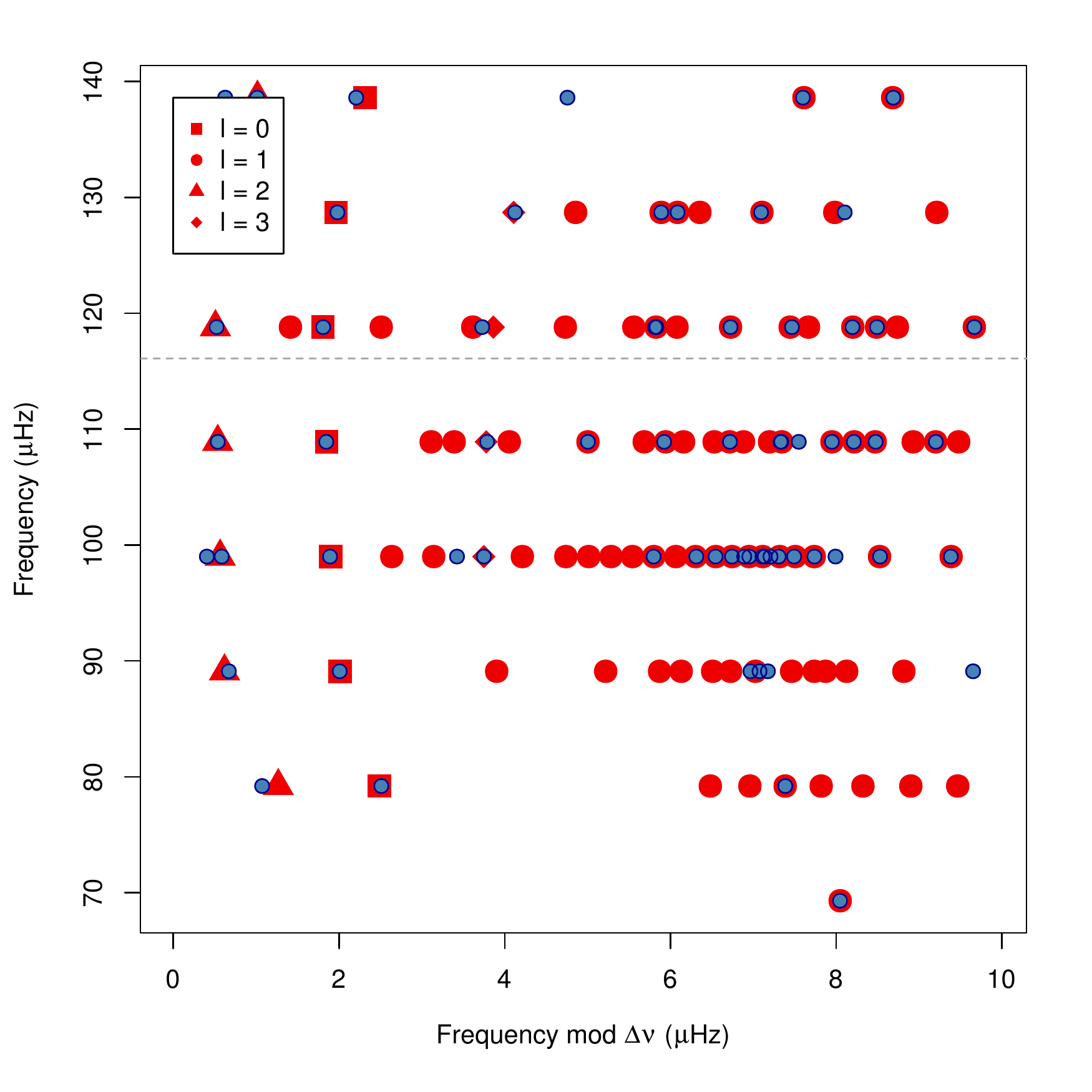}\caption{Same as Fig. \protect\ref{fig:KIC012008916echelle} for KIC 3744043.}\label{fig:KIC003744043echelle}\end{figure}\clearpage
\begin{table}
\centering
\begin{tabular}{ccccc}
  \hline
$\nu_k$ & $A_k$ & $\gamma_k$ & SNR & AIC \\ 
  \hline
67.114$\pm$0.018 & 0.423$\pm$0.148 & 0.023$\pm$0.034 & 1.52 & 4.72 \\ 
  68.496$\pm$0.038 & 0.466$\pm$0.123 & 0.053$\pm$0.04 & 1.65 & 2.85 \\ 
  89.088$\pm$0.01 & 0.718$\pm$0.132 & 0.021$\pm$0.011 & 5.64 & 44.86 \\ 
  92.379$\pm$0.005 & 0.642$\pm$0.18 & 0.007$\pm$0.007 & 4.85 & 44.95 \\ 
  93.71$\pm$0.038 & 1.073$\pm$0.115 & 0.124$\pm$0.039 & 5.36 & 69.58 \\ 
  98.595$\pm$0.006 & 0.929$\pm$0.184 & 0.013$\pm$0.007 & 12.17 & 103.93 \\ 
  99.073$\pm$0.005 & 0.674$\pm$0.174 & 0.008$\pm$0.006 & 6.96 & 47.59 \\
\hline
\end{tabular}
\caption{Resolved oscillations for KIC 6117517 .} 
\label{tab:006117517resolved}
\end{table}
\begin{table}
\centering
\begin{tabular}{ccc}
  \hline
$\nu_k$ & $H_k$ & AIC \\ 
  \hline
98.05$\pm$0.002 & 117.084$\pm$59.233 & 101.87 \\ 
  107.811$\pm$0.002 & 97.1$\pm$51.656 & 116.98 \\ 
  109.066$\pm$0.001 & 726.76$\pm$234.549 & 398.68 \\ 
  116.561$\pm$0.002 & 31.435$\pm$28.226 & 46.70 \\ 
  117.565$\pm$0.001 & 123.591$\pm$84.557 & 248.40 \\ 
  119.089$\pm$0.001 & 4803.886$\pm$1292.46 & 990.16 \\ 
  119.93$\pm$0.001 & 353.564$\pm$164.366 & 346.94 \\ 
  125.57$\pm$0.001 & 44.673$\pm$45.806 & 65.45 \\ 
  126.763$\pm$0.001 & 73.775$\pm$52.275 & 95.38 \\ 
  127.926$\pm$0.001 & 334.899$\pm$146.307 & 427.70 \\ 
  129.553$\pm$0.001 & 1054.725$\pm$327.149 & 238.62 \\ 
  130.716$\pm$0.001 & 164.473$\pm$74.261 & 172.81 \\ 
  132.286$\pm$0.002 & 81.004$\pm$61.568 & 37.35 \\ 
  133.114$\pm$0.002 & 39.786$\pm$38.407 & 22.61 \\ 
  137.434$\pm$0.001 & 93.987$\pm$54.975 & 120.10 \\ 
   \hline
\end{tabular}
\caption{Unresolved oscillations for KIC 6117517 .} 
\label{tab:006117517unresolved}
\end{table}
\begin{figure}\includegraphics[width=0.47\textwidth]{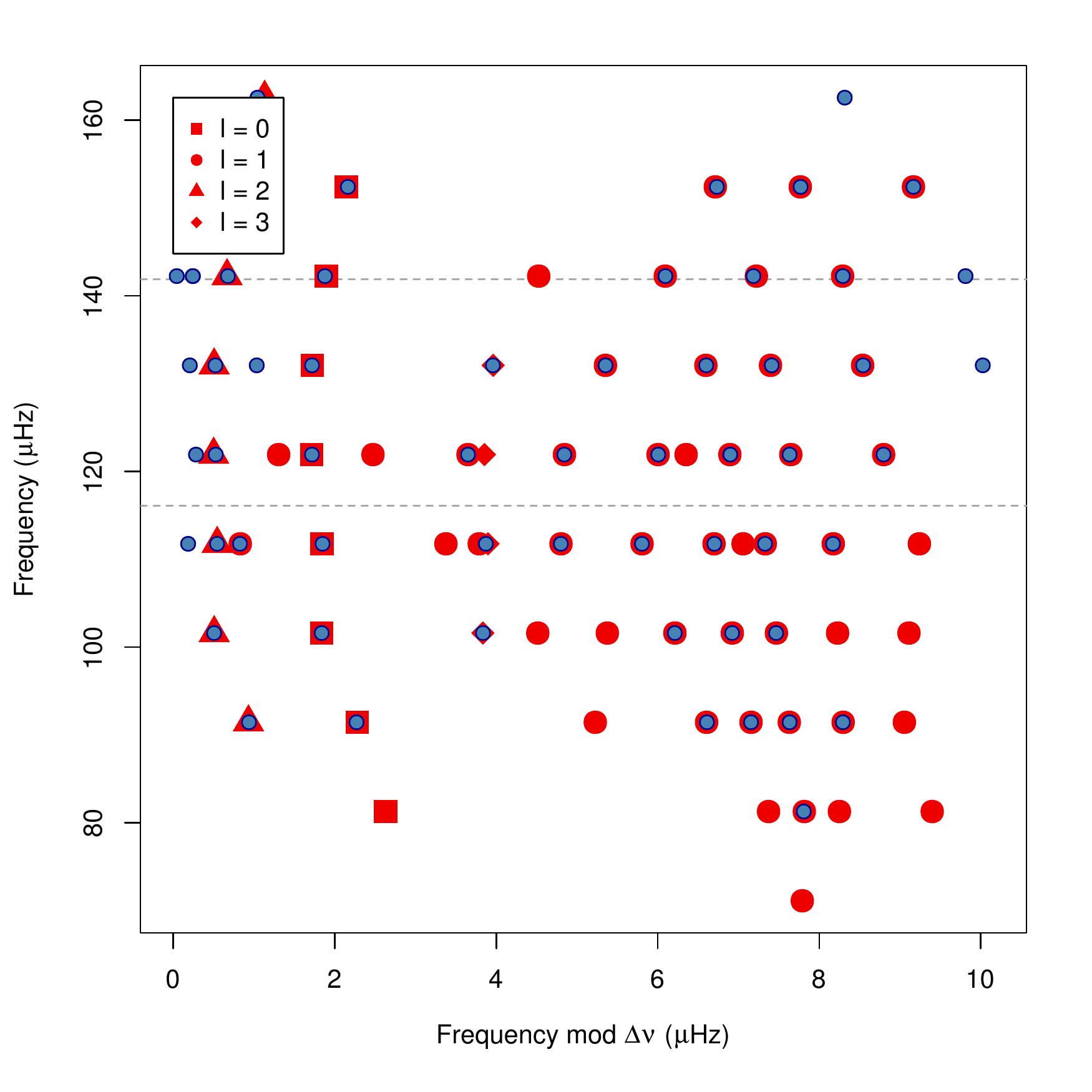}\caption{Same as Fig. \protect\ref{fig:KIC012008916echelle} for KIC 6117517.}\label{fig:KIC006117517echelle}\end{figure}\clearpage
\begin{table}
\centering
\begin{tabular}{ccccc}
  \hline
$\nu_k$ & $A_k$ & $\gamma_k$ & SNR & AIC \\ 
  \hline
86.174$\pm$0.046 & 0.587$\pm$0.159 & 0.09$\pm$0.093 & 1.94 & 7.44 \\ 
  96.907$\pm$0.042 & 0.756$\pm$0.174 & 0.116$\pm$0.112 & 2.26 & 19.81 \\ 
  97.445$\pm$0.005 & 0.314$\pm$0.141 & 0.005$\pm$0.011 & 1.65 & 1.50 \\ 
  100.483$\pm$0.032 & 0.55$\pm$0.118 & 0.054$\pm$0.033 & 1.92 & 8.64 \\ 
  101.924$\pm$0.014 & 1.048$\pm$0.136 & 0.042$\pm$0.018 & 8.07 & 102.77 \\ 
  107.172$\pm$0.067 & 1.132$\pm$0.145 & 0.214$\pm$0.081 & 1.53 & 45.56 \\ 
  107.612$\pm$0.02 & 1.036$\pm$0.144 & 0.056$\pm$0.025 & 6.15 & 52.97 \\
\hline
\end{tabular}
\caption{Resolved oscillations for KIC 6144777 .} 
\label{tab:006144777resolved}
\end{table}
\begin{table}
\centering
\begin{tabular}{ccc}
  \hline
$\nu_k$ & $H_k$ & AIC \\ 
  \hline
117.521$\pm$0.002 & 45.826$\pm$26.688 & 51.09 \\ 
  118.626$\pm$0.002 & 76.016$\pm$66.075 & 4.94 \\ 
  118.648$\pm$0.002 & 161.22$\pm$133.998 & 36.62 \\ 
  118.997$\pm$0.002 & 144.317$\pm$73.789 & 98.76 \\ 
  121.087$\pm$0.002 & 34.776$\pm$21.847 & 31.75 \\ 
  127.034$\pm$0.001 & 65.586$\pm$51.112 & 68.83 \\ 
  127.786$\pm$0.002 & 53.426$\pm$38.098 & 56.73 \\ 
  128.236$\pm$0.002 & 234.129$\pm$135.883 & 109.43 \\ 
  129.594$\pm$0.001 & 1339.976$\pm$639.958 & 158.25 \\ 
  129.955$\pm$0.001 & 625.488$\pm$293.78 & 324.77 \\ 
  130.668$\pm$0.001 & 88.336$\pm$54.587 & 90.59 \\ 
  131.119$\pm$0.002 & 203.146$\pm$124.72 & 160.28 \\ 
  132.738$\pm$0.001 & 132.713$\pm$93.244 & 87.96 \\ 
  134.974$\pm$0.002 & 37.783$\pm$28.857 & 16.03 \\ 
  138.077$\pm$0.001 & 38.795$\pm$26.616 & 43.53 \\ 
  141.021$\pm$0.002 & 101.514$\pm$63.871 & 40.41 \\ 
   \hline
\end{tabular}
\caption{Unresolved oscillations for KIC 6144777 .} 
\label{tab:006144777unresolved}
\end{table}
\begin{figure}\includegraphics[width=0.47\textwidth]{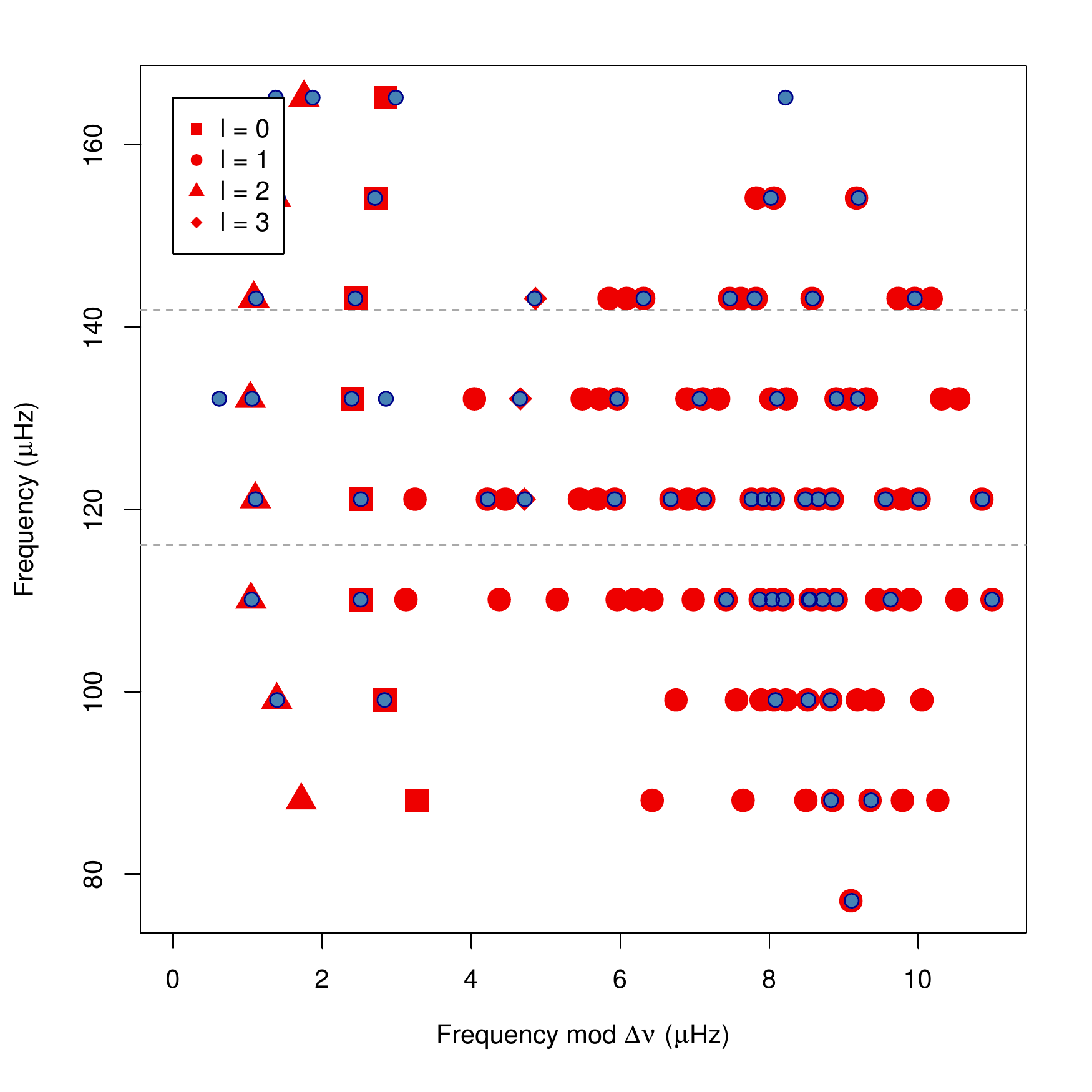}\caption{Same as Fig. \protect\ref{fig:KIC012008916echelle} for KIC 6144777.}\label{fig:KIC006144777echelle}\end{figure}\clearpage
\begin{table}
\centering
\begin{tabular}{ccccc}
  \hline
$\nu_k$ & $A_k$ & $\gamma_k$ & SNR & AIC \\ 
  \hline
90.76$\pm$0.015 & 0.557$\pm$0.116 & 0.027$\pm$0.017 & 2.96 & 17.64 \\ 
  95.679$\pm$0.004 & 0.364$\pm$0.152 & 0.005$\pm$0.011 & 2.25 & 6.26 \\ 
  96.285$\pm$0.015 & 0.384$\pm$0.113 & 0.018$\pm$0.015 & 1.88 & 2.81 \\ 
  101.327$\pm$0.03 & 0.893$\pm$0.112 & 0.077$\pm$0.026 & 5.25 & 48.70 \\ 
  106.715$\pm$0.039 & 1.158$\pm$0.125 & 0.157$\pm$0.059 & 5.31 & 78.31 \\ 
  107.157$\pm$0.004 & 0.371$\pm$0.156 & 0.005$\pm$0.011 & 2.95 & 3.13 \\ 
  107.888$\pm$0.008 & 0.34$\pm$0.118 & 0.007$\pm$0.006 & 2.78 & 4.48 \\
\hline
\end{tabular}
\caption{Resolved oscillations for KIC 7060732 .} 
\label{tab:007060732resolved}
\end{table}
\begin{table}
\centering
\begin{tabular}{ccc}
  \hline
$\nu_k$ & $H_k$ & AIC \\ 
  \hline
116.797$\pm$0.002 & 70.56$\pm$45.876 & 40.42 \\ 
  118.225$\pm$0.002 & 96.389$\pm$58.644 & 36.66 \\ 
  126.752$\pm$0.001 & 81.066$\pm$49.444 & 95.79 \\ 
  127.311$\pm$0.001 & 179.366$\pm$104.066 & 258.20 \\ 
  127.852$\pm$0.001 & 211.069$\pm$162.818 & 246.88 \\ 
  130.743$\pm$0.002 & 46.288$\pm$47.778 & 66.49 \\ 
  138.281$\pm$0.001 & 149.966$\pm$100.648 & 132.00 \\ 
   \hline
\end{tabular}
\caption{Unresolved oscillations for KIC 7060732 .} 
\label{tab:007060732unresolved}
\end{table}
\begin{figure}\includegraphics[width=0.47\textwidth]{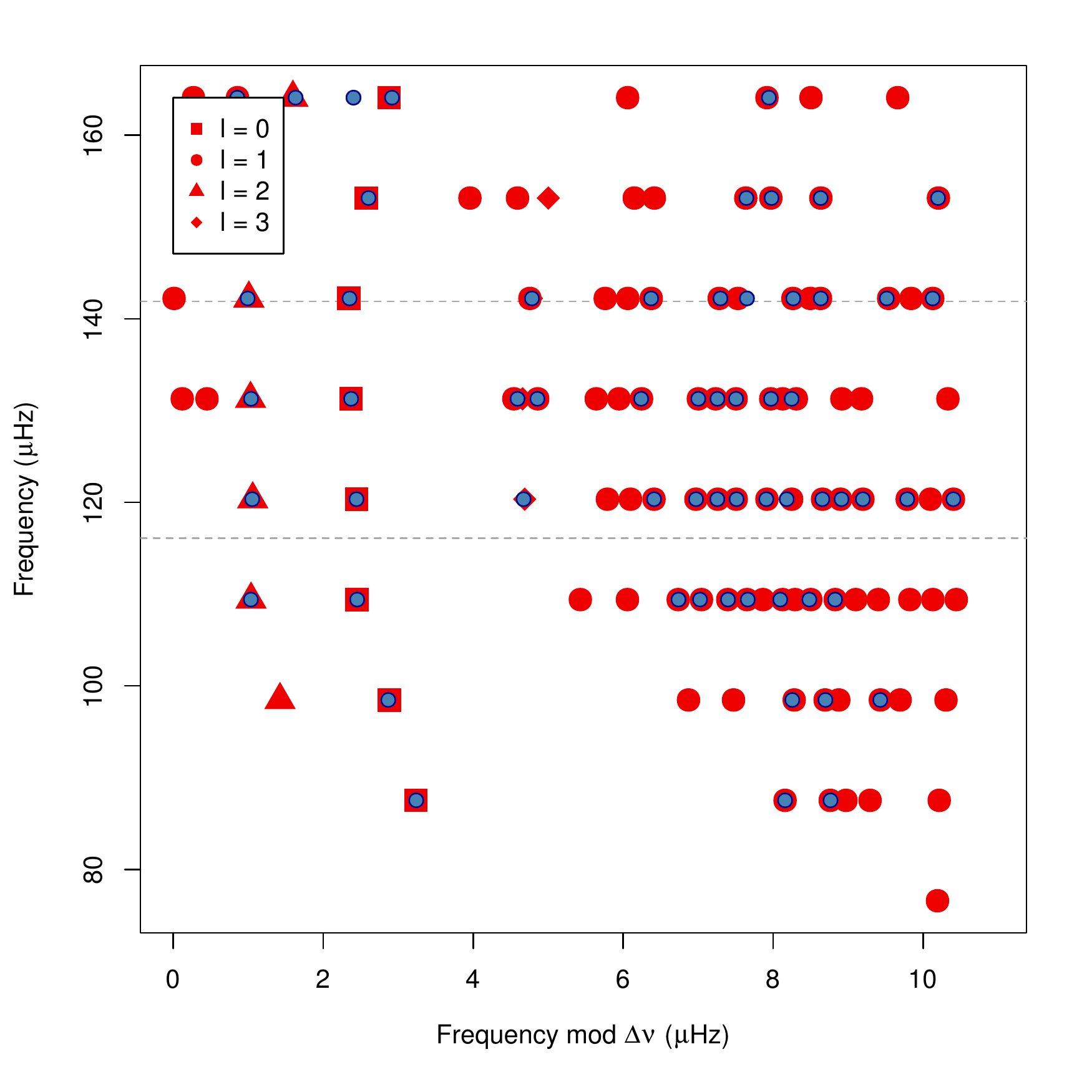}\caption{Same as Fig. \protect\ref{fig:KIC012008916echelle} for KIC 7060732.}\label{fig:KIC007060732echelle}\end{figure}\clearpage
\begin{table}
\centering
\begin{tabular}{ccccc}
  \hline
$\nu_k$ & $A_k$ & $\gamma_k$ & SNR & AIC \\ 
  \hline
122.424$\pm$0.049 & 0.546$\pm$0.125 & 0.075$\pm$0.05 & 1.76 & 5.99 \\ 
  128.42$\pm$0.018 & 0.486$\pm$0.16 & 0.013$\pm$0.022 & 3.24 & 14.07 \\ 
  128.725$\pm$0.01 & 0.545$\pm$0.123 & 0.017$\pm$0.012 & 3.18 & 18.87 \\ 
  129.195$\pm$0.023 & 0.49$\pm$0.117 & 0.029$\pm$0.021 & 2.37 & 8.58 \\ 
  133.25$\pm$0.005 & 0.472$\pm$0.171 & 0.005$\pm$0.007 & 4.31 & 15.57 \\ 
  133.449$\pm$0.019 & 0.557$\pm$0.126 & 0.036$\pm$0.028 & 1.87 & 12.72 \\ 
  135.018$\pm$0.028 & 1.124$\pm$0.117 & 0.095$\pm$0.028 & 6.41 & 91.69 \\
\hline
\end{tabular}
\caption{Resolved oscillations for KIC 7619745 .} 
\label{tab:007619745resolved}
\end{table}
\begin{table}
\centering
\begin{tabular}{ccc}
  \hline
$\nu_k$ & $H_k$ & AIC \\ 
  \hline
141.996$\pm$0.002 & 24.341$\pm$17.567 & 21.78 \\ 
  152.631$\pm$0.002 & 21.4$\pm$16.24 & 16.53 \\ 
  156.725$\pm$0.001 & 47.87$\pm$30.048 & 106.10 \\ 
  162.619$\pm$0.001 & 46.85$\pm$31.866 & 90.53 \\ 
  164.669$\pm$0.001 & 85.663$\pm$60.049 & 129.06 \\ 
  171.562$\pm$0.001 & 37.569$\pm$29.887 & 45.63 \\ 
  172.092$\pm$0.002 & 384.319$\pm$190.437 & 105.06 \\ 
  174.598$\pm$0.002 & 26.466$\pm$28.051 & 21.70 \\ 
  178.569$\pm$0.002 & 151.005$\pm$59.669 & 143.22 \\ 
  184.091$\pm$0.002 & 23.311$\pm$19.106 & 23.50 \\ 
  186.726$\pm$0.002 & 58.27$\pm$48.794 & 32.50 \\ 
   \hline
\end{tabular}
\caption{Unresolved oscillations for KIC 7619745 .} 
\label{tab:007619745unresolved}
\end{table}
\begin{figure}\includegraphics[width=0.47\textwidth]{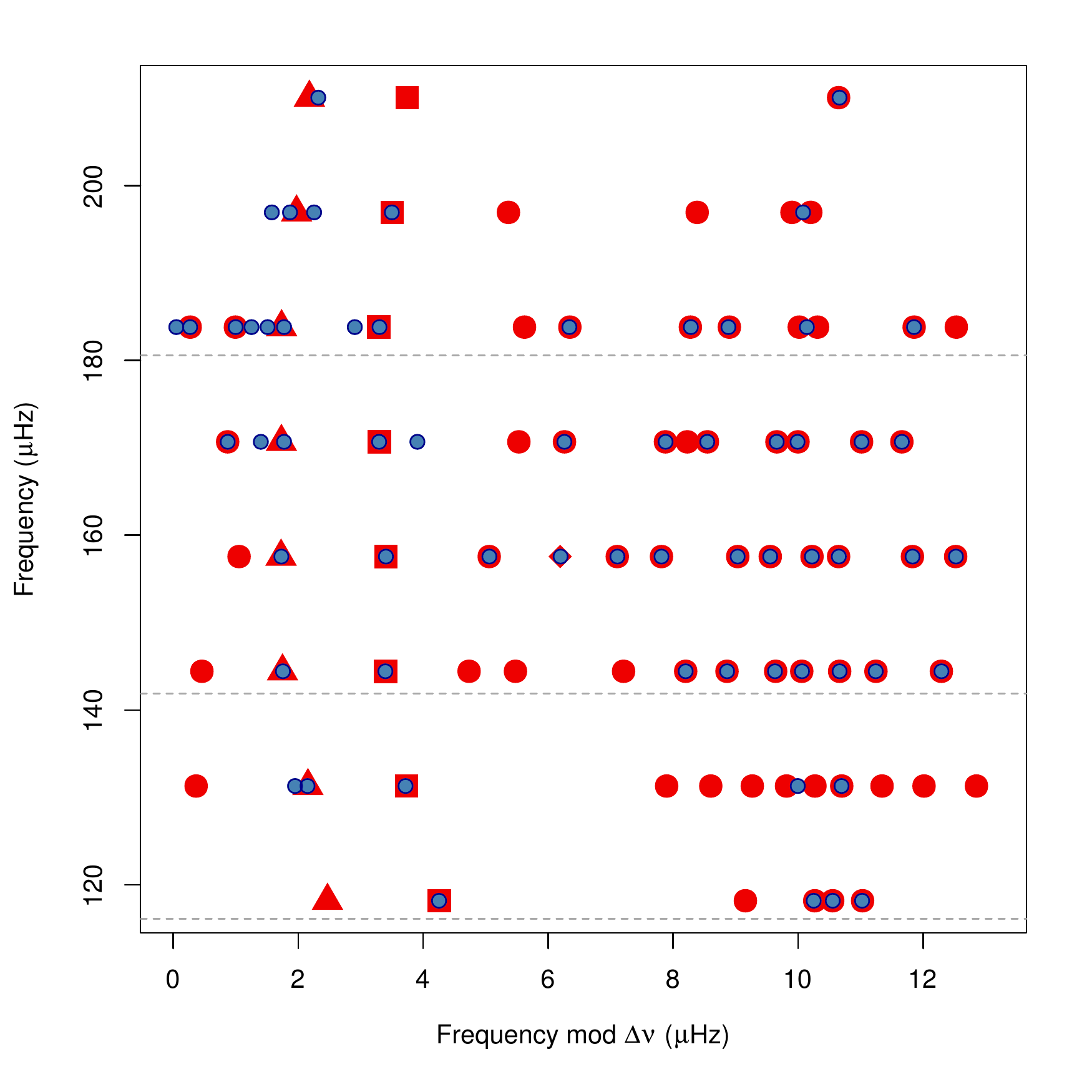}\caption{Same as Fig. \protect\ref{fig:KIC012008916echelle} for KIC 7619745.}\label{fig:KIC007619745echelle}\end{figure}\clearpage
\begin{table}
\centering
\begin{tabular}{ccccc}
  \hline
$\nu_k$ & $A_k$ & $\gamma_k$ & SNR & AIC \\ 
  \hline
134.935$\pm$0.006 & 0.463$\pm$0.135 & 0.008$\pm$0.009 & 3.21 & 14.93 \\ 
  135.937$\pm$0.01 & 0.489$\pm$0.126 & 0.014$\pm$0.012 & 3.00 & 14.83 \\ 
  137.402$\pm$0.006 & 0.322$\pm$0.153 & 0.006$\pm$0.009 & 1.85 & 3.93 \\ 
  141.663$\pm$0.024 & 0.87$\pm$0.115 & 0.062$\pm$0.023 & 4.92 & 50.75 \\ 
  147.599$\pm$0.01 & 0.611$\pm$0.134 & 0.018$\pm$0.014 & 3.92 & 26.44 \\ 
  147.898$\pm$0.01 & 0.739$\pm$0.134 & 0.021$\pm$0.011 & 5.59 & 46.42 \\ 
  148.643$\pm$0.03 & 0.455$\pm$0.121 & 0.038$\pm$0.029 & 1.68 & 3.71 \\
\hline
\end{tabular}
\caption{Resolved oscillations for KIC 8366239 .} 
\label{tab:008366239resolved}
\end{table}
\begin{table}
\centering
\begin{tabular}{ccc}
  \hline
$\nu_k$ & $H_k$ & AIC \\ 
  \hline
160.716$\pm$0.001 & 73.978$\pm$66.04 & 64.27 \\ 
  163.206$\pm$0.002 & 32.866$\pm$20.642 & 41.90 \\ 
  176.489$\pm$0.001 & 557.175$\pm$275.38 & 240.20 \\ 
  180.734$\pm$0.002 & 21.878$\pm$24.397 & 6.87 \\ 
  184.476$\pm$0.001 & 55.231$\pm$34.098 & 89.87 \\ 
  193.592$\pm$0.002 & 22.057$\pm$14.502 & 15.85 \\ 
  200.089$\pm$0.002 & 47.01$\pm$32.87 & 52.35 \\ 
   \hline
\end{tabular}
\caption{Unresolved oscillations for KIC 8366239 .} 
\label{tab:008366239unresolved}
\end{table}
\begin{figure}\includegraphics[width=0.47\textwidth]{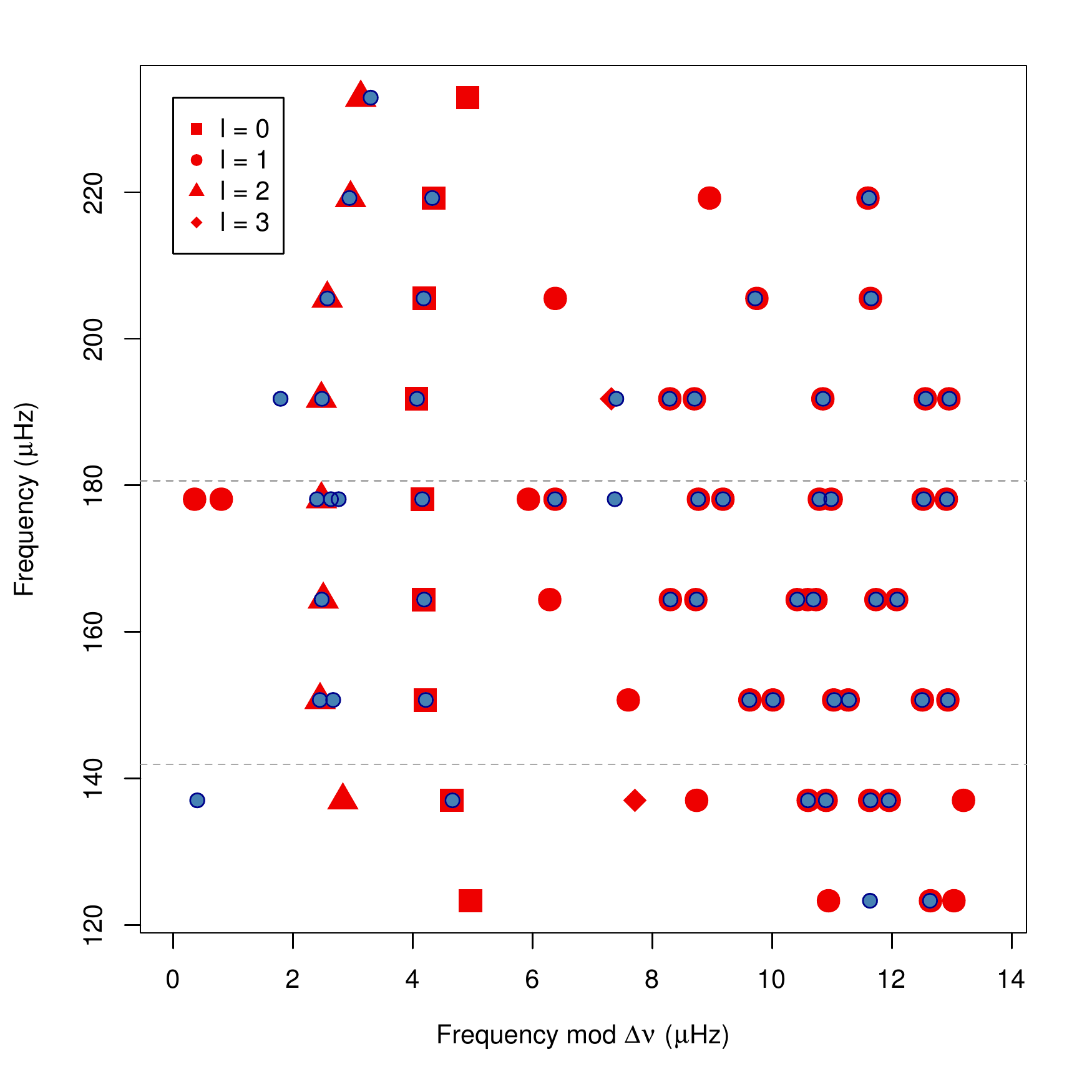}\caption{Same as Fig. \protect\ref{fig:KIC012008916echelle} for KIC 8366239.}\label{fig:KIC008366239echelle}\end{figure}\clearpage
\begin{table}
\centering
\begin{tabular}{ccccc}
  \hline
$\nu_k$ & $A_k$ & $\gamma_k$ & SNR & AIC \\ 
  \hline
79.651$\pm$0.014 & 0.714$\pm$0.122 & 0.032$\pm$0.017 & 5.01 & 36.86 \\ 
  84.28$\pm$0.034 & 0.547$\pm$0.114 & 0.051$\pm$0.028 & 2.45 & 8.76 \\ 
  84.631$\pm$0.01 & 0.641$\pm$0.153 & 0.014$\pm$0.014 & 4.60 & 33.92 \\ 
  87.684$\pm$0.053 & 0.708$\pm$0.137 & 0.101$\pm$0.068 & 2.89 & 16.09 \\ 
  88.897$\pm$0.05 & 0.968$\pm$0.113 & 0.138$\pm$0.042 & 4.14 & 44.92 \\ 
  92.566$\pm$0.004 & 0.457$\pm$0.18 & 0.005$\pm$0.007 & 3.98 & 15.80 \\ 
  93.159$\pm$0.006 & 0.631$\pm$0.152 & 0.01$\pm$0.008 & 5.66 & 36.76 \\
\hline
\end{tabular}
\caption{Resolved oscillations for KIC 8475025 .} 
\label{tab:008475025resolved}
\end{table}
\begin{table}
\centering
\begin{tabular}{ccc}
  \hline
$\nu_k$ & $H_k$ & AIC \\ 
  \hline
111.677$\pm$0.001 & 260.158$\pm$122.581 & 289.05 \\ 
  112.48$\pm$0.001 & 664.33$\pm$239.881 & 370.03 \\ 
  112.677$\pm$0 & 4016.409$\pm$2042.859 & 238.18 \\ 
  113.747$\pm$0.001 & 127.029$\pm$72.124 & 124.57 \\ 
  114.027$\pm$0.001 & 62.034$\pm$28.206 & 105.55 \\ 
  114.658$\pm$0.001 & 55.77$\pm$36.614 & 62.15 \\ 
  120.674$\pm$0.001 & 49.183$\pm$25.796 & 83.53 \\ 
  121.679$\pm$0.001 & 58.492$\pm$31.924 & 83.11 \\ 
  130.547$\pm$0.002 & 33.018$\pm$19.354 & 45.65 \\ 
   \hline
\end{tabular}
\caption{Unresolved oscillations for KIC 8475025 .} 
\label{tab:008475025unresolved}
\end{table}
\begin{figure}\includegraphics[width=0.47\textwidth]{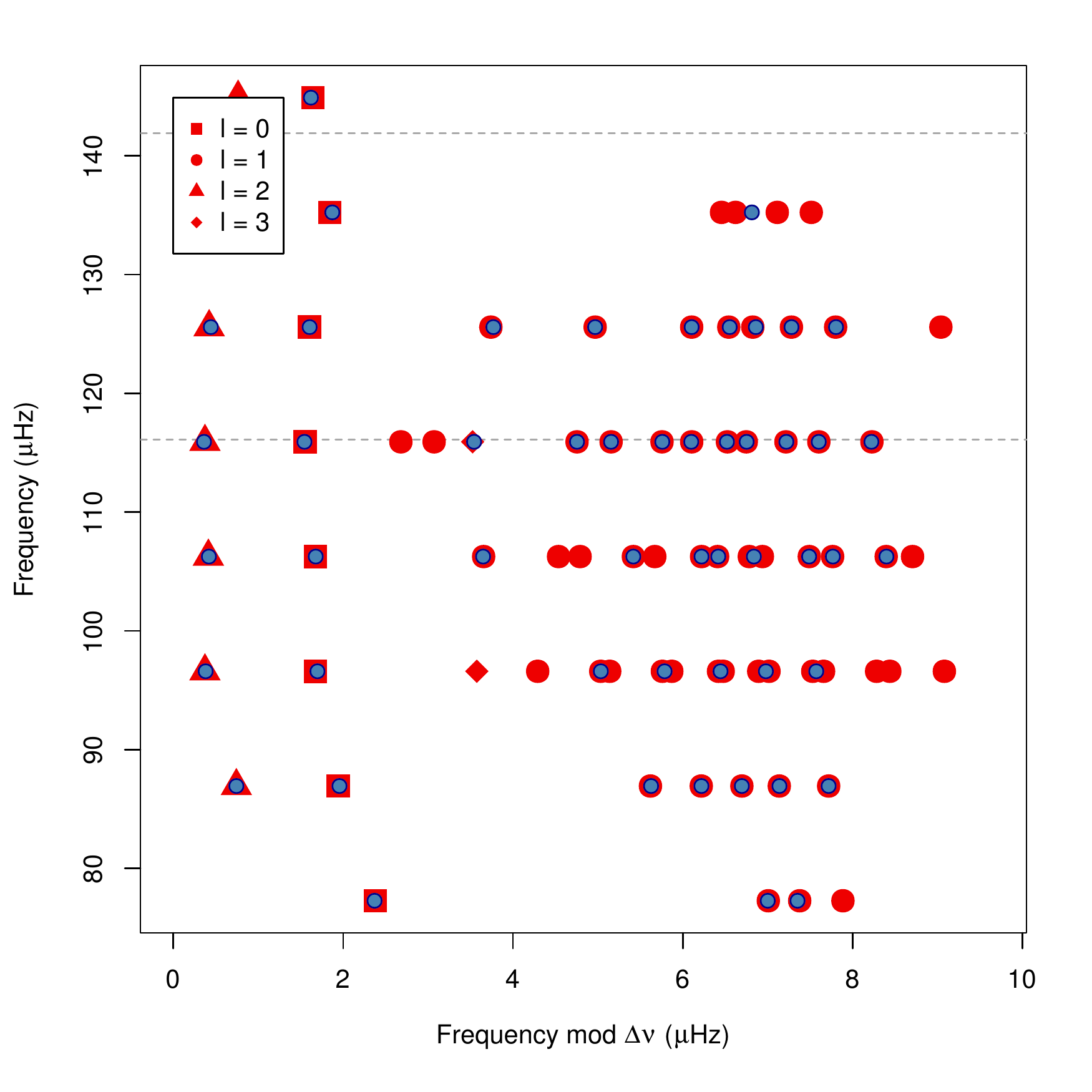}\caption{Same as Fig. \protect\ref{fig:KIC012008916echelle} for KIC 8475025.}\label{fig:KIC008475025echelle}\end{figure}\clearpage
\begin{table}
\centering
\begin{tabular}{ccccc}
  \hline
$\nu_k$ & $A_k$ & $\gamma_k$ & SNR & AIC \\ 
  \hline
92.835$\pm$0.023 & 0.538$\pm$0.148 & 0.038$\pm$0.044 & 2.41 & 11.07 \\ 
  94.452$\pm$0.014 & 0.737$\pm$0.123 & 0.03$\pm$0.015 & 4.82 & 41.77 \\ 
  98.481$\pm$0.004 & 0.462$\pm$0.175 & 0.005$\pm$0.008 & 2.86 & 19.09 \\ 
  99.855$\pm$0.008 & 0.625$\pm$0.152 & 0.012$\pm$0.012 & 4.68 & 33.27 \\ 
  100.369$\pm$0.036 & 0.682$\pm$0.131 & 0.081$\pm$0.051 & 1.75 & 16.37 \\ 
  100.708$\pm$0.027 & 0.462$\pm$0.131 & 0.021$\pm$0.02 & 2.82 & 6.70 \\ 
  103.718$\pm$0.029 & 0.942$\pm$0.114 & 0.081$\pm$0.027 & 5.08 & 56.33 \\
\hline
\end{tabular}
\caption{Resolved oscillations for KIC 8718745 .} 
\label{tab:008718745resolved}
\end{table}
\begin{table}
\centering
\begin{tabular}{ccc}
  \hline
$\nu_k$ & $H_k$ & AIC \\ 
  \hline
120.32$\pm$0.001 & 160.198$\pm$104.719 & 269.02 \\ 
  121.407$\pm$0.001 & 245.653$\pm$96.669 & 215.21 \\ 
  121.788$\pm$0.001 & 41.844$\pm$34.991 & 31.42 \\ 
  125.275$\pm$0.001 & 195.449$\pm$106.007 & 174.84 \\ 
  130.403$\pm$0.001 & 31.414$\pm$25.306 & 48.84 \\ 
  136.222$\pm$0.001 & 189.148$\pm$112.914 & 313.72 \\ 
   \hline
\end{tabular}
\caption{Unresolved oscillations for KIC 8718745 .} 
\label{tab:008718745unresolved}
\end{table}
\begin{figure}\includegraphics[width=0.47\textwidth]{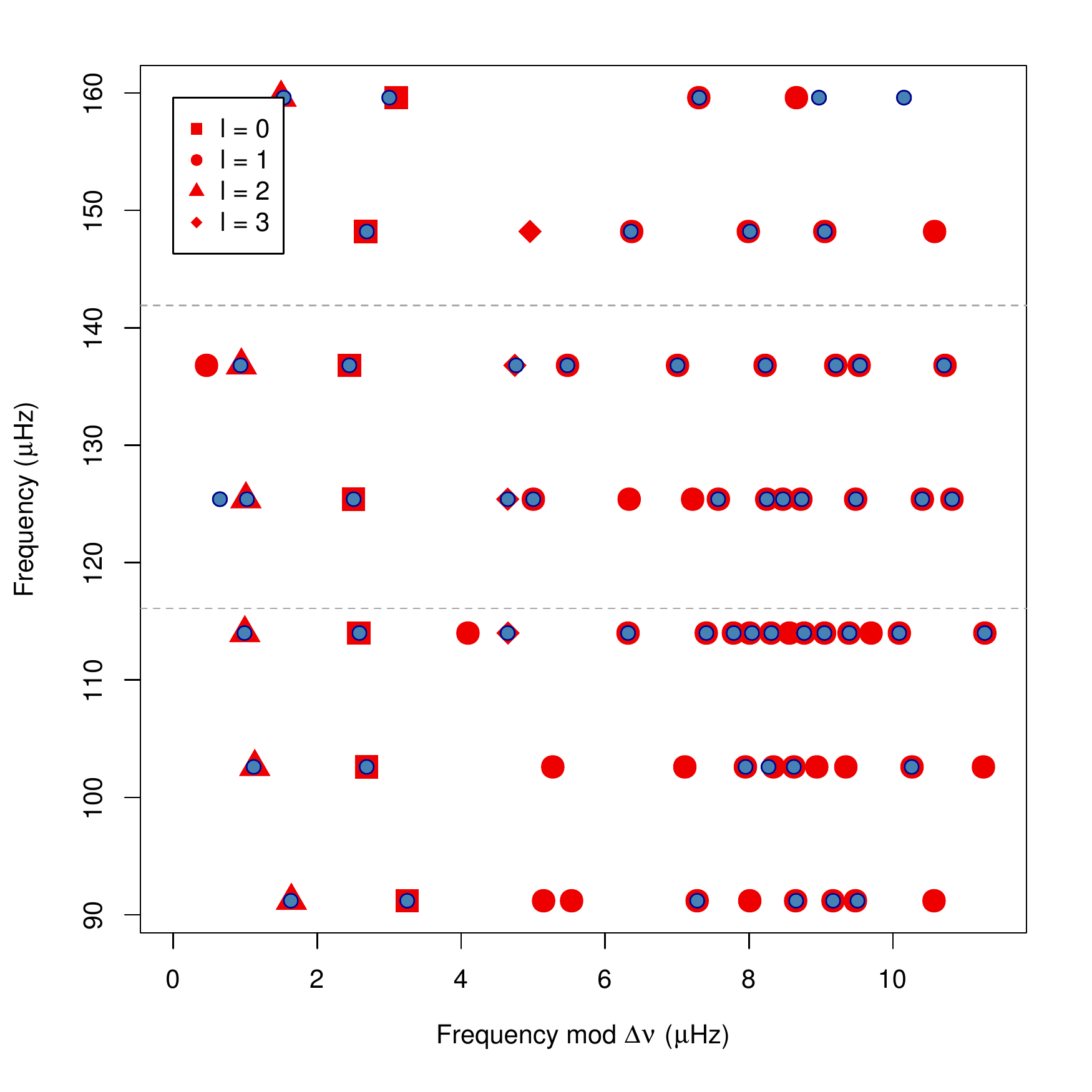}\caption{Same as Fig. \protect\ref{fig:KIC012008916echelle} for KIC 8718745.}\label{fig:KIC008718745echelle}\end{figure}\clearpage
\begin{table}
\centering
\begin{tabular}{ccccc}
  \hline
$\nu_k$ & $A_k$ & $\gamma_k$ & SNR & AIC \\ 
  \hline
80.249$\pm$0.008 & 0.378$\pm$0.121 & 0.008$\pm$0.006 & 2.21 & 6.98 \\ 
  86.045$\pm$0.014 & 0.469$\pm$0.116 & 0.02$\pm$0.015 & 2.76 & 9.61 \\ 
  89.737$\pm$0.035 & 0.844$\pm$0.12 & 0.091$\pm$0.042 & 3.91 & 36.92 \\ 
  96.93$\pm$0.006 & 0.652$\pm$0.151 & 0.01$\pm$0.007 & 6.42 & 42.99 \\ 
  100.425$\pm$0.045 & 0.936$\pm$0.116 & 0.125$\pm$0.045 & 4.26 & 42.55 \\ 
  101.87$\pm$0.019 & 1.094$\pm$0.125 & 0.059$\pm$0.018 & 7.94 & 101.51 \\ 
  105.958$\pm$0.004 & 0.434$\pm$0.13 & 0.005$\pm$0.006 & 2.81 & 31.92 \\
\hline
\end{tabular}
\caption{Resolved oscillations for KIC 9145955 .} 
\label{tab:009145955resolved}
\end{table}
\begin{table}
\centering
\begin{tabular}{ccc}
  \hline
$\nu_k$ & $H_k$ & AIC \\ 
  \hline
125.441$\pm$0.002 & 65.9$\pm$72.12 & 77.12 \\ 
  126.631$\pm$0.002 & 45.974$\pm$28.761 & 52.20 \\ 
  127.814$\pm$0.001 & 93.732$\pm$53.151 & 88.95 \\ 
  130.619$\pm$0.001 & 269.053$\pm$177.911 & 281.88 \\ 
  131.873$\pm$0.002 & 75.77$\pm$52.634 & 39.76 \\ 
  135.928$\pm$0.002 & 39.794$\pm$29.304 & 36.37 \\ 
  137.316$\pm$0.002 & 33.339$\pm$19.358 & 41.32 \\ 
  143.493$\pm$0.002 & 22.378$\pm$16.455 & 14.78 \\ 
  148.279$\pm$0.001 & 86.423$\pm$69.777 & 122.20 \\ 
   \hline
\end{tabular}
\caption{Unresolved oscillations for KIC 9145955 .} 
\label{tab:009145955unresolved}
\end{table}
\begin{figure}\includegraphics[width=0.47\textwidth]{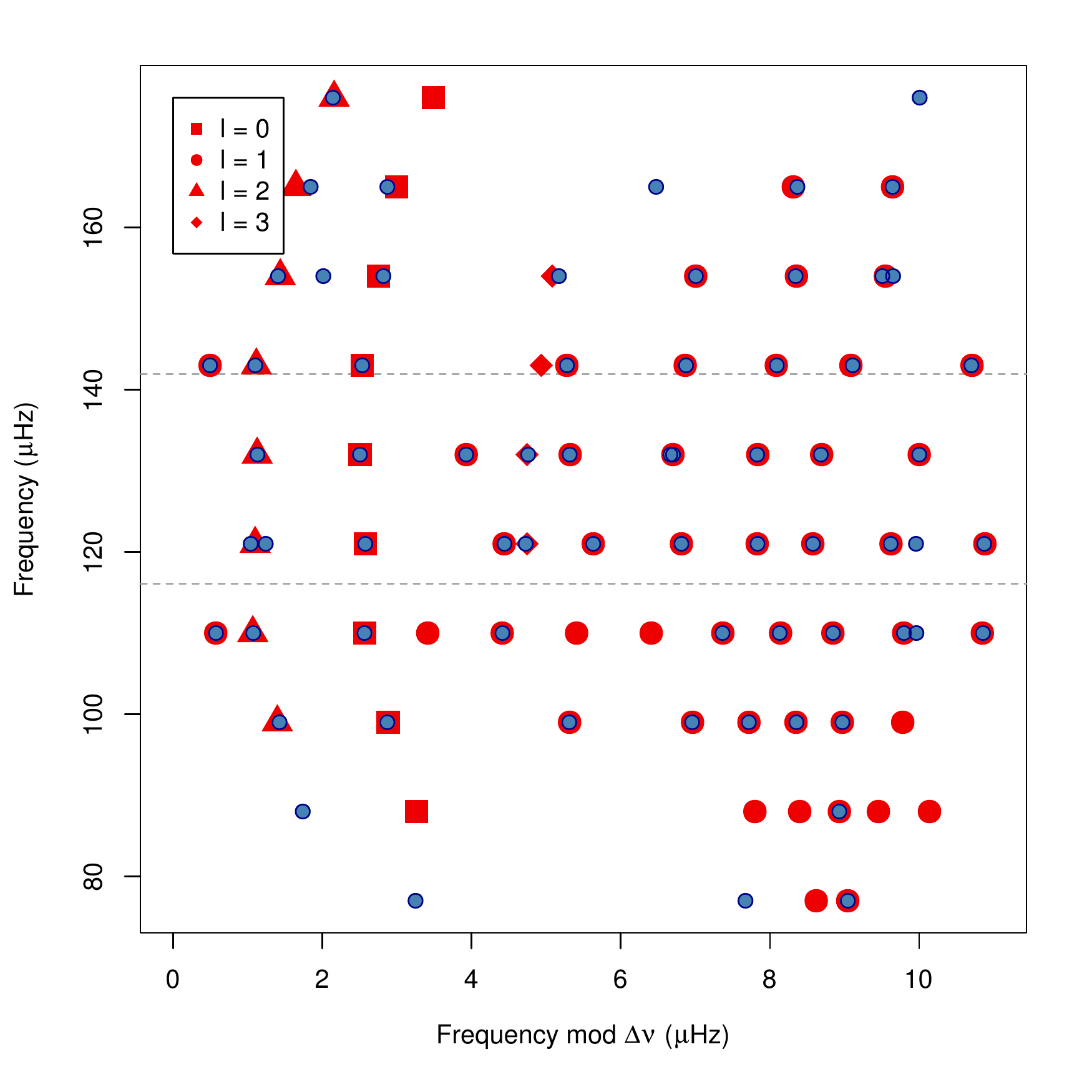}\caption{Same as Fig. \protect\ref{fig:KIC012008916echelle} for KIC 9145955.}\label{fig:KIC009145955echelle}\end{figure}\clearpage
\begin{table}
\centering
\begin{tabular}{ccccc}
  \hline
$\nu_k$ & $A_k$ & $\gamma_k$ & SNR & AIC \\ 
  \hline
75.265$\pm$0.026 & 0.445$\pm$0.11 & 0.03$\pm$0.018 & 2.27 & 4.80 \\ 
  80.586$\pm$0.02 & 0.542$\pm$0.121 & 0.032$\pm$0.024 & 2.77 & 13.09 \\ 
  85.603$\pm$0.042 & 0.53$\pm$0.113 & 0.053$\pm$0.028 & 2.32 & 7.40 \\ 
  90.617$\pm$0.082 & 0.975$\pm$0.123 & 0.22$\pm$0.08 & 3.56 & 33.87 \\ 
  93.939$\pm$0.06 & 0.771$\pm$0.117 & 0.124$\pm$0.049 & 3.17 & 19.98 \\ 
  95.38$\pm$0.015 & 1.248$\pm$0.141 & 0.05$\pm$0.017 & 10.37 & 155.80 \\ 
  99.875$\pm$0.007 & 0.781$\pm$0.157 & 0.015$\pm$0.01 & 7.88 & 59.94 \\
\hline
\end{tabular}
\caption{Resolved oscillations for KIC 9267654 .} 
\label{tab:009267654resolved}
\end{table}
\begin{table}
\centering
\begin{tabular}{ccc}
  \hline
$\nu_k$ & $H_k$ & AIC \\ 
  \hline
101.675$\pm$0.002 & 27.707$\pm$16.198 & 38.89 \\ 
  109.379$\pm$0.001 & 51.324$\pm$38.263 & 60.76 \\ 
  118.433$\pm$0.001 & 32.598$\pm$26.992 & 43.16 \\ 
  122.191$\pm$0.001 & 148.856$\pm$62.82 & 216.54 \\ 
  124.198$\pm$0.002 & 33.064$\pm$25.49 & 20.94 \\ 
  127.806$\pm$0.002 & 28.69$\pm$19.506 & 23.18 \\ 
   \hline
\end{tabular}
\caption{Unresolved oscillations for KIC 9267654 .} 
\label{tab:009267654unresolved}
\end{table}
\begin{figure}\includegraphics[width=0.47\textwidth]{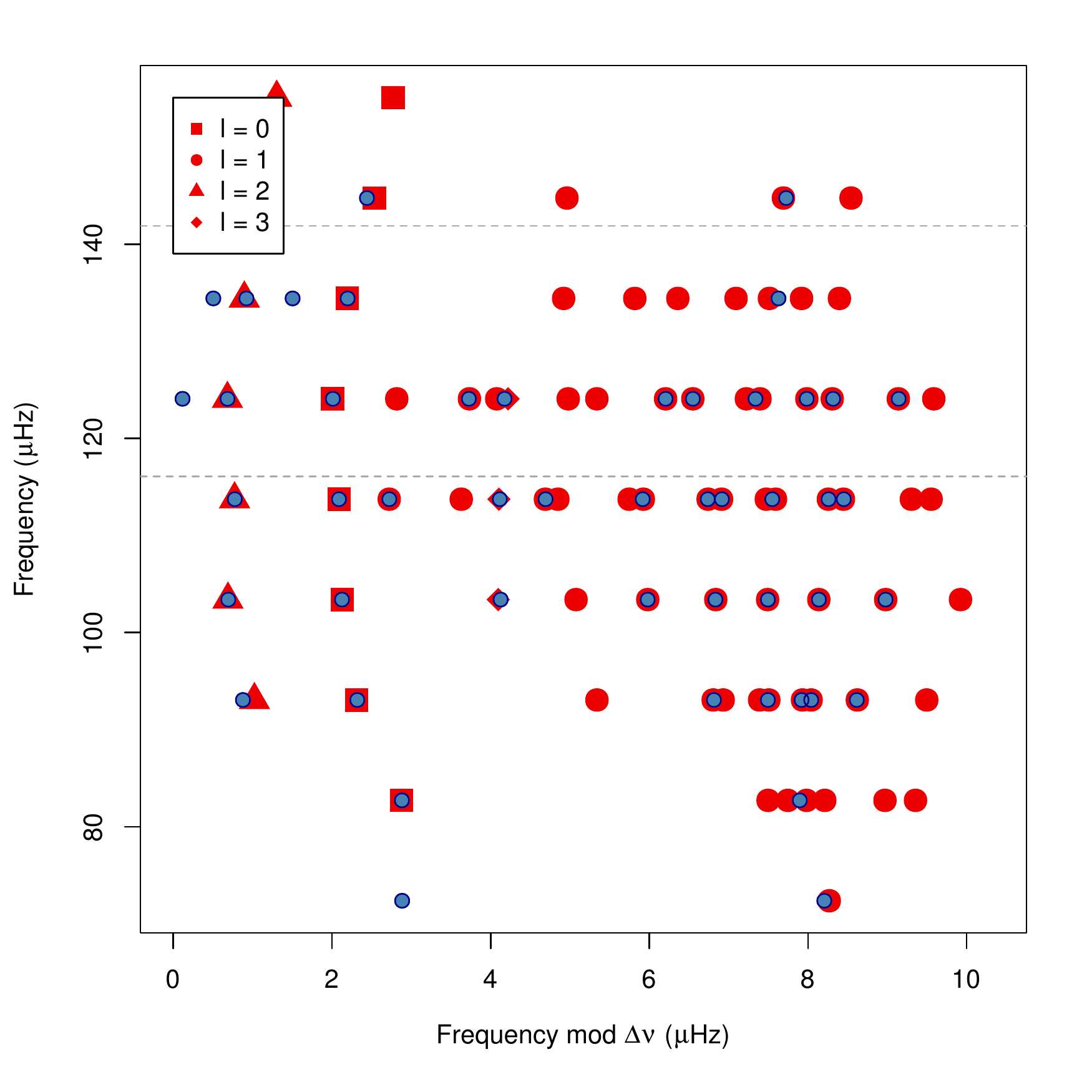}\caption{Same as Fig. \protect\ref{fig:KIC012008916echelle} for KIC 9267654.}\label{fig:KIC009267654echelle}\end{figure}\clearpage
\begin{table}
\centering
\begin{tabular}{ccccc}
  \hline
$\nu_k$ & $A_k$ & $\gamma_k$ & SNR & AIC \\ 
  \hline
80.147$\pm$0.016 & 0.414$\pm$0.114 & 0.021$\pm$0.017 & 1.77 & 4.89 \\ 
  86.425$\pm$0.017 & 0.875$\pm$0.127 & 0.049$\pm$0.026 & 5.66 & 57.64 \\ 
  89.607$\pm$0.008 & 0.572$\pm$0.145 & 0.015$\pm$0.017 & 3.57 & 25.78 \\ 
  90.965$\pm$0.018 & 1.099$\pm$0.126 & 0.056$\pm$0.017 & 8.24 & 106.28 \\ 
  93.362$\pm$0.016 & 0.374$\pm$0.109 & 0.01$\pm$0.014 & 3.38 & 6.99 \\ 
  95.806$\pm$0.023 & 0.958$\pm$0.124 & 0.068$\pm$0.028 & 4.51 & 60.84 \\ 
  96.267$\pm$0.006 & 0.979$\pm$0.201 & 0.01$\pm$0.005 & 13.95 & 110.74 \\
\hline
\end{tabular}
\caption{Resolved oscillations for KIC 9475697 .} 
\label{tab:009475697resolved}
\end{table}
\begin{table}
\centering
\begin{tabular}{ccc}
  \hline
$\nu_k$ & $H_k$ & AIC \\ 
  \hline
95.287$\pm$0.001 & 160.059$\pm$74.849 & 102.65 \\ 
  104.371$\pm$0.001 & 51.343$\pm$52.416 & 80.67 \\ 
  107.036$\pm$0.001 & 104.647$\pm$63.325 & 106.75 \\ 
  109.437$\pm$0.002 & 124.327$\pm$67.948 & 49.69 \\ 
  109.631$\pm$0.002 & 35.095$\pm$33.639 & 29.75 \\ 
  115.967$\pm$0.002 & 98.762$\pm$73.906 & 28.14 \\ 
  116.496$\pm$0.002 & 189.291$\pm$93.649 & 111.97 \\ 
  117.454$\pm$0.001 & 96.18$\pm$78.639 & 100.17 \\ 
  118.483$\pm$0.002 & 44.193$\pm$44.658 & 21.69 \\ 
  120.608$\pm$0.002 & 40.833$\pm$39.158 & 18.28 \\ 
  126.499$\pm$0.002 & 132.886$\pm$65.345 & 137.01 \\ 
   \hline
\end{tabular}
\caption{Unresolved oscillations for KIC 9475697 .} 
\label{tab:009475697unresolved}
\end{table}
\begin{figure}\includegraphics[width=0.47\textwidth]{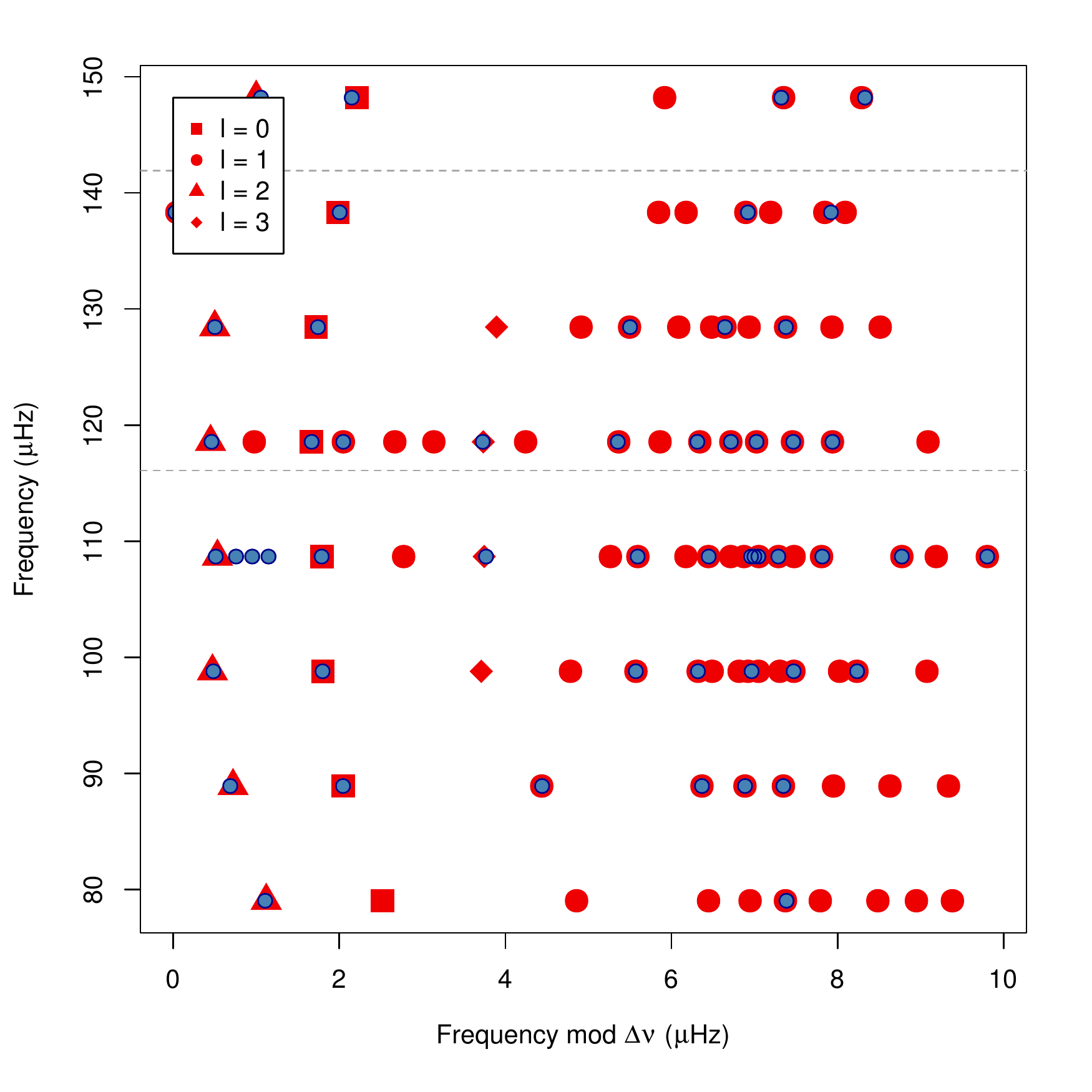}\caption{Same as Fig. \protect\ref{fig:KIC012008916echelle} for KIC 9475697.}\label{fig:KIC009475697echelle}\end{figure}\clearpage
\begin{table}
\centering
\begin{tabular}{ccccc}
  \hline
$\nu_k$ & $A_k$ & $\gamma_k$ & SNR & AIC \\ 
  \hline
127.629$\pm$0.042 & 0.649$\pm$0.116 & 0.075$\pm$0.038 & 3.07 & 15.11 \\ 
  133.897$\pm$0.01 & 0.569$\pm$0.129 & 0.017$\pm$0.013 & 3.87 & 23.28 \\ 
  134.732$\pm$0.007 & 0.549$\pm$0.135 & 0.011$\pm$0.007 & 4.52 & 25.07 \\ 
  139.092$\pm$0.049 & 0.635$\pm$0.13 & 0.098$\pm$0.063 & 1.90 & 10.41 \\ 
  140.819$\pm$0.012 & 0.831$\pm$0.132 & 0.027$\pm$0.012 & 6.63 & 62.75 \\ 
  146.627$\pm$0.005 & 1.1$\pm$0.236 & 0.01$\pm$0.005 & 17.92 & 167.18 \\ 
  147.663$\pm$0.014 & 0.937$\pm$0.132 & 0.034$\pm$0.013 & 7.41 & 81.72 \\
\hline
\end{tabular}
\caption{Resolved oscillations for KIC 9882316 .} 
\label{tab:009882316resolved}
\end{table}
\begin{table}
\centering
\begin{tabular}{ccc}
  \hline
$\nu_k$ & $H_k$ & AIC \\ 
  \hline
164.53$\pm$0.002 & 58.253$\pm$29.131 & 89.35 \\ 
  171.109$\pm$0.001 & 74.191$\pm$36.357 & 134.04 \\ 
  173.286$\pm$0.001 & 117.735$\pm$61.44 & 167.85 \\ 
  183.765$\pm$0.001 & 116.229$\pm$114.666 & 195.50 \\ 
  192.468$\pm$0.002 & 59.061$\pm$28.749 & 73.36 \\ 
  195.372$\pm$0.002 & 41.477$\pm$32.269 & 28.23 \\ 
  205.232$\pm$0.002 & 35.658$\pm$19.119 & 50.39 \\ 
   \hline
\end{tabular}
\caption{Unresolved oscillations for KIC 9882316 .} 
\label{tab:009882316unresolved}
\end{table}
\begin{figure}\includegraphics[width=0.47\textwidth]{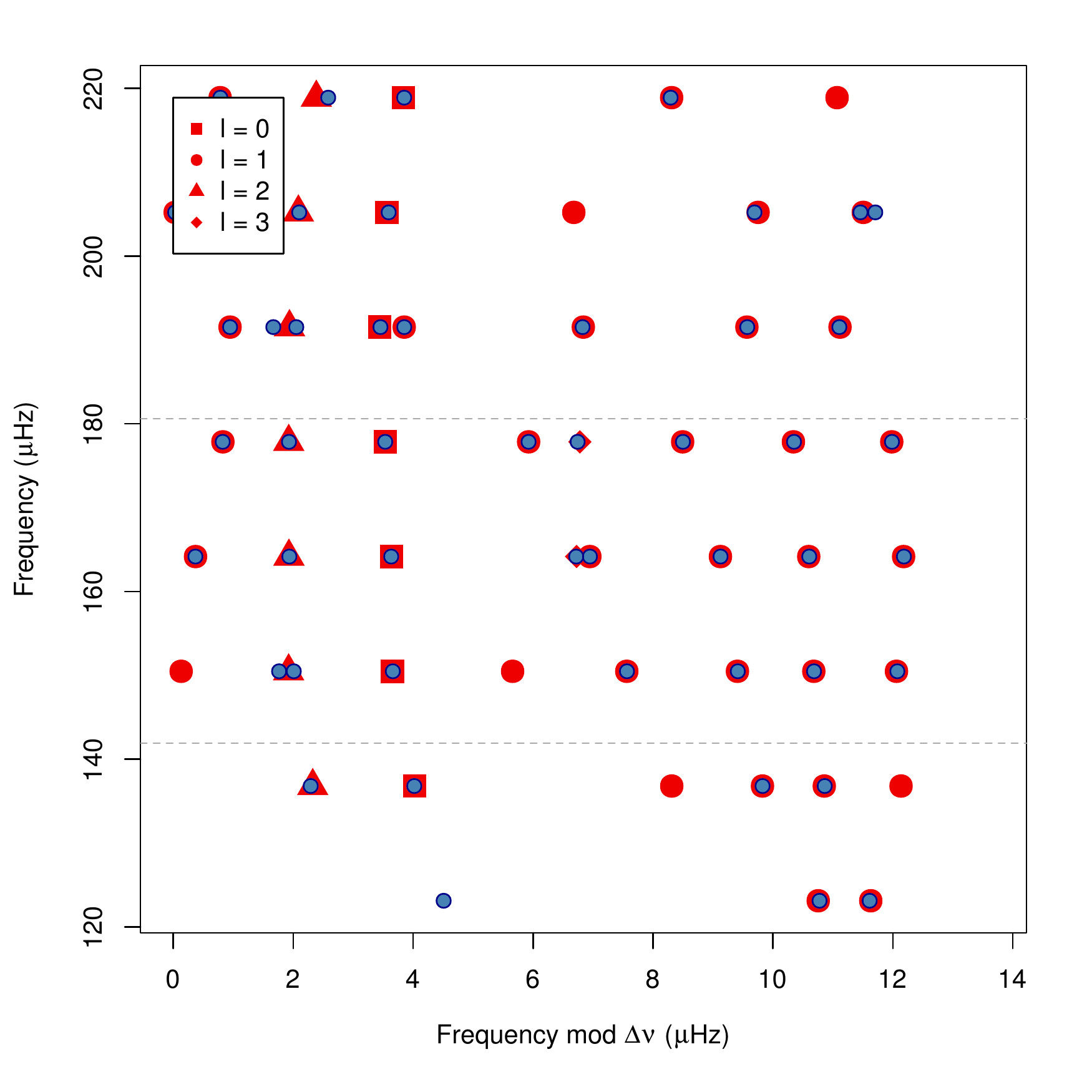}\caption{Same as Fig. \protect\ref{fig:KIC012008916echelle} for KIC 9882316.}\label{fig:KIC009882316echelle}\end{figure}\clearpage
\begin{table}
\centering
\begin{tabular}{ccccc}
  \hline
$\nu_k$ & $A_k$ & $\gamma_k$ & SNR & AIC \\ 
  \hline
113.921$\pm$0.019 & 0.594$\pm$0.113 & 0.032$\pm$0.016 & 3.43 & 19.40 \\ 
  121.413$\pm$0.005 & 0.723$\pm$0.174 & 0.008$\pm$0.005 & 8.84 & 62.91 \\ 
  125.239$\pm$0.022 & 0.688$\pm$0.115 & 0.047$\pm$0.023 & 3.50 & 25.83 \\ 
  127.056$\pm$0.045 & 0.818$\pm$0.111 & 0.103$\pm$0.036 & 3.74 & 30.19 \\ 
  132.948$\pm$0.014 & 0.516$\pm$0.115 & 0.02$\pm$0.011 & 3.63 & 14.52 \\ 
  133.965$\pm$0.007 & 1.378$\pm$0.214 & 0.019$\pm$0.007 & 19.63 & 257.34 \\ 
  135.009$\pm$0.005 & 0.758$\pm$0.183 & 0.008$\pm$0.005 & 9.38 & 68.89 \\
\hline
\end{tabular}
\caption{Resolved oscillations for KIC 10123207 .} 
\label{tab:010123207resolved}
\end{table}
\begin{table}
\centering
\begin{tabular}{ccc}
  \hline
$\nu_k$ & $H_k$ & AIC \\ 
  \hline
147.259$\pm$0.002 & 31.897$\pm$30.753 & 1.15 \\ 
  150.097$\pm$0.001 & 68.076$\pm$44.97 & 85.87 \\ 
  151.615$\pm$0.002 & 24.189$\pm$16.691 & 20.79 \\ 
  153.827$\pm$0.002 & 58.123$\pm$48.835 & 28.75 \\ 
  155.787$\pm$0.001 & 204.867$\pm$155.299 & 388.83 \\ 
  157.77$\pm$0.001 & 178.437$\pm$102.431 & 234.49 \\ 
  171.337$\pm$0.001 & 179.599$\pm$101.707 & 218.50 \\ 
  176.783$\pm$0.001 & 269.282$\pm$129.638 & 156.98 \\ 
  179.262$\pm$0.002 & 55.018$\pm$31.015 & 62.97 \\ 
   \hline
\end{tabular}
\caption{Unresolved oscillations for KIC 10123207 .} 
\label{tab:010123207unresolved}
\end{table}
\begin{figure}\includegraphics[width=0.47\textwidth]{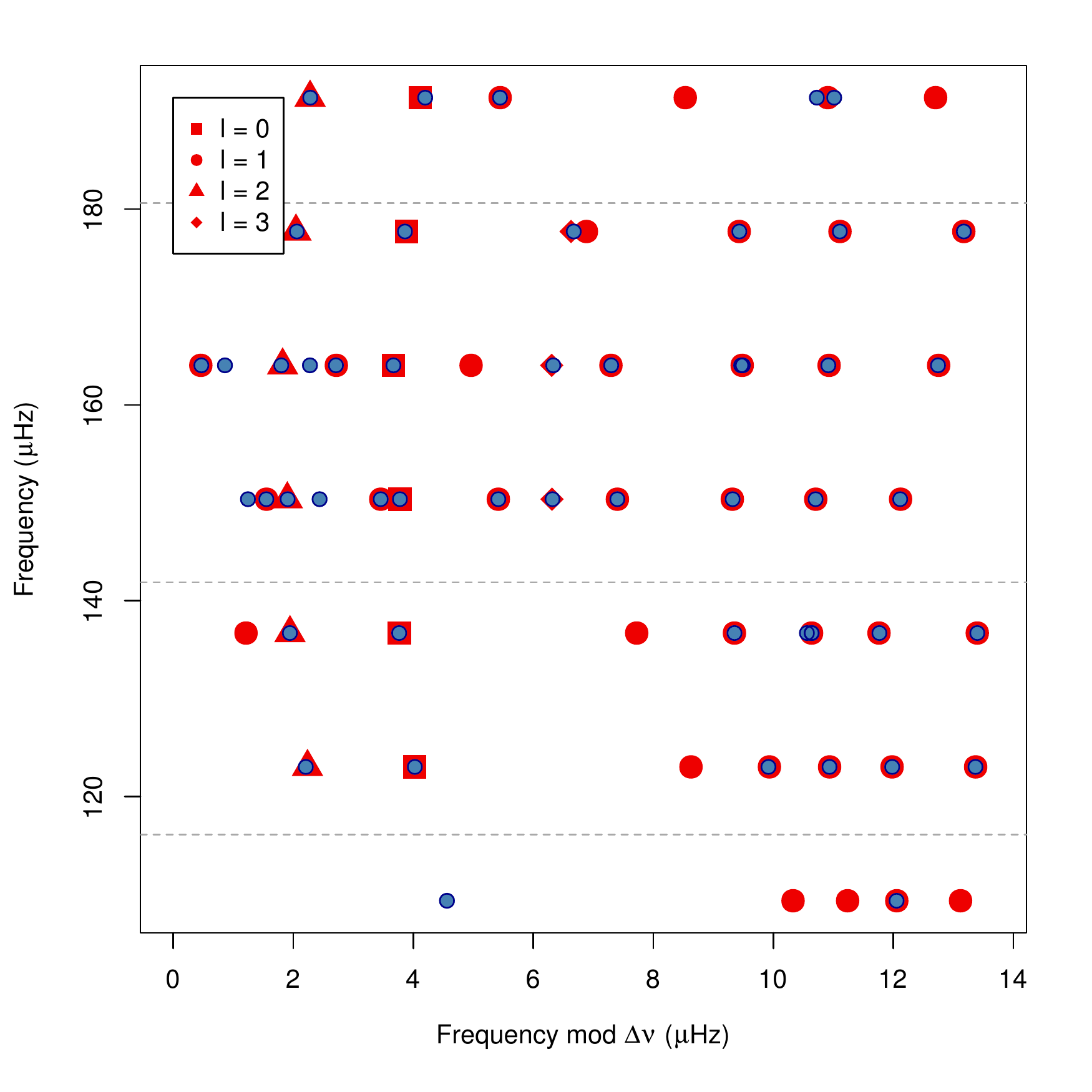}\caption{Same as Fig. \protect\ref{fig:KIC012008916echelle} for KIC 10123207.}\label{fig:KIC010123207echelle}\end{figure}\clearpage
\begin{table}
\centering
\begin{tabular}{ccccc}
  \hline
$\nu_k$ & $A_k$ & $\gamma_k$ & SNR & AIC \\ 
  \hline
102.143$\pm$0.014 & 0.489$\pm$0.12 & 0.02$\pm$0.016 & 2.75 & 11.50 \\ 
  103.863$\pm$0.038 & 0.67$\pm$0.116 & 0.079$\pm$0.039 & 2.77 & 16.41 \\ 
  109.669$\pm$0.005 & 0.521$\pm$0.164 & 0.006$\pm$0.007 & 3.70 & 23.01 \\ 
  110.356$\pm$0.015 & 0.883$\pm$0.125 & 0.04$\pm$0.016 & 6.34 & 64.22 \\ 
  114.092$\pm$0.06 & 1.025$\pm$0.12 & 0.188$\pm$0.065 & 3.81 & 45.00 \\ 
  115.826$\pm$0.021 & 1.193$\pm$0.125 & 0.073$\pm$0.022 & 8.49 & 119.84 \\ 
  119.522$\pm$0.006 & 0.413$\pm$0.13 & 0.008$\pm$0.008 & 2.93 & 9.82 \\
\hline
\end{tabular}
\caption{Resolved oscillations for KIC 10200377 .} 
\label{tab:010200377resolved}
\end{table}
\begin{table}
\centering
\begin{tabular}{ccc}
  \hline
$\nu_k$ & $H_k$ & AIC \\ 
  \hline
108.856$\pm$0.002 & 55.79$\pm$32.438 & 41.14 \\ 
  121.649$\pm$0.002 & 93.282$\pm$58.418 & 67.63 \\ 
  132.384$\pm$0.002 & 22.15$\pm$14.244 & 24.95 \\ 
  137.025$\pm$0.002 & 68.52$\pm$43.003 & 79.00 \\ 
  138.529$\pm$0.002 & 138.229$\pm$75.43 & 84.18 \\ 
  139.512$\pm$0.002 & 35.226$\pm$28.378 & 36.92 \\ 
  140.054$\pm$0.002 & 33.989$\pm$29.921 & 30.67 \\ 
  144.859$\pm$0.001 & 172.249$\pm$110.713 & 276.67 \\ 
  150.399$\pm$0.001 & 166.524$\pm$132.412 & 126.34 \\ 
  152.199$\pm$0.002 & 76.12$\pm$49.178 & 57.73 \\ 
  152.496$\pm$0.002 & 22.868$\pm$21.707 & 17.25 \\ 
  162.361$\pm$0.002 & 63.45$\pm$38.877 & 60.74 \\ 
   \hline
\end{tabular}
\caption{Unresolved oscillations for KIC 10200377 .} 
\label{tab:010200377unresolved}
\end{table}
\begin{figure}\includegraphics[width=0.47\textwidth]{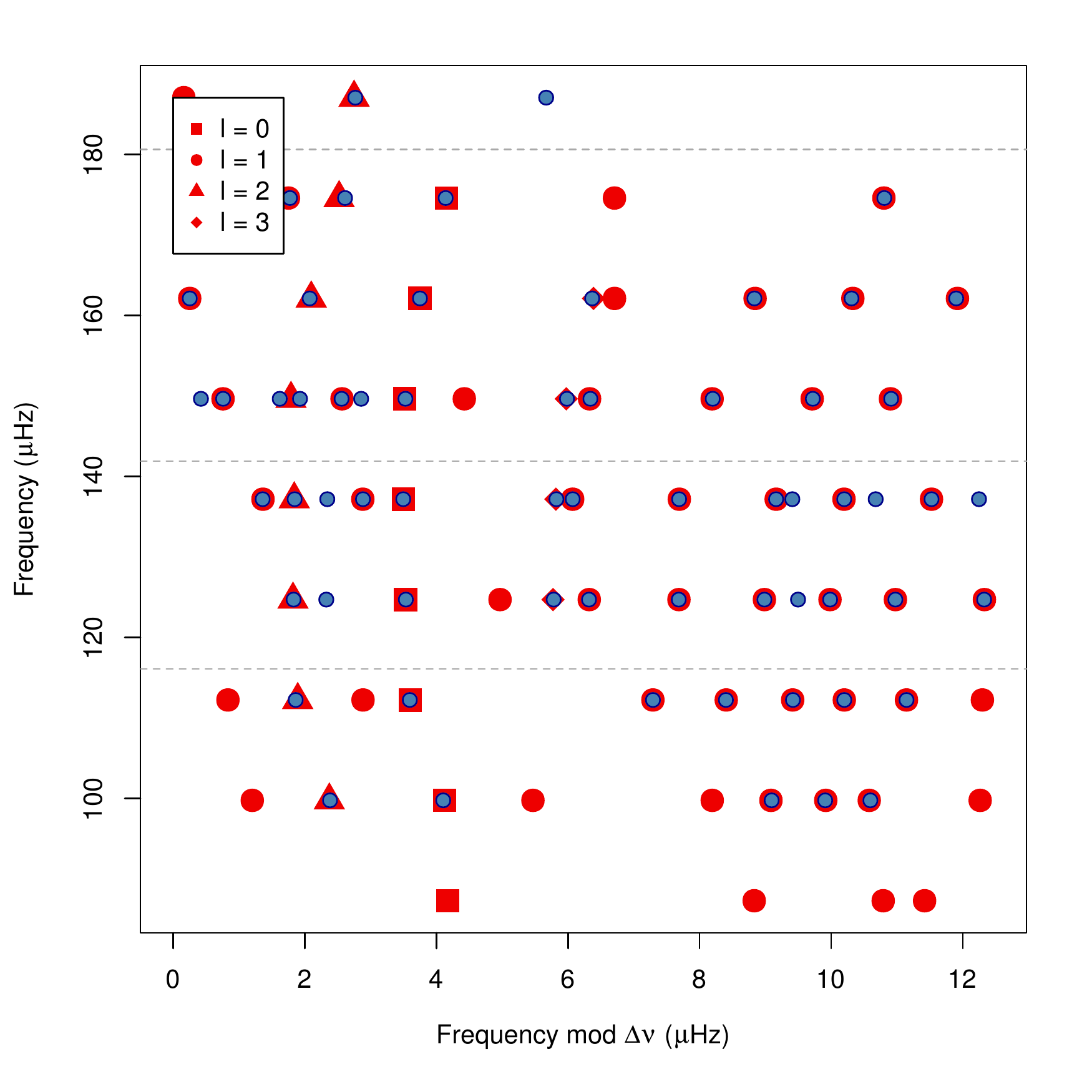}\caption{Same as Fig. \protect\ref{fig:KIC012008916echelle} for KIC 10200377.}\label{fig:KIC010200377echelle}\end{figure}\clearpage
\begin{table}
\centering
\begin{tabular}{ccccc}
  \hline
$\nu_k$ & $A_k$ & $\gamma_k$ & SNR & AIC \\ 
  \hline
107.603$\pm$0.025 & 0.529$\pm$0.113 & 0.039$\pm$0.023 & 2.86 & 9.97 \\ 
  113.211$\pm$0.041 & 0.661$\pm$0.112 & 0.073$\pm$0.032 & 2.75 & 16.56 \\ 
  119.019$\pm$0.013 & 0.839$\pm$0.131 & 0.032$\pm$0.016 & 5.55 & 59.45 \\ 
  119.657$\pm$0.01 & 0.899$\pm$0.149 & 0.021$\pm$0.01 & 8.26 & 83.19 \\ 
  120.603$\pm$0.01 & 0.385$\pm$0.118 & 0.011$\pm$0.01 & 2.24 & 5.28 \\ 
  123.358$\pm$0.024 & 0.905$\pm$0.121 & 0.066$\pm$0.029 & 4.72 & 53.98 \\ 
  124.884$\pm$0.018 & 1.348$\pm$0.134 & 0.067$\pm$0.019 & 10.85 & 176.92 \\
\hline
\end{tabular}
\caption{Resolved oscillations for KIC 10257278 .} 
\label{tab:010257278resolved}
\end{table}
\begin{table}
\centering
\begin{tabular}{ccc}
  \hline
$\nu_k$ & $H_k$ & AIC \\ 
  \hline
128.989$\pm$0.002 & 31.395$\pm$19.874 & 25.95 \\ 
  130.487$\pm$0.002 & 76.194$\pm$46.358 & 50.92 \\ 
  133.246$\pm$0.002 & 25.07$\pm$17.63 & 29.03 \\ 
  141.006$\pm$0.002 & 80.675$\pm$88.066 & 140.31 \\ 
  144.578$\pm$0.001 & 26.934$\pm$20.047 & 34.64 \\ 
  145.492$\pm$0.002 & 22.57$\pm$17.844 & 25.24 \\ 
  146.141$\pm$0.001 & 57.948$\pm$37.929 & 74.29 \\ 
  150.548$\pm$0.002 & 22.209$\pm$19.329 & 26.91 \\ 
  152.312$\pm$0.001 & 41.453$\pm$37.006 & 67.27 \\ 
  158.193$\pm$0.002 & 59.328$\pm$33.223 & 35.50 \\ 
  163.205$\pm$0.002 & 30.553$\pm$19.126 & 37.07 \\ 
  165.266$\pm$0.002 & 22.45$\pm$21.499 & 28.07 \\ 
  166.178$\pm$0.002 & 27.108$\pm$19.989 & 28.98 \\ 
  170.055$\pm$0.002 & 22.437$\pm$15.358 & 21.06 \\ 
   \hline
\end{tabular}
\caption{Unresolved oscillations for KIC 10257278 .} 
\label{tab:010257278unresolved}
\end{table}
\begin{figure}\includegraphics[width=0.47\textwidth]{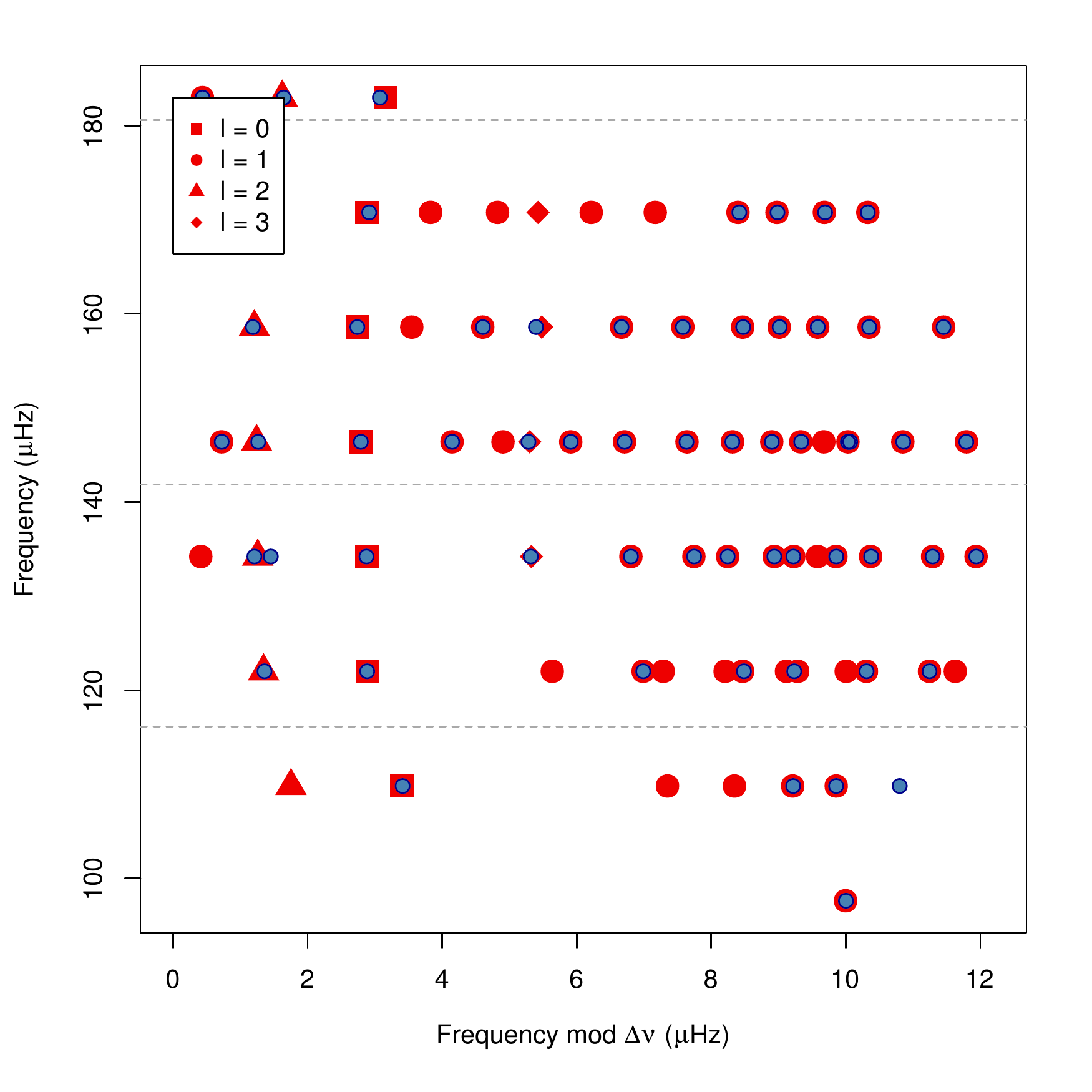}\caption{Same as Fig. \protect\ref{fig:KIC012008916echelle} for KIC 10257278.}\label{fig:KIC010257278echelle}\end{figure}\clearpage
\begin{table}
\centering
\begin{tabular}{ccccc}
  \hline
$\nu_k$ & $A_k$ & $\gamma_k$ & SNR & AIC \\ 
  \hline
94.344$\pm$0.043 & 0.801$\pm$0.129 & 0.11$\pm$0.061 & 2.64 & 26.67 \\ 
  99.265$\pm$0.029 & 1.108$\pm$0.117 & 0.096$\pm$0.029 & 6.45 & 87.31 \\ 
  103.647$\pm$0.025 & 0.706$\pm$0.113 & 0.042$\pm$0.016 & 4.97 & 28.46 \\ 
  104.345$\pm$0.013 & 0.993$\pm$0.138 & 0.034$\pm$0.013 & 8.62 & 86.13 \\ 
  104.852$\pm$0.014 & 1.085$\pm$0.142 & 0.039$\pm$0.016 & 9.16 & 107.74 \\ 
  105.581$\pm$0.023 & 0.651$\pm$0.122 & 0.036$\pm$0.022 & 3.72 & 22.83 \\ 
  108.281$\pm$0.026 & 1.172$\pm$0.125 & 0.095$\pm$0.034 & 6.55 & 99.25 \\
\hline
\end{tabular}
\caption{Resolved oscillations for KIC 11353313 .} 
\label{tab:011353313resolved}
\end{table}
\begin{table}
\centering
\begin{tabular}{ccc}
  \hline
$\nu_k$ & $H_k$ & AIC \\ 
  \hline
125.929$\pm$0.002 & 52.115$\pm$44.343 & 15.65 \\ 
  127.395$\pm$0.002 & 43.174$\pm$31.602 & 40.18 \\ 
  135.931$\pm$0.001 & 509.74$\pm$267.253 & 262.75 \\ 
   \hline
\end{tabular}
\caption{Unresolved oscillations for KIC 11353313 .} 
\label{tab:011353313unresolved}
\end{table}
\begin{figure}\includegraphics[width=0.47\textwidth]{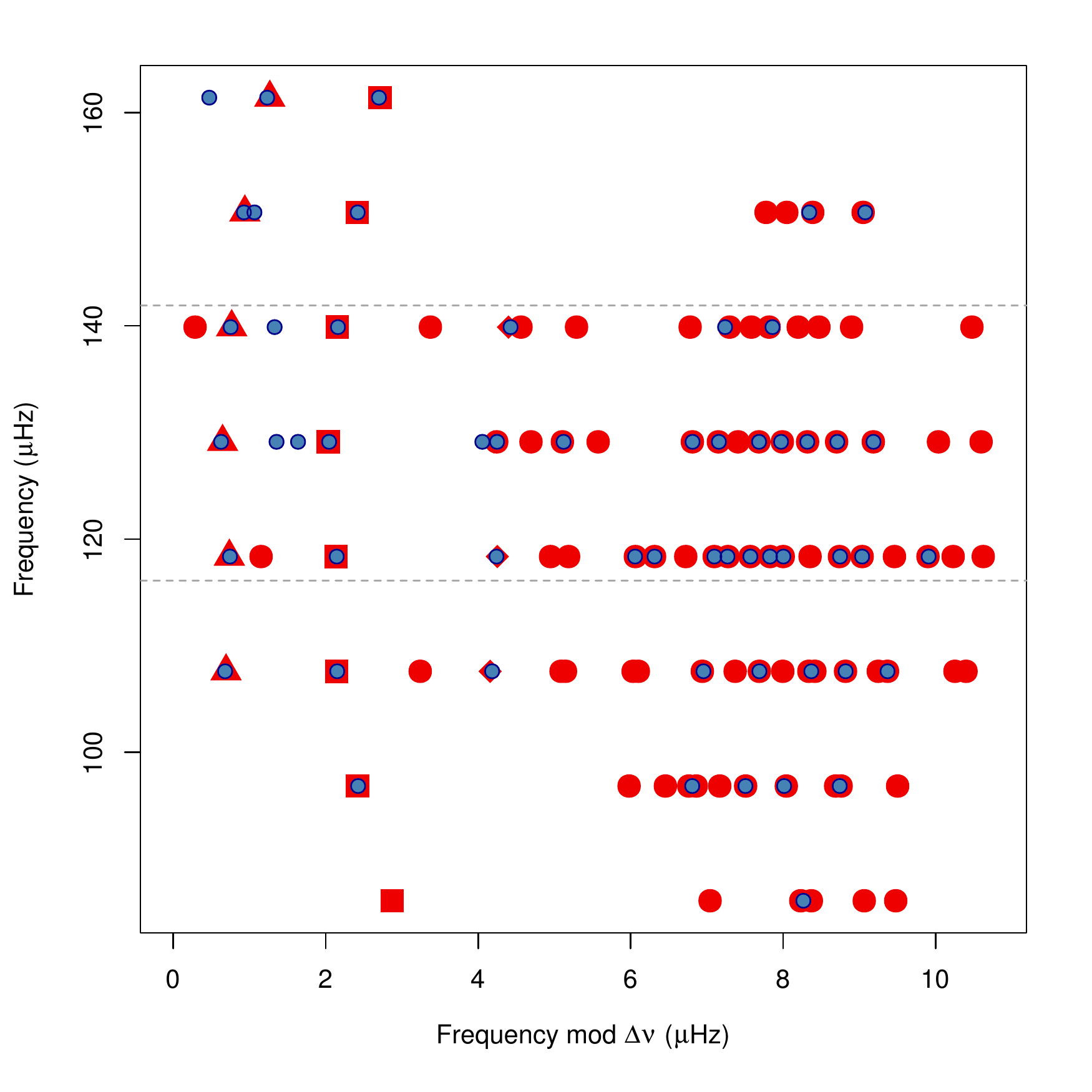}\caption{Same as Fig. \protect\ref{fig:KIC012008916echelle} for KIC 11353313.}\label{fig:KIC011353313echelle}\end{figure}\clearpage
\begin{table}
\centering
\begin{tabular}{ccccc}
  \hline
$\nu_k$ & $A_k$ & $\gamma_k$ & SNR & AIC \\ 
  \hline
84.097$\pm$0.032 & 0.576$\pm$0.114 & 0.054$\pm$0.029 & 2.54 & 11.45 \\ 
  89.012$\pm$0.03 & 0.617$\pm$0.122 & 0.062$\pm$0.039 & 2.76 & 13.93 \\ 
  89.974$\pm$0.015 & 0.397$\pm$0.112 & 0.017$\pm$0.012 & 2.17 & 4.26 \\ 
  92.358$\pm$0.014 & 1.127$\pm$0.14 & 0.04$\pm$0.014 & 9.30 & 126.34 \\ 
  92.691$\pm$0.013 & 0.454$\pm$0.123 & 0.015$\pm$0.011 & 2.96 & 7.29 \\ 
  93.649$\pm$0.012 & 0.751$\pm$0.145 & 0.022$\pm$0.017 & 6.47 & 42.76 \\ 
  93.811$\pm$0.014 & 0.566$\pm$0.126 & 0.022$\pm$0.014 & 2.98 & 14.41 \\
\hline
\end{tabular}
\caption{Resolved oscillations for KIC 11913545 .} 
\label{tab:011913545resolved}
\end{table}
\begin{table}
\centering
\begin{tabular}{ccc}
  \hline
$\nu_k$ & $H_k$ & AIC \\ 
  \hline
108.773$\pm$0.002 & 371.821$\pm$242.555 & 92.18 \\ 
  110.129$\pm$0.001 & 65.433$\pm$43.77 & 59.16 \\ 
  110.66$\pm$0.001 & 39.651$\pm$27.856 & 71.86 \\ 
  111.581$\pm$0.001 & 35.493$\pm$36.682 & 46.76 \\ 
  117.009$\pm$0.001 & 92.303$\pm$40.359 & 175.35 \\ 
  118.024$\pm$0.001 & 48.979$\pm$26.489 & 84.12 \\ 
  119.159$\pm$0.001 & 1469.197$\pm$534.294 & 391.85 \\ 
  120.462$\pm$0.001 & 106.578$\pm$73.478 & 134.75 \\ 
  120.996$\pm$0.001 & 67.169$\pm$31.529 & 101.84 \\ 
  127.471$\pm$0.002 & 40.57$\pm$34.03 & 54.41 \\ 
   \hline
\end{tabular}
\caption{Unresolved oscillations for KIC 11913545 .} 
\label{tab:011913545unresolved}
\end{table}
\begin{figure}\includegraphics[width=0.47\textwidth]{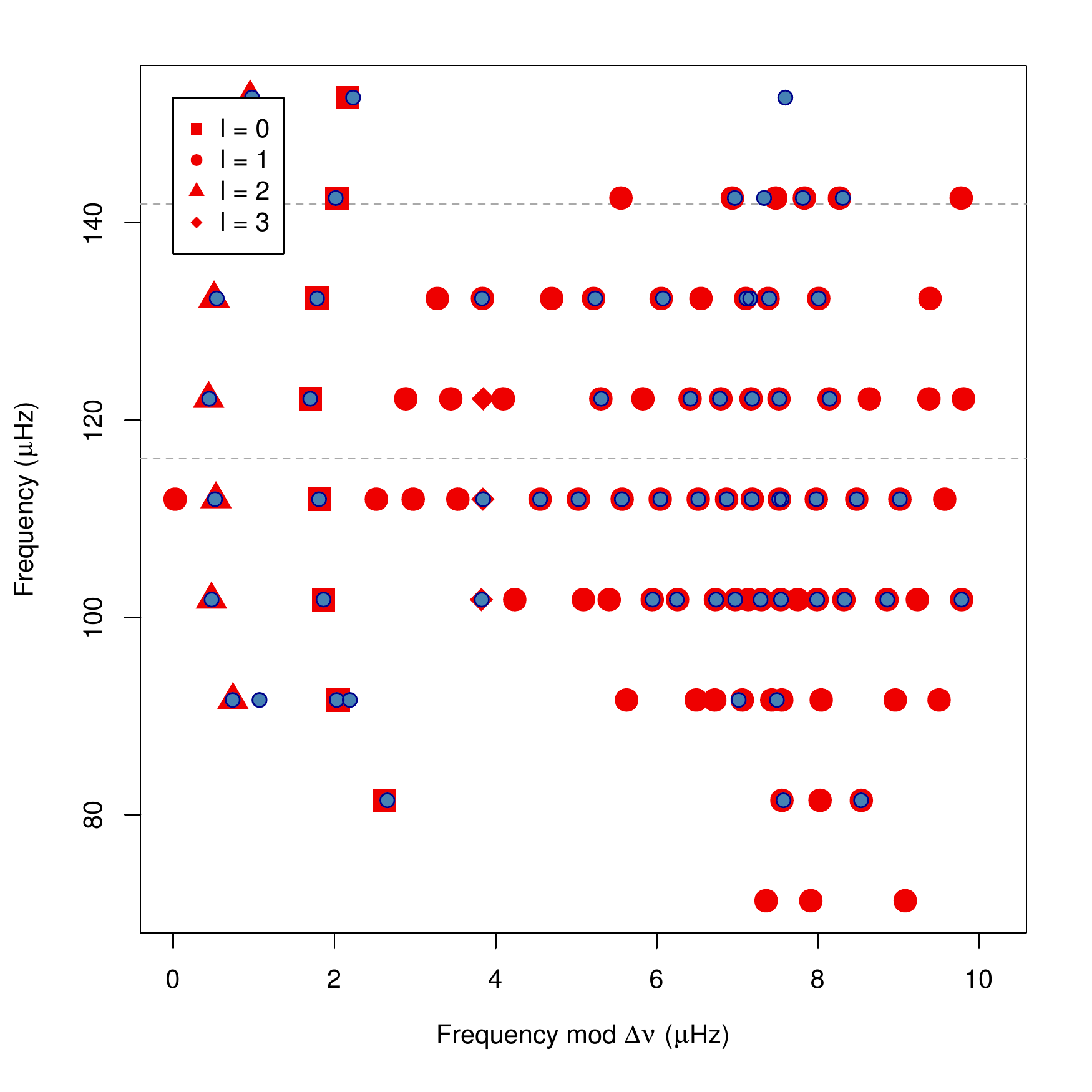}\caption{Same as Fig. \protect\ref{fig:KIC012008916echelle} for KIC 11913545.}\label{fig:KIC011913545echelle}\end{figure}\clearpage
\begin{table}
\centering
\begin{tabular}{ccccc}
  \hline
$\nu_k$ & $A_k$ & $\gamma_k$ & SNR & AIC \\ 
  \hline
104.421$\pm$0.026 & 0.598$\pm$0.111 & 0.044$\pm$0.021 & 3.32 & 15.80 \\ 
  105.853$\pm$0.029 & 0.658$\pm$0.114 & 0.057$\pm$0.027 & 3.18 & 19.47 \\ 
  111.071$\pm$0.018 & 0.464$\pm$0.118 & 0.024$\pm$0.018 & 2.33 & 6.37 \\ 
  111.668$\pm$0.065 & 0.775$\pm$0.15 & 0.125$\pm$0.086 & 2.48 & 20.19 \\ 
  112.064$\pm$0.009 & 0.487$\pm$0.137 & 0.011$\pm$0.01 & 3.88 & 12.89 \\ 
  115.261$\pm$0.026 & 1.033$\pm$0.12 & 0.08$\pm$0.029 & 5.08 & 75.94 \\ 
  116.775$\pm$0.019 & 1.06$\pm$0.124 & 0.056$\pm$0.018 & 7.75 & 94.83 \\
\hline
\end{tabular}
\caption{Resolved oscillations for KIC 11968334 .} 
\label{tab:011968334resolved}
\end{table}
\begin{table}
\centering
\begin{tabular}{ccc}
  \hline
$\nu_k$ & $H_k$ & AIC \\ 
  \hline
122.895$\pm$0.002 & 66.213$\pm$45.672 & 68.73 \\ 
  132.204$\pm$0.002 & 38.296$\pm$19.552 & 62.24 \\ 
  132.562$\pm$0.001 & 92.73$\pm$80.836 & 138.46 \\ 
  132.906$\pm$0.002 & 42.868$\pm$22.898 & 65.58 \\ 
  133.406$\pm$0.002 & 61.114$\pm$51.565 & 38.05 \\ 
  133.856$\pm$0.004 & 38.933$\pm$44.564 & 3.61 \\ 
  134.191$\pm$0.002 & 304.135$\pm$168.703 & 44.08 \\ 
  134.693$\pm$0.001 & 195.375$\pm$100.549 & 158.83 \\ 
  142.585$\pm$0.001 & 40.139$\pm$27.954 & 66.26 \\ 
  144.085$\pm$0.001 & 123.51$\pm$73.516 & 150.67 \\ 
  146.143$\pm$0.002 & 151.484$\pm$79.545 & 104.09 \\ 
  146.814$\pm$0.002 & 34.801$\pm$20.522 & 43.84 \\ 
   \hline
\end{tabular}
\caption{Unresolved oscillations for KIC 11968334 .} 
\label{tab:011968334unresolved}
\end{table}
\begin{figure}\includegraphics[width=0.47\textwidth]{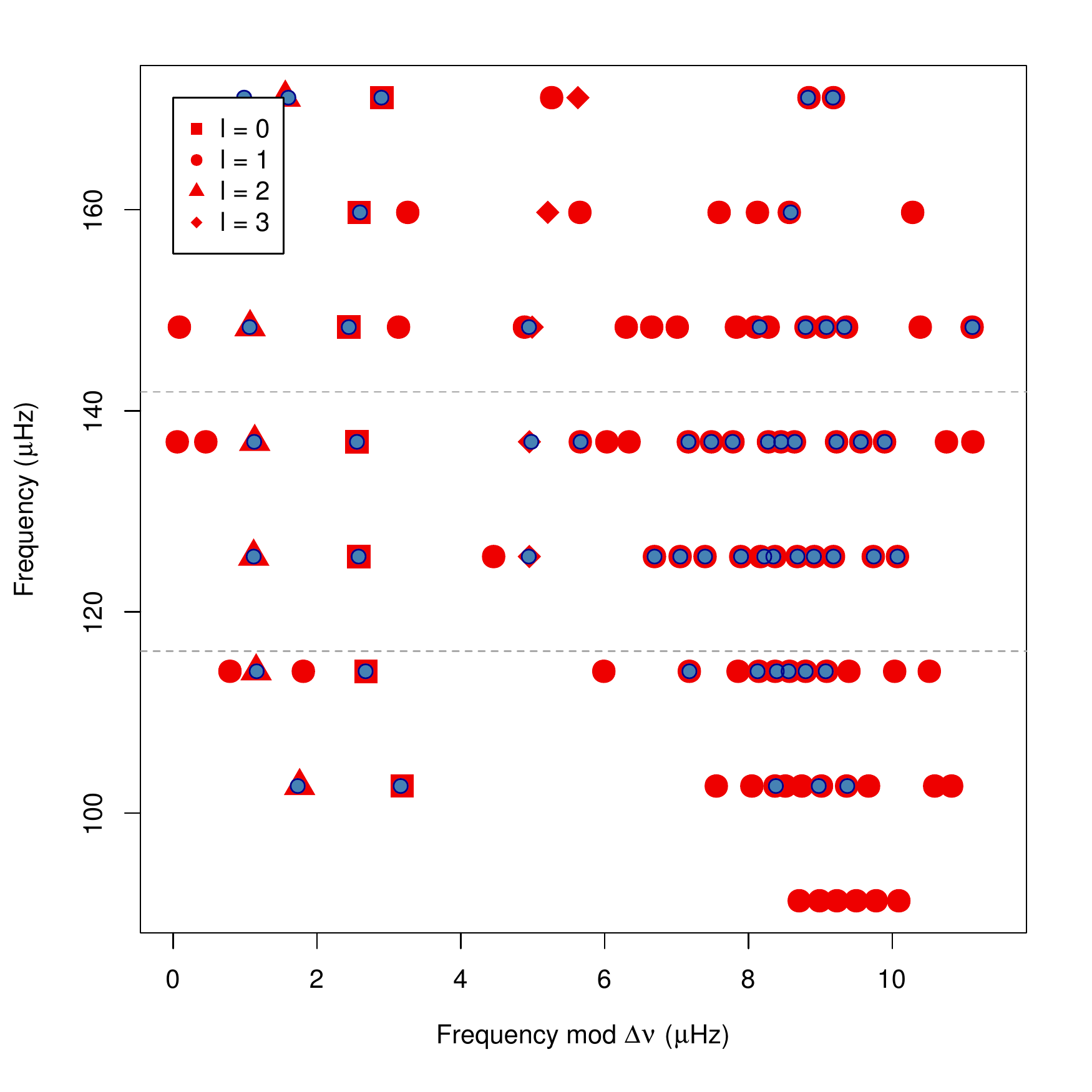}\caption{Same as Fig. \protect\ref{fig:KIC012008916echelle} for KIC 11968334.}\label{fig:KIC011968334echelle}\end{figure}\clearpage
\begin{table}
\centering
\begin{tabular}{ccccc}
  \hline
$\nu_k$ & $A_k$ & $\gamma_k$ & SNR & AIC \\ 
  \hline
120.126$\pm$0.019 & 0.758$\pm$0.117 & 0.044$\pm$0.02 & 4.17 & 38.49 \\ 
  125.802$\pm$0.017 & 0.463$\pm$0.116 & 0.025$\pm$0.018 & 2.38 & 7.31 \\ 
  126.334$\pm$0.014 & 0.529$\pm$0.12 & 0.019$\pm$0.011 & 3.60 & 13.88 \\ 
  126.573$\pm$0.017 & 0.727$\pm$0.119 & 0.035$\pm$0.015 & 4.18 & 35.40 \\ 
  127.146$\pm$0.005 & 0.416$\pm$0.182 & 0.005$\pm$0.008 & 3.71 & 16.11 \\ 
  130.754$\pm$0.051 & 0.845$\pm$0.118 & 0.132$\pm$0.052 & 3.40 & 28.24 \\ 
  132.422$\pm$0.014 & 1.328$\pm$0.15 & 0.045$\pm$0.015 & 12.10 & 191.02 \\
\hline
\end{tabular}
\caption{Resolved oscillations for KIC 12008916 .} 
\label{tab:012008916resolved}
\end{table}
\begin{table}
\centering
\begin{tabular}{ccc}
  \hline
$\nu_k$ & $H_k$ & AIC \\ 
  \hline
138.297$\pm$0.001 & 42.283$\pm$38.461 & 49.67 \\ 
  139.777$\pm$0.002 & 28.635$\pm$24.084 & 24.74 \\ 
  150.157$\pm$0.001 & 291.229$\pm$202.156 & 320.05 \\ 
  154.164$\pm$0.001 & 42.179$\pm$33.057 & 61.94 \\ 
  154.981$\pm$0.001 & 64.75$\pm$36.938 & 90.01 \\ 
  166.578$\pm$0.002 & 81.346$\pm$44.237 & 91.20 \\ 
  169.582$\pm$0.002 & 374.07$\pm$199.403 & 59.65 \\ 
  175.577$\pm$0.002 & 161.837$\pm$76.891 & 128.42 \\ 
  176.347$\pm$0.002 & 24.322$\pm$23.96 & 5.27 \\ 
  178.679$\pm$0.002 & 108.316$\pm$68.092 & 67.41 \\ 
   \hline
\end{tabular}
\caption{Unresolved oscillations for KIC 12008916 .} 
\label{tab:012008916unresolved}
\end{table}
\begin{figure}\includegraphics[width=0.47\textwidth]{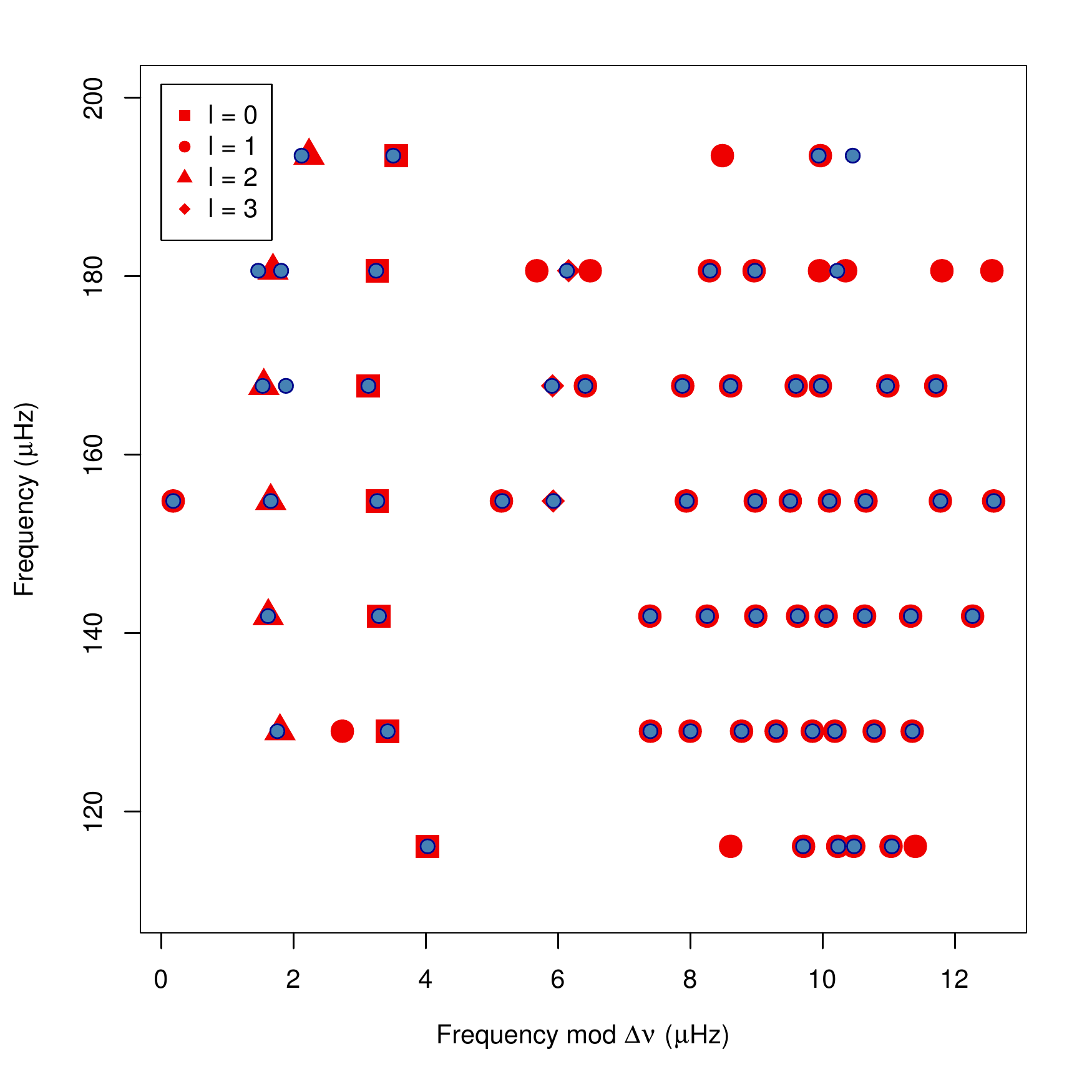}\caption{Same as Fig. \protect\ref{fig:KIC012008916echelle} repeated here for completeness.}\label{fig:KIC011968334echellebis}\end{figure}\clearpage

\bsp	
\label{lastpage}
\end{document}